\documentclass{aa}  

\usepackage{multirow}
\usepackage{graphicx}
\usepackage{pdflscape}
\usepackage{caption}
\usepackage{subcaption}

\usepackage{natbib}  
\bibpunct{(}{)}{;}{a}{}{,} 

\usepackage{longtable}
\usepackage{txfonts}
\usepackage[colorlinks,linkcolor=blue,anchorcolor=blue,citecolor=blue]{hyperref}
\usepackage{xcolor, soul}
\sethlcolor{green}
\usepackage{placeins}

\newcommand{\ros}{ROSAT\xspace}
\newcommand{\xmm}{XMM-Newton\xspace}
\newcommand{\swift}{Swift\xspace}
\newcommand{\mission}[1]{\textit{#1}}

\begin{document}

   \title{eRO-ExTra: eROSITA extragalactic non-AGN X-ray transients and variables in eRASS1 and eRASS2}

   \author{I. Grotova\inst{\ref{inst1}}
          \and A. Rau\inst{\ref{inst1}}
          \and M. Salvato\inst{\ref{inst1}}
          \and J. Buchner\inst{\ref{inst1}}
          \and A. J. Goodwin\inst{\ref{inst2}}
          \and Z. Liu\inst{\ref{inst1}}
          \and A. Malyali\inst{\ref{inst1}}
          \and A. Merloni\inst{\ref{inst1}}
          \and D. Tub{\'\i}n-Arenas\inst{\ref{inst6}}
          \and D. Homan\inst{\ref{inst6}}
          \and M. Krumpe\inst{\ref{inst6}}
          \and K. Nandra\inst{\ref{inst1}}
          \and R. Shirley\inst{\ref{inst1}}
          \and G. E. Anderson\inst{\ref{inst2}}
          \and R. Arcodia\inst{\ref{inst9}}
          \and S. Bahic\inst{\ref{inst6}}
          \and P. Baldini\inst{\ref{inst1}}
          \and D. A. H. Buckley\inst{\ref{inst3},\ref{inst4},\ref{inst5}}
          \and S. Ciroi\inst{\ref{inst7}}
          \and A. Kawka\inst{\ref{inst2}}
          \and M. Masterson\inst{\ref{inst9}}
          \and J. C. A. Miller-Jones\inst{\ref{inst2}}
          \and F. Di Mille\inst{\ref{inst8}}   
        }
   \institute{Max-Planck-Institut f\"ur extraterrestrische Physik, Giessenbachstrasse 1, 85748 Garching, Germany\label{inst1}\\\email{grotova@mpe.mpg.de}
   \and Leibniz-Institut f\"ur Astrophysik Potsdam, An der Sternwarte 16, 14482 Potsdam, Germany\label{inst6}
   \and International Center for Radio Astronomy Research, Curtin University, GPO Box U1987, Perth, WA 6845, Australia\label{inst2}
   \and South African Astronomical Observatory, PO Box 9, Observatory Rd, 7935 Observatory, Cape Town, South Africa\label{inst3}
   \and Department of Astronomy, University of Cape Town, Private Bag X3, Rondebosch 7701, South Africa\label{inst4}
   \and Department of Physics, University of the Free State, PO Box 339, Bloemfontein 9300, South Africa\label{inst5}
   \and Dipartimento di fisica e Astronomia, Università degli Studi di Padova, Vicolo dell’Osservatorio 3, 35122 Padova, Italy\label{inst7}
   \and Las Campanas Observatory - Carnegie Institution for Science, Colina el Pino, Casilla 601, La Serena, Chile \label{inst8}
   \and MIT Kavli Institute for Astrophysics and Space Research, 70 Vassar Street, Cambridge, MA 02139, USA \label{inst9}
    }

    \date{Received September 15, 1996; accepted March 16, 1997}

 
  \abstract
   {}
   {The eROSITA telescope aboard the {\it Spectrum Roentgen Gamma} (SRG) satellite provides an unprecedented opportunity to explore the transient and variable extragalactic X-ray sky due to the sensitivity, sky coverage, and cadence of the all-sky survey. While previous studies showed the dominance of regular active galactic nuclei (AGN) variability, a small fraction of sources expected in such a survey arise from more exotic phenomena such as tidal disruption events (TDEs), quasi-periodic eruptions, or other short-lived events associated with supermassive black hole accretion. This paper describes the systematic selection of X-ray extragalactic transients found in the first two eROSITA all-sky surveys (eRASS) that are not associated with known AGN prior to eROSITA observations.
   }
   {We generated a variability sample using the data from the first and second eRASS, which includes sources with a variability significance and a fractional amplitude larger than four in the 0.2--2.3 keV energy band. The sources were discovered between December 2019 and December 2020, and are located in the Legacy Survey DR10 (LS10) footprint. When possible, transients were associated with optical LS10 counterparts. The properties of these counterparts were used to exclude stars and known active galaxies. The sample was additionally cleaned from known AGN using pre-eROSITA SIMBAD and the Million Quasars Catalog (Milliquas) classifications, archival optical spectra, and archival X-ray data. We explored archival X-ray variability, long-term (2--2.5 years) eROSITA light curves, and peak X-ray spectra to characterize the X-ray properties of the sample. Sources with radio counterparts were identified using the Rapid ASKAP Continuum Survey (RACS) and the Karl G. Jansky Very Large Array Sky Survey (VLASS). 
   }
   {We present a catalog of 304 extragalactic eROSITA transients and variables not associated with known AGN, called eRO-ExTra. More than 90\,\% of sources are associated with reliable LS10 optical counterparts. For each source, we provide archival X-ray data from {\it \swift}, {\it \ros}, and {\it \xmm}; the eROSITA long-term light curve with a light curve classification; as well as the best power law fit spectral results at the peak eROSITA epoch. Reliable spectroscopic and photometric redshifts, which are both archival and from follow-up data, are provided for more than 80\,\% of the sample. Several sources in the catalog are known TDE candidates discovered by eROSITA. In addition, 31 sources are radio detected. The origin of radio emission needs to be further identified.
   }
   {The eRO-ExTra transients constitute a relatively clean parent sample of non-AGN variability phenomena associated with massive black holes. The eRO-ExTra catalog includes more than 95\,\% of sources discovered in X-rays with eROSITA for the first time, which makes it a valuable resource for studying unique nuclear transients.}


   \maketitle

\section{Introduction}
The extended ROentgen Survey with an Imaging Telescope Array (eROSITA; \citealt{predehl_2021}) X-ray telescope was launched in 2019 on board the Russian/German \mission{Spektrum Roentgen Gamma} mission \citep{2021A&A...656A.132S}. During its operational phase, eROSITA significantly increased the number of known X-ray sources, with nearly 930,000 sources detected in the German half of the eROSITA sky (eROSITA\_DE; $359.9442 ^{\circ}>l> 179.9442 ^{\circ}$) during the first all-sky survey (eRASS1) and presented in the first data release (DR1) catalog \citep{Merloni_2024}. Before entering safe mode in February 2022, eROSITA completed a total of 4.3 all-sky surveys (eRASS1--5), each lasting six months, providing a valuable resource to explore the variable and transient X-ray sky over a period of 2.5 years. 

The X-ray sky harbors a wide variety of variable and transient sources. In our Galaxy and nearby galaxies, the detected populations predominantly consist of stellar phenomena and compact objects, such as flaring coronal stars, cataclysmic variables, and X-ray binaries. The X-ray extragalactic variable and transient sky is dominated by events associated with accretion onto supermassive black holes (SMBHs). Perturbations in the prolonged accretion flow in active galactic nuclei (AGN) \citep{hawkins2002,noda2018}, corona instabilities \citep{1998MNRAS.299L..15D,2015MNRAS.449..129W,2020ApJ...898L...1R}, and obscuration \citep{2002ApJ...571..234R,matt_2003,2024A&A...684A.101M} can lead to variability on a wide range of timescales and amplitudes. In short-term accretion events such as tidal disruption events (TDEs), on the other hand, the X-ray transient is thought to be the result of a star being disrupted in the gravitational potential of a SMBH \citep{1975Natur.254..295H,1988Natur.333..523R,1999A&A...343..775K,2020SSRv..216...85S}. In some cases a star is only partially disrupted, called a partial TDE (pTDE), resulting in repeating flares on scales from months \citep{Liu_2023} to decades \citep{malyali_rep}. Most recently, quasi-periodic eruptions (QPEs) have been added to the list \citep{2019Natur.573..381M,2020A&A...636L...2G,Arcodia_2021,2024A&A...684A..64A}. The origin of these quasi-periodic X-ray flares with timescales from hours to days is still debated, but an observational link to TDEs has been found in some events \citep{2021Chakraborty,2023miniuttu,2023Quintin,Evans_2023,2024A&A...684A..64A}.

Systematic studies of the variable X-ray sky have been performed previously. For example, \citet{Fuhrmeister} classified 1207 variable sources in the \mission{ROSAT} all-sky survey. Long-term variability studies of large samples of AGN were performed in {\it Chandra} Deep field-South \citep{10.1093/mnras/stx1761} and in the COSMOS field with \mission{XMM-Newton} \citep{2014ApJ...781..105L}. Unsurprisingly, in the latter study the majority of the variable sources in this deep extra-galactic field were associated with AGN, based either on their spectral energy distribution or spectral classification. It was shown that most AGN exhibit variability in their X-ray emission. The first systematic look at the variable X-ray sky with eROSITA was performed in the 140-square-degree final Equatorial-Depth Survey (eFEDS) field by \cite{boller_efeds}. Having explored variability on the timescale of days, 65 significantly varying sources were identified, most of which are consistent with originating from Galactic stellar flares or variable AGN. The continuation of this study for the whole eROSITA\_DE sky was performed in \citet{boller2024erosita} for eRASS1 and they identified more than a thousand variable sources. Similarly to the eFEDS field, the majority of objects were associated with flaring stars, and 10~\% of sources were associated with AGN. The methodology of characterizing eROSITA X-ray variability close to survey poles is also discussed in \citet{2024arXiv240117278B}.

In this paper, we present a systematically selected sample of extragalactic X-ray transients and variables discovered in eROSITA-DE during the first two all-sky surveys (eRASS1: 12 December 2019--11 June 2020 and eRASS2: 11 June 2020--14 December 2020). We specifically focus on sources that are not associated with known AGN prior to the detection of X-ray variability with eROSITA, do not show AGN-typical multiwavelength properties, and are not associated with X-ray binaries in nearby, well-resolved galaxies. However, there might still be a fraction of AGN in the sample, in particular, low-luminosity AGN, about which previous knowledge was not available. Based on the selection criteria, sampled volume, luminosities, and timescales probed by the eROSITA data, the majority of the selected events are expected to originate from accretion events onto otherwise inactive SMBHs in the centers of galaxies. eROSITA's variable AGN population will be published in separate papers.

The paper is structured as follows: in Sect.~\ref{sec:selection_all} we present the selection of the eRO-ExTra catalog, including the primary selection of the eROSITA X-ray variability catalog, the optical counterpart identification and exclusion of galactic and AGN contaminants.
In Sect.~\ref{sec:xrayanalysis} we present the eROSITA light curve and X-ray archival data analysis, X-ray spectral modeling and redshift compilation. We report on radio properties of the catalog in Sect.~\ref{sec:radio} and several included sources detected as transients at other wavelengths in Sect.~\ref{sec:tns}. We conclude with a discussion in Sect.~\ref{sec:discussion} and a summary in Sect.~\ref{sec:summary}. We adopt a flat $\mathrm{\Lambda}$CDM cosmology throughout this paper, with $H_{0}=67.7$\,km s$^{-1}$ Mpc$^{-1}$ and $\Omega_m=0.309$ \citep{2016A&A...594A..13P}. Most uncertainties quoted in the paper are based on 68\,\% confidence intervals.

\section{The selection of extragalactic transients and variables }
\label{sec:selection_all}
\subsection{The eRASS1-- eRASS2 X-ray variability sample}
\label{sec:varcat}

\begin{figure}
  \includegraphics[width=\linewidth]{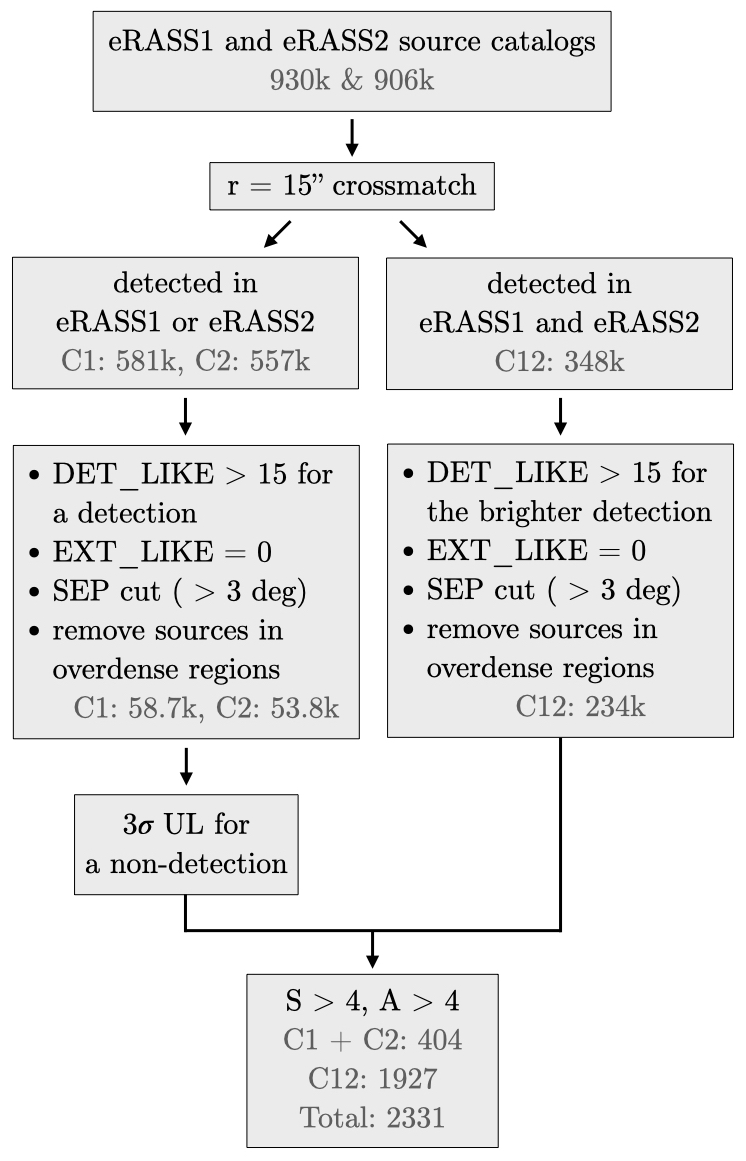}
  \caption{eRASS1-eRASS2 variability sample selection steps. First, the eRASS1 and eRASS2 source catalogs are crossmatched to identify sources detected in both (C12) or only in one eRASS (C1, C2). Then, cuts are applied to exclude spurious and extended sources. Next, $3\sigma$ upper limits for nondetections are calculated for C1 and C2 subsamples. Finally, the variability significance, $S$, and fractional amplitude, $A$,  are computed. In total, 2331 sources (C1+C2: 404; C12: 1927) with $S>4$ and $A>4$ remain.}
  \label{fig:varcat}
\end{figure}

To build the sample of extragalactic transients, we first compiled the sample of variable sources selected based on the amplitude and the significance of the variability between eRASS1 and eRASS2. This allows the selection of all sources that brightened or faded significantly between eRASS1 and eRASS2. The flowchart of the included selection steps is presented in Fig.~\ref{fig:varcat} and described in more detail in the following.

\begin{figure*}
\centering
    \includegraphics[width=\linewidth]{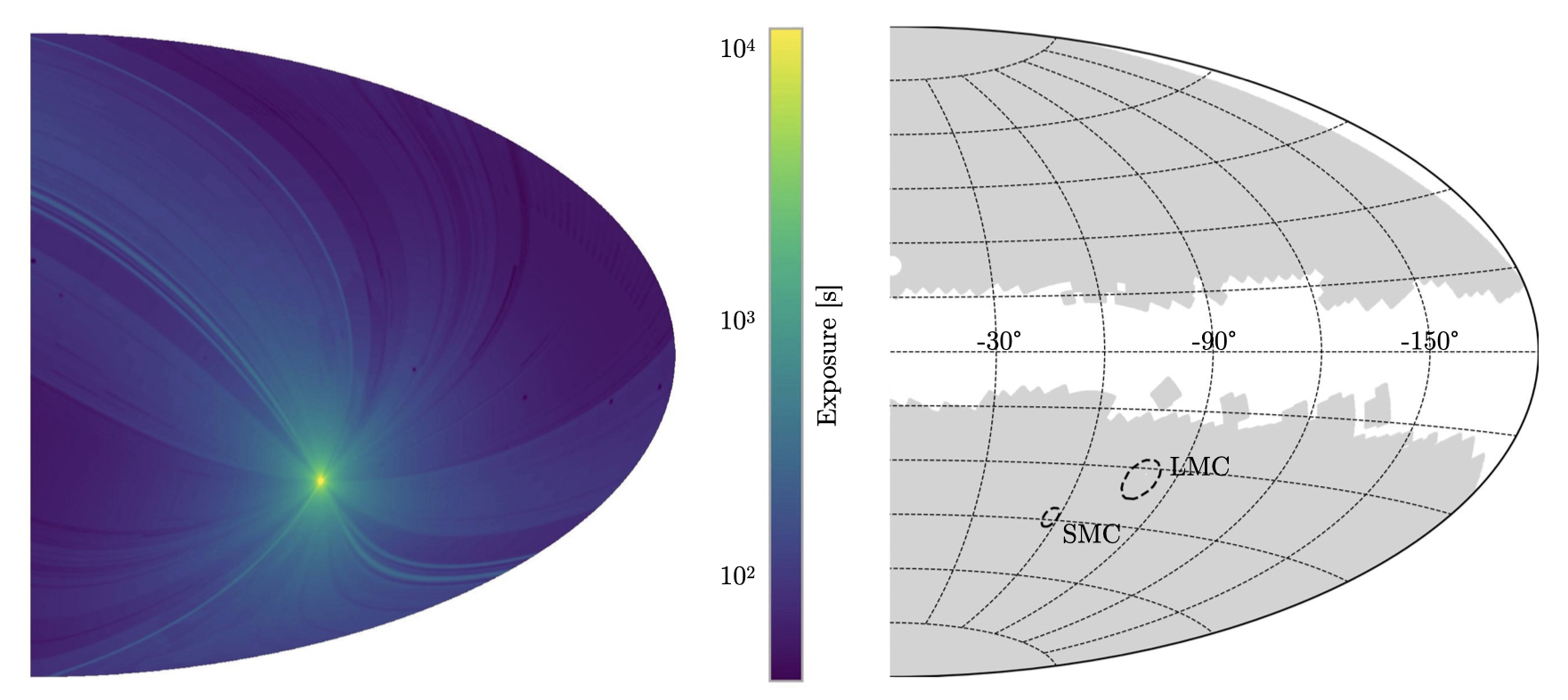}  

\caption{Representation of eROSITA\_DE sky in Aitoff projection in Galactic coordinates. The left panel shows the (vignetted) effective eRASS1 exposure map with values ranging from $\approx100$\,s (blue) at the ecliptic equator to more than 10,000\,s (yellow) close to the ecliptic pole. The right panel shows the LS10 footprint in the same projection and coordinates, covering 76\,\% of eROSITA\_DE. The locations of Large Magellanic Cloud (LMC) and Small Magellanic Cloud (SMC) are shown for illustrative purposes.}
\label{fig:hemispheres1}
\end{figure*}

First, we crossmatched the eRASS1 \citep{Merloni_2024} and
eRASS2 (unpublished, 1B: 0.2--2.3 keV, ver.221031, includes sources with $\mathrm{DET\_LIKE>6}$) source catalogs with a radius of 15\arcsec\ and determined subsamples of sources which were detected in both eRASS (C12), only in eRASS1 (C1), or only in eRASS2 (C2). The eRASS2 catalog was processed with an updated version of the eROSITA standard data processing pipeline (ver.020). Comparing to that used for the DR1 \citep[][ver.010]{Merloni_2024}, the new data processing version includes updates on the pattern and energy tasks, an improved boresight correction, a change in flaregti parameters, an improved low-energy detector noise suppression, and a better computation of the subpixel position. Subsequently, we applied a detection likelihood cut $\mathrm{DET\_LIKE>15}$ on the brighter detection of a source in either eRASS1 or eRASS2 to minimize spurious events. The chosen threshold corresponds to $\approx 0.03\%$ of false positives \citep[see][Fig.~3]{seppi2022}. Extended sources were removed using the extension likelihood cut $\mathrm{EXT\_LIKE=0}$. The comprehensive description of the source detection parameters is provided in \citet{brunner22}. The crossmatching radius was chosen based on eROSITA's typical positional uncertainty \citep[see][Fig.~5]{Merloni_2024}. Provided that a brighter detection has a detection likelihood of at least $\mathrm{DET\_LIKE = 15}$, and that of a fainter detection can be as low as $\mathrm{DET\_LIKE = 6}$, the chosen radius accounts for the combined positional error of more than 90\,\% of point sources. Since we searched for the closest match, we checked if there are cases in C12 where the crossmatch would have multiple counterparts within 15\arcsec\ and found 98/348k sources. After applying the selection cuts (see Fig.~\ref{fig:varcat}), there remained only four potential spurious matches, for which we manually confirmed that the closest match is correct.

eROSITA's scanning strategy \citep{Merloni_2024} results in a nonuniform exposure across the sky (see Fig. \ref{fig:hemispheres1}, left) with the highest exposure being reached near the ecliptic poles and the lowest near the ecliptic equator. For this study, we removed sources within 3\degr\ from the South Ecliptic Pole (SEP). At the SEP, the significantly larger exposure (up to 30\,ks in each eRASS compared to $\sim100$\,s near the equator) and depth, leading to high source density and source confusion, require a very different data analysis from the rest of the sky. Results of the variability analysis of the SEP are presented in \citet{2024arXiv240117278B}. To reduce the number of spurious sources, we also excluded regions in selected overdensities in the eRASS source catalogs. This includes, for example, regions near supernova remnants, very bright point sources, stellar clusters, and large local galaxies. The summary of the flagging procedure is provided in \citet{Merloni_2024} in Table 5. Regions around nearby galaxy clusters were not excluded, since eROSITA’s resolving power is sufficient to distinguish between X-ray point sources associated with individual galaxies and diffuse cluster emission, and cluster member galaxies can host extragalactic transients and variables.

For sources in the C1 and C2 samples, we calculated the 3$\sigma$ upper limits for the nondetections at the X-ray detection positions in the 0.2--2.3\,keV band, assuming the same spectral model as used in the eRASS catalogs (absorbed power law with $\Gamma = 2$, N$_{\rm H}=3\times10^{20}$\,cm$^{-2}$). The details of the upper limit computation can be found in Appendix~\ref{appendix_ul}.

To characterize the variability, we determined the fractional amplitude, $\mathrm A$, and significance of the variability $\mathrm{S}$ for each source as:

\begin{equation}
\label{eq:a}
\mathrm{A = F_\mathrm{max}/F_\mathrm{min}}
\end{equation}
\begin{equation}
\label{eq:sigma}
\mathrm{S = \frac{F_{\rm max}-F_{\rm min}}{\sqrt{\mathrm{F\_ERR_\mathrm{max}^2+F\_ERR_\mathrm{min}^2}}}},
\end{equation}

where $F_{\rm max}$ and $F_{\rm min}$ are the maximum and minimum fluxes, corresponding to the integrated fluxes either in eRASS1 or eRASS2, and $F\_ERR\_\mathrm{max}$ and $F\_ERR\_\mathrm{min}$ are their 1$\sigma$ symmetric errors. For detections we used the catalog fluxes and their uncertainties. For sources for which the catalogs did not include error estimates (see the discussion in \citealt{Merloni_2024} for details), we assigned the median of the count error of all sources with counts and exposure time within 10\,\% of those of the source. This threshold provided at least 20 similar sources for each error estimation, with the exception of one high-count source, for which we assumed a $\mathrm{\sqrt{N}}$ error. To calculate the flux uncertainties in both cases, count rates were converted to fluxes using the standard catalog energy conversion factor ($\mathrm{ECF = 1.074\times10^{12}}$). For the nondetections in C1 and C2, $F_{\rm min}$ is the 3$\sigma$ upper flux limit and $\mathrm{F\_ERR\_\mathrm{min}=0}$. The final sample includes  2331 variable sources (C12: 1927; C1+C2: 404), which vary with $\mathrm{A>4}$ and $\mathrm{S>4}$, where 6\,\% (i.e., 147) have both amplitude and significance higher than 10. 

\subsection{Counterpart identification}
\label{sec:extra_cat}

\begin{figure}
  \centering
   \includegraphics[width=\linewidth]{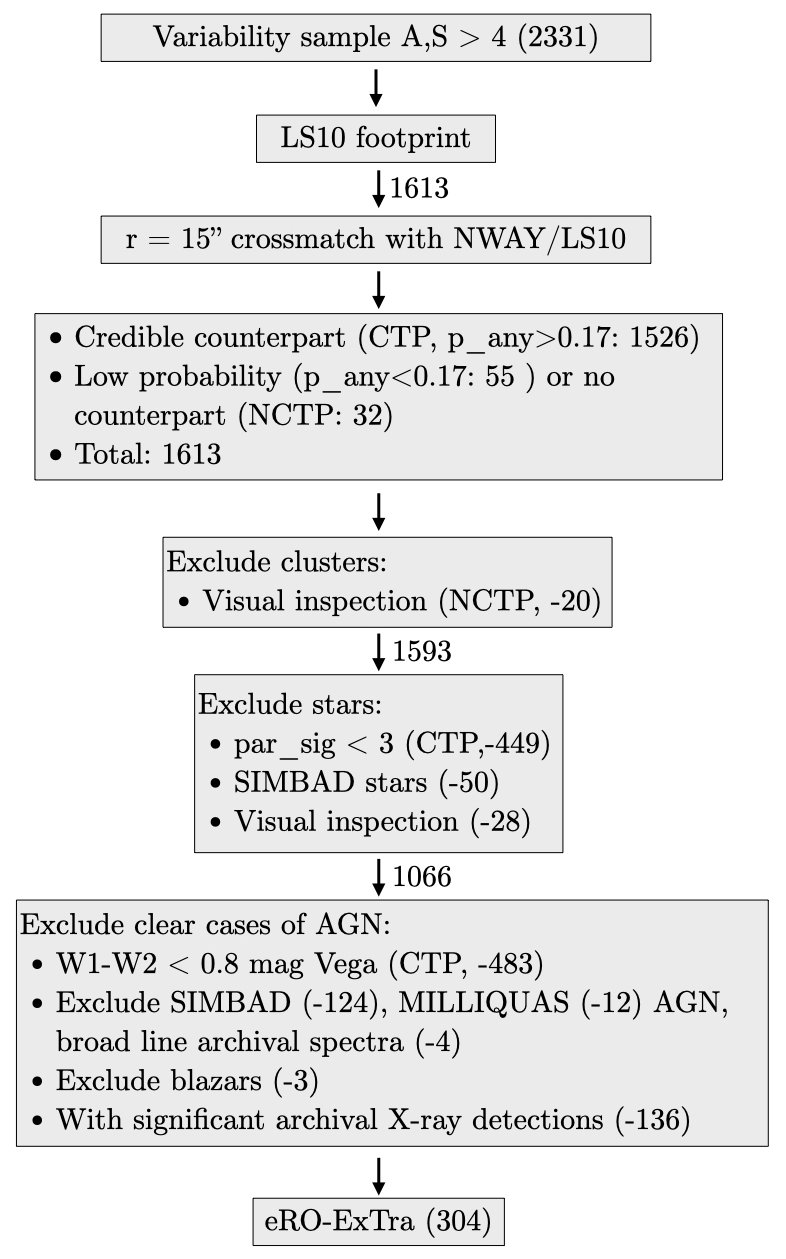}
  \caption{Selection of extragalactic transients from variability sample. Firstly, we assign NWAY LS10 optical counterparts: 1526 have a counterpart within 15\arcsec and 87 are not associated with an NWAY/LS10 optical source or do not have a reliable counterpart. Then we apply numerous cuts to exclude contaminating clusters, stars and known pre-eROSITA AGN. The final catalog of Extragalactic X-ray transients (eRO-ExTra) includes 304 sources (93\,\% with a reliable counterpart).}

  \label{fig:selection_flowchart}
\end{figure}

An important step to determine the nature of an X-ray transient is to identify its multiwavelength counterpart. In this section, we describe the criteria by which optical counterparts were associated with the X-ray detections. The flowchart of the selection steps is shown in Fig.~\ref{fig:selection_flowchart}. 

For the identification of the optical and near-infrared counterparts, we concentrated on the Legacy Survey DR10 area (LS10\footnote{\url{https://www.legacysurvey.org/dr10/}}; \citealt{2019AJ....157..168D}), since it provides extensive coverage of the eROSITA\_DE sky (76\,\%) except in the Galactic Plane and several other small sky regions (see Fig. \ref{fig:hemispheres1}, right). From the 2331 sources in the variability sample, 1613 ($\approx70$\,\%) are within the LS10 footprint. Sources not covered by the LS10 footprint are primarily concentrated in the Galactic Plane ($\mathrm{|b| \lessapprox 15 ^{\circ}}$) and were excluded from further study. As variability for sources in the Galactic Plane is heavily dominated by stellar phenomena, the exclusion of this region is expected to have minor impact on identifying the extragalactic transient and variable population presented in this work.

\begin{figure}
  \includegraphics[width=\linewidth]{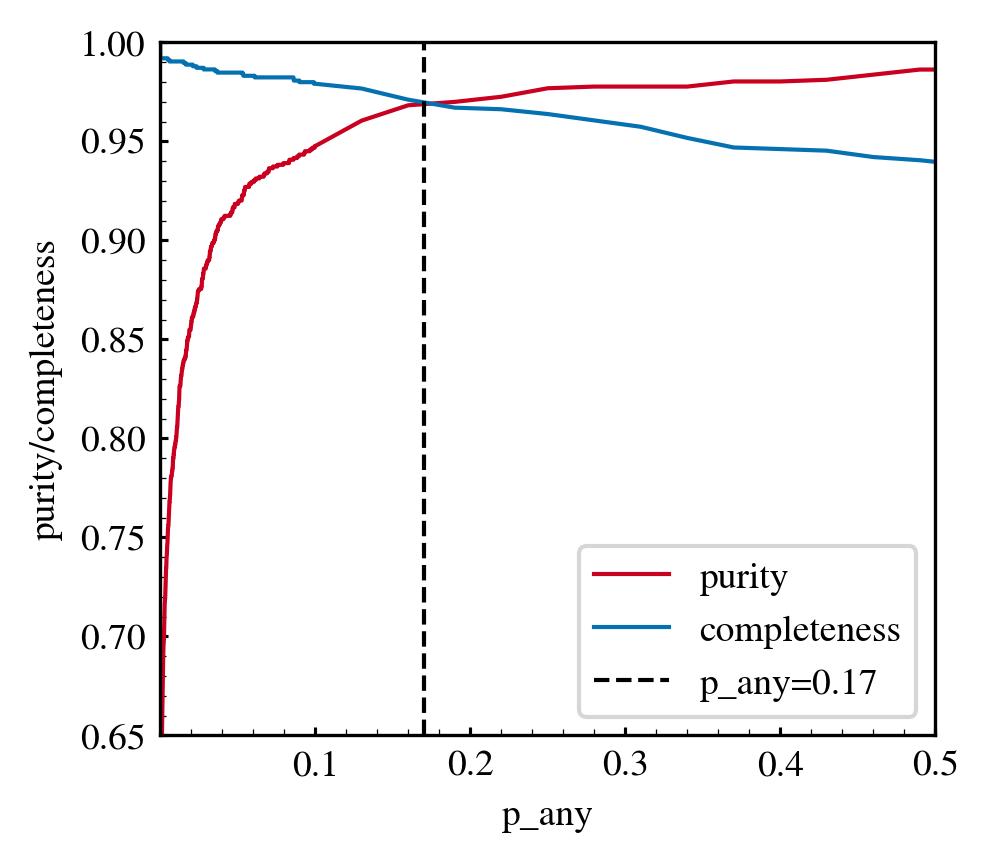}
  \caption{Functions of purity and completeness vs. \texttt{p\_any}. The functions were calculated for our sample after the NWAY match, given detection likelihood values of $\mathrm{DET\_LIKE>15}$. The value of \texttt{p\_any}=0.17 at the intersection of purity and completeness functions at 0.97 is chosen as the optimal minimum probability value for the selection of reliable NWAY counterparts for eRO-ExTra.}
  \label{fig:pany_choice}
\end{figure}

LS10 provides deep and homogenized photometry in the {\it g, r, i, z} and WISE W[1,2,3,4] bands. For sources located in areas within the LS10 mission footprint but where catalog data is not yet available in this data release, we used Gaia DR3 \citep{gaia2016,gaia2023} and CatWise2020 \citep{catwise2020} independently to replace missing photometry results. However, we note that Gaia (G$<$20\,mag) is shallower than LS10 and is biased against extended sources, such as nearby galaxies. While Catwise2020 is deeper than the LS10 W[1,2,3,4] bands (six times as many exposures), it suffers from blending.
The counterparts were identified using NWAY \citep{Salvato2018} by applying the procedure as described for the eROSITA eFEDS field \citep[see][Sect.~3.1]{Salvato_2022}. Specifically, the geometrical distribution of sources (i.e., the distance from the X-ray position and positional uncertainty, the number density, and the similarity of their SEDs compared to those in a training sample) are taken into account. For details on the priors for LS10, Gaia DR3, and CatWise2020 and the agreement between the three surveys see Salvato et al. (in prep.), where the counterpart identification process for eRASS1 is described. The training sample was not specifically selected for the counterpart identification of X-ray transients and variables, but rather for typical X-ray emitters, due to the lack of previously discovered sources with suitable properties. The most complete to-date version of the NWAY/LS10 counterpart catalog for eROSITA is eRASS:5, a not-yet-published stacked source catalog for all eRASS. We associated NWAY/LS10 counterparts to the sources in the variability sample by crossmatching the X-ray position of the brighter detection in eRASS1 or eRASS2 with the eRASS:5 X-ray position with a radius of $\mathrm{r = 15}$\arcsec. 

Of the 1613 sources within the LS10 footprint, 1581 ($\approx98$\,\%) have at least one possible counterpart in the NWAY catalog within 15\arcsec. The probability that for a given X-ray source one or more counterparts with properties typical for an X-ray emitter were found, is defined by the  \texttt{\texttt{p\_any}} parameter \citep{Salvato2018}. The optimal minimal value of \texttt{p\_any}=0.17 was determined as the intersection of purity and completeness functions calculated for various \texttt{p\_any} for our sample after the NWAY match. Henceforth, the counterparts with \texttt{p\_any}>0.17 are considered reliable. As shown in Fig.~\ref{fig:pany_choice}, the threshold of \texttt{p\_any}>0.17 corresponds to chance coincidences of $<5$\,\%. There are 1526 ($97\,\%$) sources with a counterpart with \texttt{p\_any}$>$0.17; the remaining 55 sources ($3\,\%$) have a potential counterpart with a lower probability (0$<$\texttt{p\_any}$<$0.17). The \texttt{p\_any} values are provided in our final catalog, and users are advised to consider the association probability for their analysis. For 32 sources, no NWAY counterpart was found within $15^{\prime\prime}$ from the X-ray position. In addition, NWAY identified multiple counterparts in LS10 for 14/1581 sources. Multiple potential counterparts were kept during all further selection steps. For simplicity, we cited the numbers of unique X-ray sources in the description of the selection flow and in Fig.~\ref{fig:selection_flowchart}. After all cuts were applied, 11/14 X-ray sources were excluded as either stars or AGN (see Sect.~\ref{sec:exclude_stars} and Sect.~\ref{sec:exclude_agn}). For another source, one LS10 counterpart was excluded as an AGN and another one remained in the catalog. For two remaining sources, multiple matches were associated with LS10 duplicates, which pointed to the same galaxy. In these cases, the closest match was selected for the catalog. 

Finally, we visually inspected the X-ray images of all sources without an NWAY/LS10 counterpart. Here, 20/32 sources were removed from the sample by their association with extended X-ray cluster emission. These sources were not excluded by the EXT\_LIKE cut earlier, since the extended emission is clearly identified in the compiled 4(5) eRASS only. Additionally, three sources were manually identified with NWAY/LS10 counterparts with \texttt{p\_any}$>$0.17 using a larger crossmatching radius ($\mathrm{16\arcsec<r<20\arcsec}$). After the exclusion of clusters, the sample included 1593 unique X-ray sources.

\subsection{Exclusion of galactic objects}
\label{sec:exclude_stars}
To further reduce the contamination of the sample by Galactic objects, we made use of (1) the parallax significance estimates from Gaia DR3 (if \texttt{p\_any}>0.17) and (2) available classifications provided in the SIMBAD database \citep{wenger_simbad}. The summary is shown in the sixth box in Fig.~\ref{fig:selection_flowchart}.  First, the parallax significance cut of $\mathrm{par\_sig<3}$ was applied, where $\mathrm{par\_sig = parallax\times \sqrt{\mathrm{parallax\_ivar}}}$ with parallax\_ivar being the inverse-variance of the parallax. This cut removed 449 sources.

Next, we excluded 50 known stars by crossmatching the remaining sources with SIMBAD, which contains names, positions, and known properties of more than 16 million objects and provides their classifications. For sources with reliable NWAY/LS10 counterparts (\texttt{p\_any}$>$0.17), we
searched around the LS10 counterpart position for SIMBAD sources, otherwise (\texttt{p\_any}$<$0.17) around the X-ray position. In both cases, if a star
is present in SIMBAD within 15\arcsec, we excluded the sources. Finally, we visually inspected the LS10 images and excluded 28 obvious stellar contaminants such as saturated bright stars and sources in densely populated stellar fields. One of these sources was identified as a Galactic nova in a fireball phase \citep[][MGAB-V207]{2022Natur.605..248K}. In total, additional 527 ( $\approx30\%$) sources were removed from the eRASS1-eRASS2 variability sample as likely to be stars (see Fig. \ref{fig:selection_flowchart} for step-by-step numbers). At this step 1066 variable X-ray sources were left in the sample.

\subsection{Exclusion of AGN}
\label{sec:exclude_agn}
A large contribution to the extragalactic X-ray variable sky comes from AGN. In this section, we describe the identification of sources with properties matching those of regular AGN and how they were excluded. The applied selection criteria are shown in the second to last step of the flowchart in Fig.~\ref{fig:selection_flowchart}. 

Firstly, we removed sources with major AGN contribution by applying a mid-IR color ($W1-W2$<$0.8$\,mag$_{\rm Vega}$) cut from \cite{stern2012} for sources with reliable counterparts. This discarded 483 sources ($\approx30\,\%$ of all sources in LS10 footprint).

Secondly, we excluded sources classified as active galaxies in SIMBAD (124 sources), using the crossmatch described in Sect.~\ref{sec:extra_cat}. In addition, The Million Quasars Catalog (Milliquas; \citealt{flesch2023million}) helped us to identify more contaminating AGN and QSOs in the sample. Milliquas contains 907,144 type-I QSOs and AGN, largely complete from the literature to 30 June 2023. We crossmatched our catalog and Milliquas using the same approach as for the SIMBAD catalog and excluded additional 12 AGN. To ensure the non-AGN nature of sources based on archival pre-flare data, we also visually inspected available archival optical spectra and excluded four sources which show characteristics of broad-line AGN. Finally, we crossmatched our sample with the eROSITA blazar catalog (Hämmerich in prep.\footnote{The eRASS1 blazar catalog includes known sources from the 4FGL \citep{ballet2024}, ROMA BZCAT \citep{2016yCat.7274....0M}, 3HSP \citep{2019A&A...632A..77C}, HighZ (Sbarrato et al., in prep), Milliquas \citep{flesch2023million}, KDEBLLACS and WIBRaLS2 \citep{2019ApJS..242....4D}, ABC \citep{2020yCat..36410062P}, and BROS \citep{2021yCat..19010003I} catalogs, found within 8\arcsec from the eROSITA position.}) and excluded three sources. 

\subsection{Historical variability from X-ray archives}

\label{sec:archival_xray}

Archival X-ray observations helped us further to exclude possible AGN which are not yet removed in previous steps, in particular, sources with significant archival X-ray detections, since their flaring is likely caused by long-term AGN variability. To characterize the history of the eROSITA transients and variables, we explored the archives of previous and operating X-ray missions. We used the web-based High-Energy light curve Generator \citep[HILIGT\footnote{\url{https://xmmuls.esac.esa.int/upperlimitserver}};][]{2022A&C....3800531S,2022A&C....3800529K} to retrieve flux and upper limits values estimated via aperture photometry with the Bayesian approach of \citet{1991ApJ...374..344} for archival observations from Neil Gehrels {\it \swift} X-ray Telescope (XRT; \citealt{swiftxrt}), {\it \ros} \citep{voges1999rosat}, and for the {\it \xmm } Slew Survey \citep{Saxton_2008_xmmslew}. For the flux estimation in the energy band 0.2-2.0\,keV we chose a uniform spectral model of an absorbed power law with $\Gamma=2$ (same as in the eROSITA source catalogs), and a confidence interval (CL) of 99.87\,\%, corresponding to a one-sided 3$\sigma$ level. 

All presented eROSITA transients and variables were covered by at least one of the above missions. Only upper limits estimates could be derived for 296 of sources. If an upper limit, $F_{\rm ul}$, constrained the historical flux to be fainter than the peak measured by eROSITA, $F$, so that $F_{\rm ul}\leq(F - F_{\rm err})_{\rm peak}$, where $F_{\rm peak}$ and $(F_{\rm err})_{\rm peak}$ refer to the maximum flux of the eROSITA light curve and its error, the source received $\mathrm{arch\_flag=1}$ (“constraining") in our catalog. Otherwise, $\mathrm{arch\_flag=0}$, (“not constraining”). 

In total 144 sources have archival detections. These detections can also constrain whether a source shows long-term variability or the eROSITA detection marks a brightening from a previously lower flux level. Archival detections performed at least one year before the eROSITA peak indicate X-ray long-term variability if one of the following criteria is true: 1) the significance of the source variability S (Eq.~\ref{eq:sigma}) between the archival detection and the fainter of the eRASS1 or eRASS2 flux constraints is $\geq3$ ; 2) the significance of the variability S between the archival detection and the eROSITA peak is $\leq3$; 3) the archival detection is brighter than the eROSITA peak. A source that fulfills at least one of these conditions received $\mathrm{arch\_flag=-1}$ ("variable") in the catalog. Otherwise, the sources are flagged with $\mathrm{arch\_flag = 2}$ (“not variable”). The latter group includes sources with archival detections fainter or comparable with the eROSITA minimum since they might indicate the pre-flare quiescent state of the system.

Overall, 136 sources identified as variable ($\mathrm{arch\_flag=-1}$) were excluded from the extragalactic variability catalog. Whereas all of these excluded sources were in the {\it \ros} footprint, 86\,\% of sources were also covered by {\it \xmm} and 23\,\% by {\it \swift} pointed or slew observations. Although some bright archival detections might indicate a repeating transient, for example, as it was shown in \citet{malyali_rep} with a TDE candidate re-brightening on a scale of decades, a more likely explanation is long-term AGN variability. One the other hand, some AGN, which do not have significant archival detections, might have been missed in this cut due to the limitations in sensitivity and coverage of previous missions. All steps combined, 760/1068 sources were classified as AGN and were excluded to form the eRO-ExTra catalog.

A summary of the historical light curve classification is presented in Table~\ref{tab:archive_classification}. The archival fluxes with their uncertainties, instruments, and observing dates are provided in the catalog. Overall, 296/304 sources are not detected in archival observations and have only archival upper limits and, therefore, can be considered as genuinely new eROSITA transients. 

\begin{table}[ht]
    \centering
    \caption{Summary of source classification based on X-ray-archival data.}
    \label{tab:archive_classification}
    \begin{tabular}{cccc}
        \hline
        \hline
        
       \multicolumn{2}{c}{Detected} &
\multicolumn{2}{c}{Not Detected} \\
        \text{Variable} &  \text{Not variable}& \text{Constraining} & \text{Not constraining} \\
        \hline
        136 & 8 & 68 & 228 \\
        \hline
    \end{tabular}
\end{table}

\subsection{The final catalog}
The resulting catalog eRO-ExTra contains 304 transients and variables without a known pre-eROSITA AGN origin. All sources are associated with an optical LS10 counterpart. Since this study is focused on non-AGN sources, all further analysis was performed for the eRO-ExTra catalog. The summary of the catalog and its properties are discussed further in Sect.~\ref{sec:results}.

\section{X-ray light curve and spectral analysis}
 \label{sec:xrayanalysis}
\subsection{eROSITA light curves}
 \label{sec:light curves}

The eROSITA X-ray light curves were analyzed to characterize the long-term flux evolution. Each survey observes a source's position multiple times ("visits") with a cadence of $\approx 4$ hours (referred to as one "eroday"). The number of visits, hence the accumulated exposure, depends on the proximity to the survey poles. It ranges from at least six visits near the survey equator to several tens near the poles. The long-term behavior can be assessed using the data of all four complete all-sky surveys (eRASS1-eRASS4) and eRASS5, if available. With each
successive eRASS being $\approx 6$ months after the previous one, this provides a coverage of 1.5--2\,years.
 
To analyze the long-term eROSITA behavior, we first crossmatched the eRASS:5 X-ray position with the detections in the eRASS3, eRASS4, and eRASS5 (v.221031, v.221031 and v.21101, respectively, unpublished, 1B: 0.2-2.3 keV, processing version 020) source catalogs using a matching radius of r=15$^{\prime\prime}$. The eRASS:5 coordinates are used for the crossmatch since the accumulated catalog contains the most accurate X-ray positions. Upper limits were computed following the procedure described in Appendix~\ref{appendix_ul} if a source was not detected in an eRASS.

\begin{figure}[h!]
    \centering
  \includegraphics[width=0.9\linewidth]{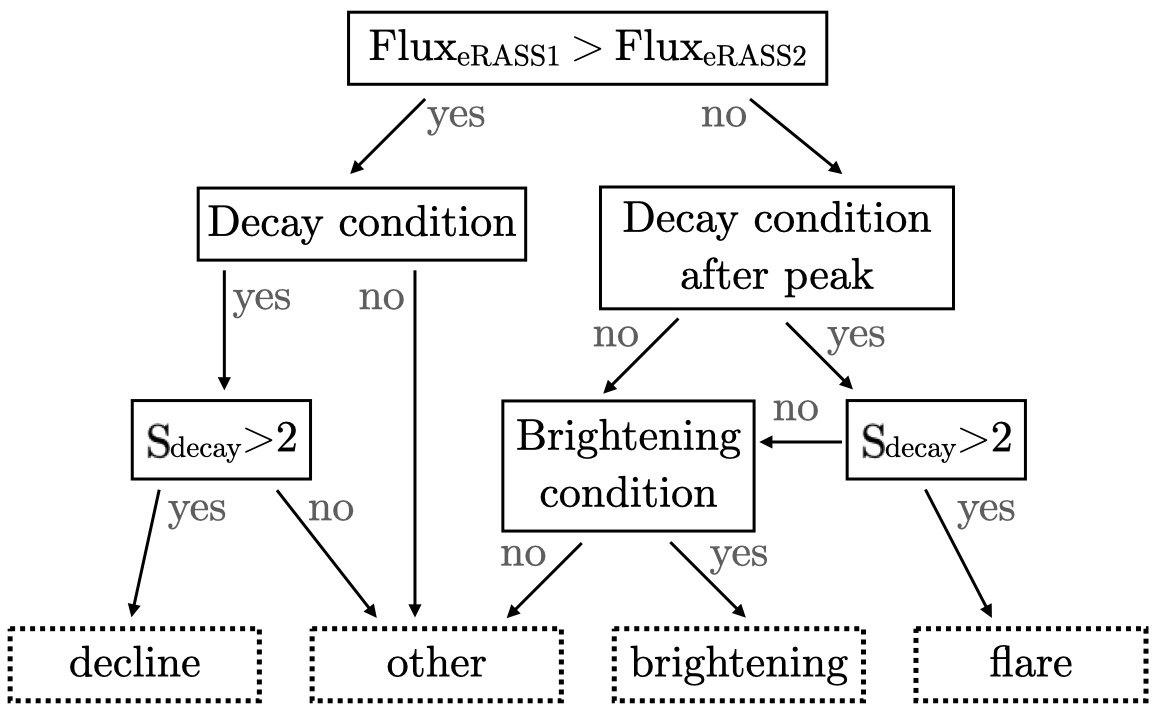}
  \caption{Decision tree of light curve classification into four classes: decline, flare, brightening, and other.}
  \label{fig:decision_tree}
\end{figure}

The light curves were then divided into four classes: "flare," "decline," "brightening," and "other," using an automated procedure considering general light curve trends and variability significance across eRASS measurements. The decision tree for the light curve classification is detailed in Fig.~\ref{fig:decision_tree}. 

 \begin{figure}[h!]
  \includegraphics[width=\linewidth]{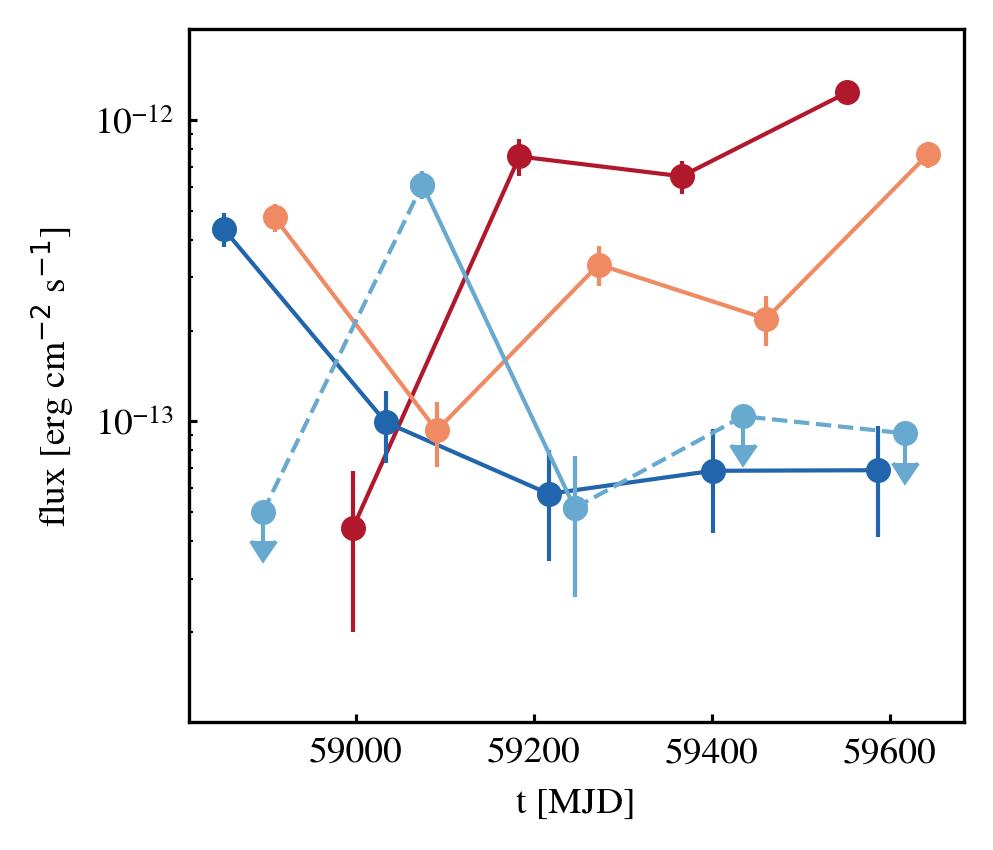}
  \caption{Light curve class examples: decline (dark blue), flare (light blue), brightening (red), and other (orange). Solid lines connect detections. Dashed lines connect to upper limits.}
  \label{fig:lc_examples}
\end{figure}

Sources peaking in eRASS1 are classified as either decline or other. The decline class includes sources, which peak in eRASS1 and decay onward. The light curve (or its part) is considered decaying if one of the following condition is satisfied (i.e., decay condition): for each pair of light curve points (m,n) after the peak, where m < n: (1) flux(m) > flux(n); or (2) flux(n) and flux(m) are consistent within 2$S$, where $S$ is defined by Eq.~\ref{eq:sigma}. Also, we require $S_{\rm decay}$, defined as the significance of the total decay from the peak to the last light curve point eRASS4(5), to be larger than two. Otherwise, the source is classified as other.

Sources peaking in eRASS2 or in later epochs belong to either brightening, flare or other light curve class. The brightening class shows a flux rise from eRASS1 to eRASS2 and then continues brightening, in agreement with the following condition (i.e., brightening condition): for each pair of light curve points (m,n), where m < n: (1) flux(m) < flux(n); or (2) flux(n) and flux(m) are consistent within 2$S$. Otherwise, the source is classified as other. The light curve has a flare if it first brightens from eRASS1 to eRASS2 and later decays. In order to have a flare class, the source should satisfy the decay condition after the peak epoch as well as have $S_{\rm decay}$>2. If the latter condition is not satisfied (in other words if light curve decay is not sufficient) the source is classified as either brightening or other, depending on the agreement with the brightening condition.

\begin{figure*}
\centering
\includegraphics[width=\linewidth]{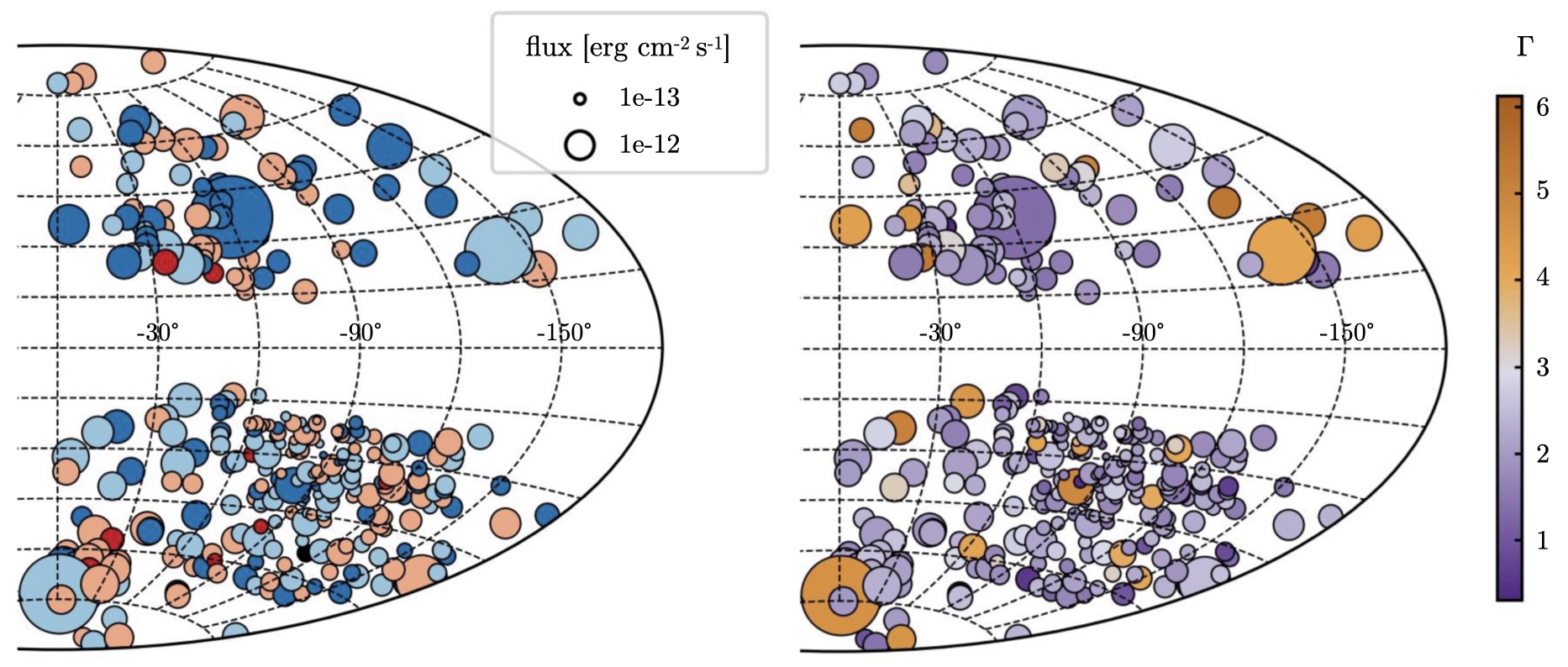}  
    
\caption{eRO-ExTra catalog in Aitoff projection in Galactic coordinates. Left: sources color coded by light curve type: dark blue - decline, light blue - flare, red - brightening, orange -  other (see Fig.~\ref{fig:lc_examples}). Right: color coded by the peak photon index from the hardest (purple) to the softest (orange). In both plots, the symbol size is scaled by the peak flux. The higher concentration of faint sources in the southern hemisphere is caused by higher exposure times near the SEP.}
\label{fig:hemispheres}
\end{figure*}

Table~\ref{tab:lcclassification} provides the light curve classification summary. Examples for each class are shown in Fig.~\ref{fig:lc_examples}. The light curve classification is included as flag LC\_type in the final catalog. The distribution of sources in the sky color coded by their light curve class and scaled by their peak eRASS1 or eRASS2 flux is shown in the left panel of Fig.~\ref{fig:hemispheres}. A noticeable concentration of sources, including faint transients, in the southern Galactic hemisphere is caused by higher eROSITA exposure times near the SEP.

In addition, we separately searched for available X-ray observations by other missions, performed slightly earlier or during eROSITA epochs. The information about these observations is included in the catalog as well. Detections within a year before the eROSITA peak may have already been associated with the start of the event later found by eROSITA: 1eRASS J004058.3-683816 and 1eRASS J132252.1+182253 were detected by \xmm Slew Survey a few days before the eROSITA peak, with the archival points being in agreement within 1.5$S$ with the eROSITA peak flux and the light curve decay trend. The archival classification of these sources ($\mathrm{arch\_flag=0}$) in Sect.~\ref{sec:archival_xray} is based on their earlier archival observations. The X-ray mission archives also contained 150 eRO-ExTra sources where detections or upper limits constrain the behavior during or after the eROSITA observations. Mostly, additional points during or after eROSITA observations are nonconstraining upper limits. In addition, several detections and constraining upper limits are consistent with the expectations from the eROSITA light curve and their classifications. Only two sources should belong to a different light curve class taking into account the additional data, namely, 1eRASS J045457.8-652846 changes from a flare to other, and 1eRASS J102851.3+251439 - from decline to flare. For consistency, the catalog includes light curve classification based on the eROSITA points only.

\begin{table}[ht]
    \centering
    \caption{Summary of the eROSITA-based light curve classification for 304 sources in eRO-ExTra.}
    \label{tab:lcclassification}
    \begin{tabular}{cccc}
        \hline
        \hline
        \text{Flare} &  \text{Decline}& \text{Brightening} & \text{Other} \\
        \hline
        101 & 79 & 8 & 116 \\
        \hline
    \end{tabular}
\end{table}

\subsection{eROSITA spectral modeling}
\label{sec:spectralmodelling}

For each source, we extracted spectra for the peak epoch using the \texttt{eSASS} task \texttt{SRCTOOL} (version eSASSusers\_211214) with circular source and annular background regions centered on the X-ray positions. Utilizing \texttt{SRCTOOL}'s AUTO mode, the radii of the source and background regions were selected taking into account the source counts, the level of the background map model at the source position, the best fitting source extent model radius, as well as a model of the effective eROSITA PSF at the source position (averaged over the integration, and summed over all operating telescope modules).
\begin{figure}[h!]
  \includegraphics[width=\linewidth]{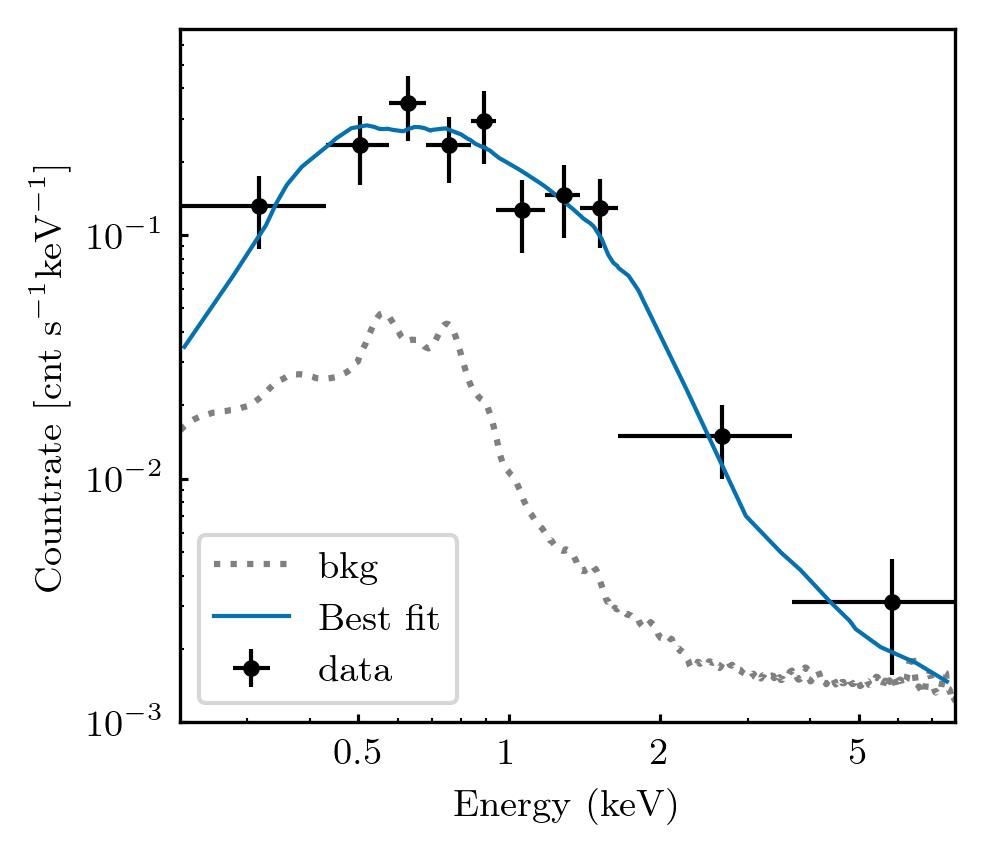}
  \caption{Example BXA fit to eRASS2 data of 1eRASS J143045.4-332705. This is a typical source from the sample with $\mathrm{A=8.2}$ and $\mathrm{S = 7}$, peak flux $\mathrm{f=5.05\times10^{-13}}$\,erg cm$^{-2}$ s$^{-1}$ and the best fit value of photon index $\mathrm{\Gamma = 1.97\pm 0.24}$. Black markers are the observed data (binned for illustration purposes only). The blue line shows the best-fit convolved model for an absorbed power law. The background model is shown as a black dotted line.}
  \label{fig:bxa_example}
\end{figure}

X-ray spectra were analyzed using the Bayesian X-ray Analysis software \citep[BXA;][]{2014A&A...564A.125B}, which connects the nested sampling algorithm UltraNest \citep{buchner2021ultranest} with the fitting environment XSPEC \citep{1996ASPC..101...17A}. The spectra were fitted unbinned using C-statistic \citep{1976cash}, and the eROSITA background was modeled using the principal component analysis (PCA) technique described in \citet{2018A&A...618A..66S} and \citet{2022A&A...661A...5L}. Each extracted eROSITA spectrum, which contains a contribution from both the source and background emission, was jointly fitted. The background spectrum was modeled with the fixed derived background model, and the source spectrum was modeled with a chosen source model convolved with the X-ray responses plus the background model convolved with a diagonal matrix response. Here, an absorbed power law model $\tt{tbabs*powerlaw}$ was used with fixed Galactic equivalent neutral hydrogen column density ($N_{\rm H,gal}$) extracted from the HI4PI survey \citep{HI4PICollaboration} individually for each source. In addition to the free parameters of the source model (photon index and normalization), the background model has a free normalization parameter, which should equal to unity in the simultaneous source and background fitting. We assumed wide flat priors on the photon index and the logarithm of the normalization. The best-fit values of the photon index gamma $\Gamma$ and the corresponding 0.2--2.3 keV fluxes were both derived as medians from the BXA posterior distributions with uncertainties calculated as 68\% percentile interval around the median (1$\sigma$). An example is shown in Fig.~\ref{fig:bxa_example}. The values of $\mathrm{N_{H,gal}}$, $\Gamma$, fluxes with errors and goodness of fit can be found in the eRO-ExTra catalog.

\begin{figure}
  \includegraphics[width=\linewidth]{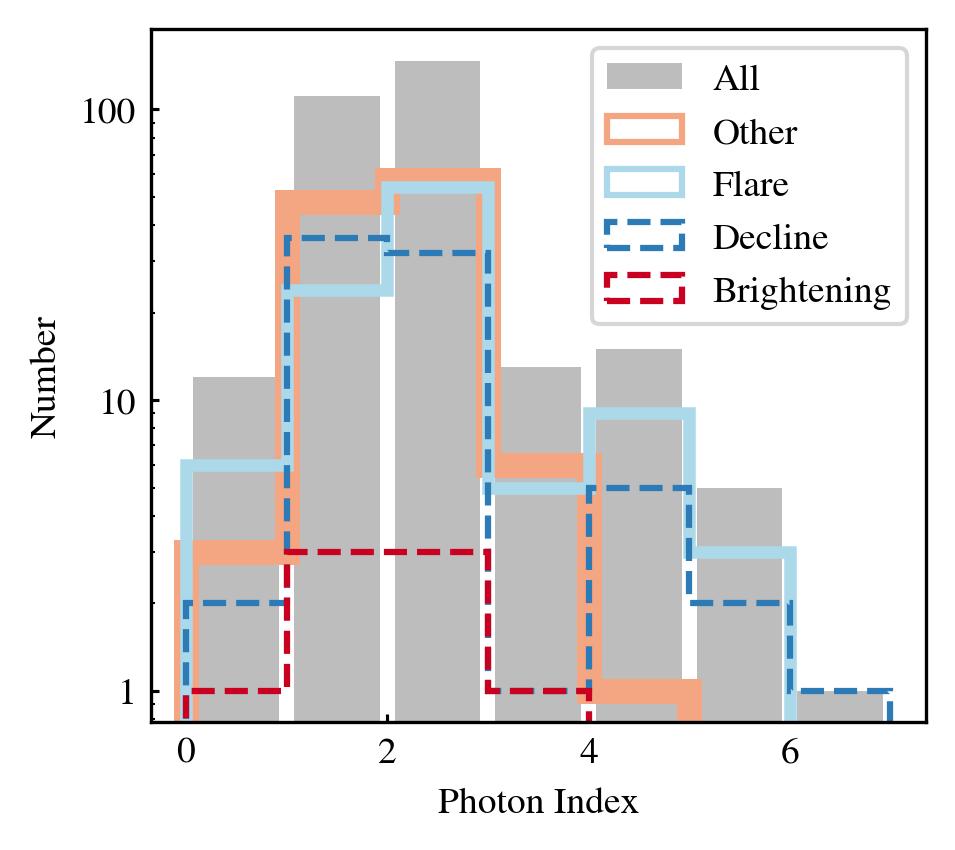}
  \caption{Distribution of best fit photon indices for peak X-ray spectrum of each source in eRO-ExTra. The different histograms show the distributions separated by light curve class.}
  \label{fig:gamma_distr}
\end{figure}

The sample includes sources with photon indices in the range $0.3<\Gamma<6.1$ (see  Fig.~\ref{fig:gamma_distr}) and a mean of $\Gamma=2.2$, dominated by 85\,\% of sources having $1<\Gamma<3$. The soft tail of the distribution includes 33 sources (11\,\%) with $\Gamma>3$, based on the mean posterior values. It is important to note that the best-fit results of the hard tail of the distribution with $\Gamma<1$ (4\,\% of sources) may be affected by the choice of the spectral model, particularly an underestimation of the intrinsic obscuration introduced by using a fixed $N_{\rm H,gal}$. However, due to the low photon counts for most sources, a more complex spectral modeling cannot be performed consistently. Similarly, the extremely steep photon index, namely, $\Gamma>6$, for one faint source (1eRASS J064449.4-603704), is due to the simple spectral model assumption. The distribution of sources in the sky color coded by $\Gamma$ is shown in the right panel of Fig.~\ref{fig:hemispheres}, illustrating how sources with various photon indexes are homogeneously distributed.

\subsection{Redshifts}
\label{sec:redshifts}
Spectroscopic redshifts are available for 79 sources, 27 of which are archival and 52 were obtained in our dedicated follow-up campaign of eROSITA X-ray transients. The summary of the follow-up observations is provided in Appendix \ref{optical_fu}. For sources without spectroscopic redshifts, we provided reliable photometric redshifts (photo-z), which have accuracy 6\% with a fraction of outliers estimated to be around 12\%, see Salvato et al. (2024a, submitted) for more details. Also for a reliable photo-z we required all LS10 photometry bands ({\it griz}) to be available. For even more accuracy, one could also filter photo-z using LS10 parameters ANYMASK and ALLMASK, which denote sources that touch bad pixels; however, we did not exclude such photo-z from consideration. Also the values for sources \texttt{p\_any}<0.17 should be used only after confirming the counterpart. As a result, using only reliable redshifts and counterparts, the redshift completeness of the sample reaches 84\,\%. 

\begin{figure}
  \includegraphics[width=\linewidth]{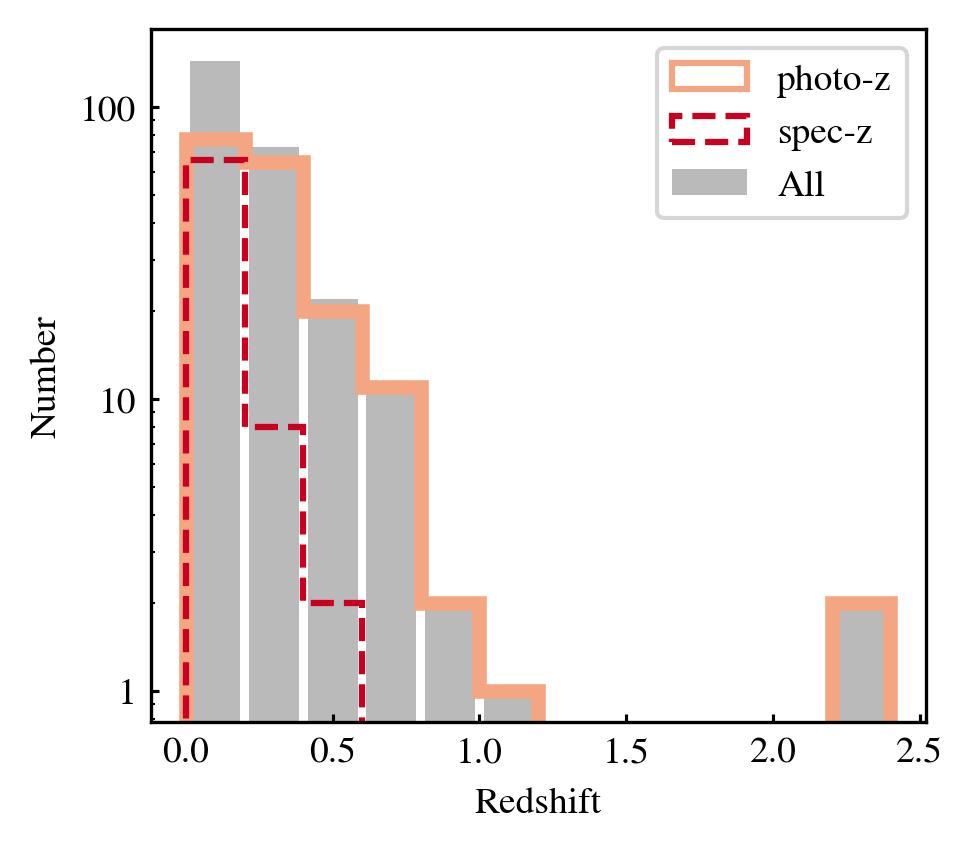}
  \caption{Distribution of reliable spectroscopic and photometric redshifts available for 255 eRO-ExTra sources.}
  \label{fig:z_distr}
\end{figure}

Fig.~\ref{fig:z_distr} shows the total, spectroscopic, and photometric redshift distributions. The majority of sources are at $\mathrm{z<0.3}$ (64\,\%), and several sources (1\,\%) have $\mathrm{z>1}$. Photometric redshifts are typically higher than spectroscopic ones due to the limitations of optical spectroscopic follow-up redshift measurements.

\section{Radio properties}
\label{sec:radio}
We investigated the radio properties of the eRO-ExTra sources using the Rapid Australian Square Kilometre Array Pathfinder (ASKAP; \citealt{Hotan_2021}) Continuum Survey (RACS; \citealt{RACS}) and the Karl G. Jansky Very Large Array Sky Survey (VLASS; \citealt{Lacy2020}). RACS is the first large-area radio survey completed with the full 36-dish ASKAP radio telescope, which is part of the Australia Telescope National Facility (ATNF). It provides detections and upper limits at 0.88\,GHz (RACS-low) and 1.39\,GHz (RACS-mid). VLASS is a high angular resolution (2.5\arcsec), high sensitivity (rms $\approx 120\,\mu$Jy/bm) survey of the sky visible to the VLA (Dec $>$ -40\,deg) at a central frequency of 3\,GHz. Overall, three VLASS epochs are foreseen, each 32 months apart. VLASS 1 began in 2017, and thus, most eRO-ExTra visible to the VLA were observed before the eRASS scans in VLASS 1 and after/during the eROSITA surveys in VLASS 2. This makes the catalogs complementary for radio transient searches associated with the eRO-ExTra sources.    

Crossmatching the eRO-ExTra catalog with RACS-low resulted in 30 matches. Here, a radius of $\mathrm{5\arcsec}$ was used for sources with reliable LS10 counterparts, and a larger radius of $\mathrm{15\arcsec}$ otherwise. Additionally, we crossmatched the catalog with VLASS2.1. We found eight matches using a radius of $\mathrm{2.5\arcsec}$ based on the positional uncertainty of VLASS for sources with reliable counterparts and $\mathrm{15\arcsec}$ in other cases. All VLASS sources overlap with sources in the RACS catalog, providing spectral information between 0.8--3\,GHz. We used RACS images to perform visual morphology checks of the radio sources. All sources are compact point sources consistent with both non-AGN and AGN hypotheses. In addition, we searched for radio variability between VLASS 1 and VLASS 2. Only one source, 1eRASS J092753.8+162005, showed a significant brightening from VLASS 1 to VLASS 2. No significant variability was detected for the remaining objects. 

\begin{figure}
  \includegraphics[width=0.96\linewidth]{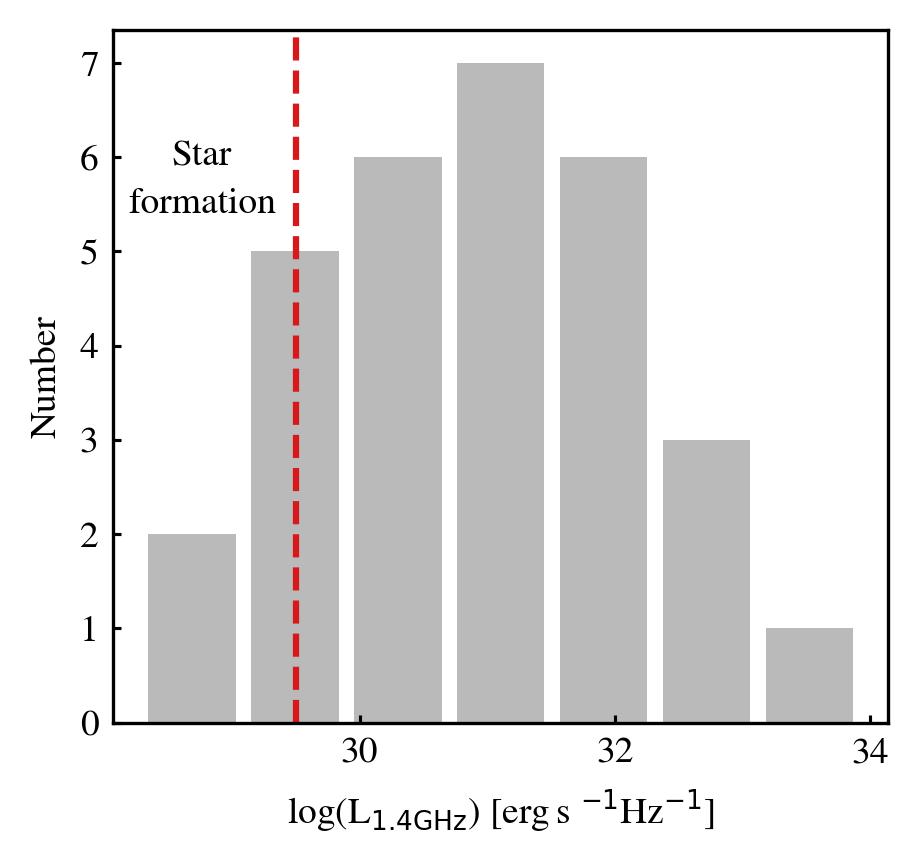}
  \caption{Radio luminosity distribution at 1.4\,GHz for 30 RACS-detected sources. The dashed red line shows the radio luminosity corresponding to star formation rate SFR = 20 M\textsubscript{\(\odot\)}$\mathrm{yr^{-1}}$. Radio emission of $\approx20$\,\% of sources can be explained by star formation, while higher luminosities for remaining sources suggest other emission processes.}
  \label{fig:sfr_distr}
\end{figure}

Overall, the eRO-ExTra catalog contains 31 known radio sources, including one source not in RACS or VLASS for which a radio detection is reported in SIMBAD. To explore the origin of the radio emission, we compared the RACS radio luminosities with the typical luminosities associated with star formation (see Appendix~\ref{appendix_sfr} for details; the results for the VLASS detections are consistent with those from RACS). As can be seen in (Fig.~\ref{fig:sfr_distr}), the large majority of the eRO-ExTra sources have radio luminosities significantly above the expectation from star formation (<20 M\textsubscript{\(\odot\)}$\mathrm{yr^{-1}}$, \citealt{1983ApJ...272...54K}), suggesting another dominating emission mechanism. This can be, for example, an underlying activity of the SMBH or directly related to the transient event detected by eROSITA. As a result of the overlap of the radio and eROSITA observing periods, it is challenging to distinguish between these scenarios without further radio observations. Therefore, radio properties were not used as a criterion in the sample selection.

\section{Known transients}
\label{sec:tns}
Consistent with the selection criteria, eRO-ExTra includes three previously reported eROSITA TDE candidates: 1eRASS J082336.8+042303 (AT2019avd, \citealt{Malyali_AT2019avd}), eRASSt J074426.3+291606 \citep{Malyali_2023}, and  eRASSt J234403.1-352640 \citep{homan2344, Goodwin_2023}. These transients show no archival X-ray detections, a flaring light curve and a soft X-ray spectrum at the peak, as expected from canonical TDEs \citep{Gezari_2021}. In addition, 1eRASS J123822.2-253210 was reported in \citet{2020ATel13416....1W} as a bright, transient X-ray source with no identified origin.

To further explore the nature of unknown X-ray events, we crossmatched eRO-ExTra with Transient Name Server\footnote{\url{ https://www.wis-tns.org}} (TNS). We found 17 (6\,\%) sources reported in TNS as new optical transients. Two of have been classified as AGN (1eRASS J141904.3-215330\footnote{\url{https://www.wis-tns.org/object/2020csk/}}, 1eRASS J082336.8+042303\footnote{\url{https://www.wis-tns.org/object/2019avd}}), while the remaining are unclassified. The TNS entry information is provided in the catalog.

\section{The eRO-ExTra catalog}
\label{sec:results}

The eRO-ExTra catalog includes 304 transient and variable sources varying with $\mathrm{S>4}$ and $\mathrm{A>4}$ between eRASS1 and eRASS2. Following the meticulous selection procedure described in Sect.~\ref{sec:selection_all}, the sample was cleaned from known AGN and Galactic objects and predominantly comprises events of, at this stage, unconstrained origin. A fraction of unknown to-date AGN might still be present in the catalog.

The example columns of the first five entries of the catalog are shown in Table~\ref{tab:cat_example}, and the full catalog column description is provided in Table~\ref{tab:columns}.
The eRO-ExTra includes 31 radio-detected sources (Sect.~\ref{sec:radio}) and 3 previously reported eROSITA TDE candidates (Sect.~\ref{sec:tns}).

\begin{figure}
  \includegraphics[width=\linewidth]{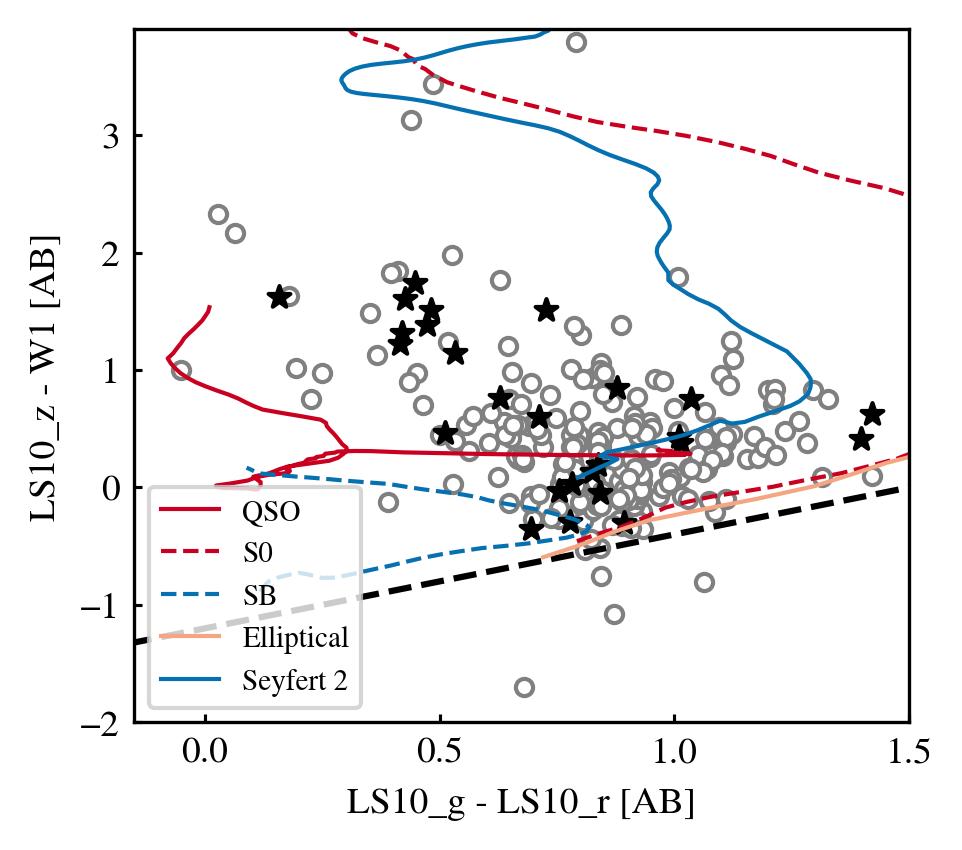}
  \caption{Distribution of 226 eRO-ExTra sources with reliable counterparts (\texttt{p\_any} > 0.17) and accurately measured LS10 fluxes in {\it z}-W1 vs. { \it g-r} plane. Black stars indicate the radio-detected sources. Tracks of active and inactive galaxies are included and described in the figure legend. The black dashed line ($z-$W1-0.8$\times(g-r)$+1.2=0) marks the division between Galactic (below) and extragalactic (above) sources. The theoretical tracks and the marked division between Galactic and extragalactic sources are both from \citet{Salvato_2022}.
}
  \label{fig:color_plot}
\end{figure}

 Fig.~\ref{fig:color_plot} shows the distribution of all eRo-ExTra sources with a reliable counterpart (\texttt{p\_any}>0.17) in the {\it g-r} vs. {\it z}-W1 photometry plane. The overplotted lines are the division between Galactic and extragalactic objects \citep{Salvato_2022}, and several representative tracks of different classes of extragalactic objects: QSO, obscured Seyfert, and inactive galaxies (S0, SB, Elliptical). The computation of the theoretical color-redshift tracks and used SED templates for each object type are discussed in \citet{Salvato_2022}. The plot testifies to the successful removal of Galactic sources from the sample, with only four falling below the dashed black separation line. Visual inspection of the LS10 images confirmed that these sources are galaxies. The majority of the eRO-ExTra population is located in the area where active and inactive galaxies can be found. 
 
 Most of the sources in the catalog have a reliable ($\texttt{p\_any}>0.17$) NWAY/LS10 optical counterpart. The remaining 7\,\% with less reliable associations may include hostless transients, or events originating from higher redshifts or faint galaxies. 

\section{Discussion}
\label{sec:discussion}
\begin{figure}
  \includegraphics[width=\linewidth]{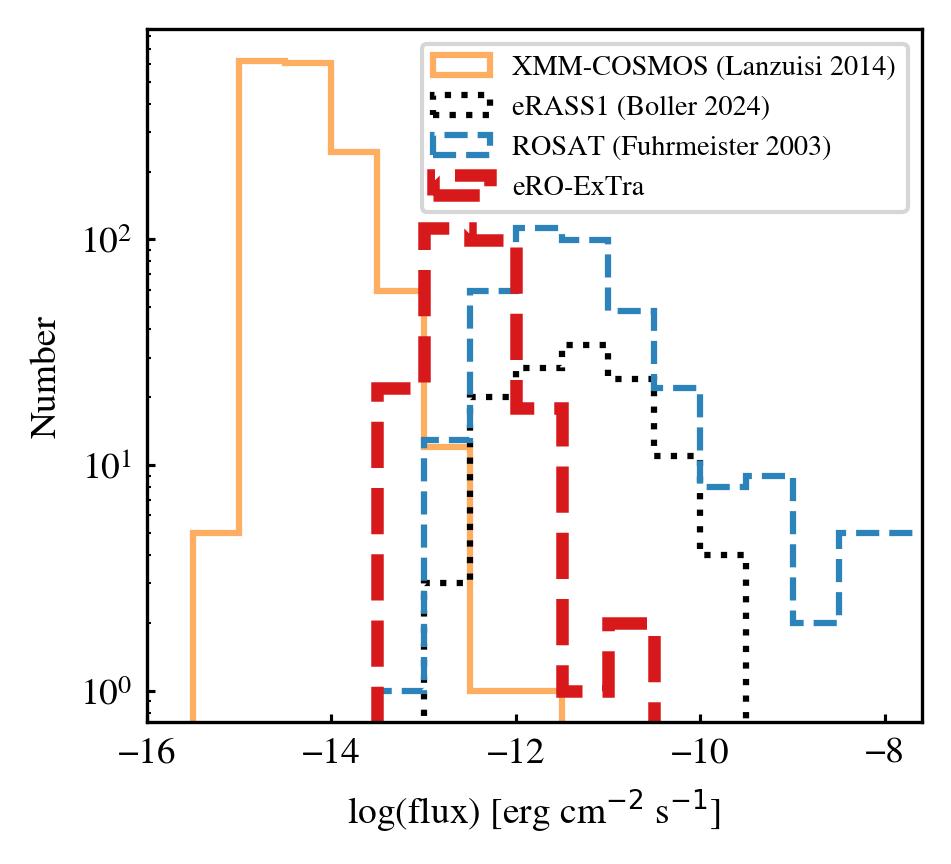}
  \caption{Comparison of 0.2--2.3\,keV flux distributions of eRO-ExTra with those of variable or transient extragalactic sources in previous studies. The dashed red line shows this study, the orange line shows XMM-COSMOS \citep{2014ApJ...781..105L}, the dotted black line shows eRASS1 \citep{boller2024erosita}, and the dashed blue line shows ROSAT \citep{Fuhrmeister}. }
  \label{fig:flux_distr}
\end{figure}

A detailed classification of all sources included in the eRO-ExTra catalog is beyond the scope of this paper, and the reader is encouraged to use the catalog to explore the nature of the sources. The eRO-ExTra catalog probes the non-AGN extragalactic population at X-ray energies to an extent not reached by previous studies, due to the eROSITA sensitivity, all sky coverage and consistent cadence over 2.5 years. The catalog is suitable for studying long-term transients and variables thanks to the presented multistep selection: only $\approx$0.001\,\% of all eRASS1 and eRASS2 significant detections (306/389k sources with $\mathrm{DET\_LIKE>15}$) remain. With the sky density of unknown X-ray variables reaching 0.03 $\mathrm{source/deg^{2}/year}$, the catalog can be used to find rare astronomical events as well as to perform systematic population studies. A systematic study of the eROSITA canonical TDE population using eRO-ExTra will be presented in a forthcoming publication (Grotova et al. in prep).

While the eRO-ExTra sample selection took great care to minimize the contamination by AGN, a fraction of the remaining sources is likely to be associated with regular accretion events onto SMBHs. Future multiwavelength follow-up is needed to confirm the classification of the sources.
As can be seen in the photon index distribution (Fig.~\ref{fig:gamma_distr}), 41\,\% of eRO-ExTra sources have a spectral slope of $\Gamma<2$. Although AGN typically have similar values of $\Gamma$ and some $\Gamma<2$ sources in the sample might have this origin, hard X-ray spectra are also not uncommon in other source populations. For example, the accretion onto the SMBH during a TDE can lead to the formation of a corona \citep{guolo2024systematic} while the TDE is still X-ray bright. Due to the time sampling of eROSITA observations (one visit per 6 months), such events may be detected only in the phase when the source is X-ray hard. Another example is eROSITA TDE candidate eRASSt J045650.3--203750 \citep{Liu_2023}. This source had a hard spectrum during the X-ray plateau phase, reaching $\Gamma=2.5$. Other subclasses of TDE candidates often show harder spectra as well. In a recent mid-infrared-selected TDE sample \citep{masterson2024new}, representing a new population of dusty TDEs, WTP15abymdq showed $\Gamma = 0.7-2.0$ across all presented epochs. Another class of hard sources are jetted TDEs, for example, Swift J164449.3+573451 \citep{2011Natur.476..421B,Saxton_2012} and Swift J2058.4+0516 \citep{Bradley_Cenko_2012} which both had hard ($\Gamma<2$) spectra even at peak. 

Similarly to the spectral slope not being a decisive criterion to select against AGN, the X-ray light curve class (see Sect.~\ref{sec:light curves}) cannot be used directly to exclude AGN. A decline or flare light curve is expected for a variety of non-AGN transient source populations. Sources with brightening or other light curves may look like variable AGN. However,  given the limited temporal coverage provided by the eROSITA data alone, also transients such as TDEs or partial TDEs with X-ray flares, rebrightenings, or even QPEs can appear different than monotonically decaying \citep{Malyali_AT2019avd, Liu_2023,Malyali_2023}. Therefore, non-AGN transients can belong to any presented light curve class.

Next, we compared the eRO-ExTra catalog with previously performed studies of the X-ray variable sky. \cite{Fuhrmeister} performed a systematic study of the X-ray variability in the {\it ROSAT} all-sky survey. Among the 1207 sources in their sample, they identified more than 60\,\% with stars, 10\,\% with an extragalactic origin, while 25\,\% remained without a counterpart association. In the eRASS1-eRASS2 variability sample within the LS10 footprint (Fig.~\ref{fig:selection_flowchart}), the fraction of stars was only 33\,\%, likely as a result of the removal of the Galactic Plane. Only 4\,\% of the sources have no reliable counterpart, thanks to NWAY and the available multiwavelength data. Moreover, 66\,\% of the sources are extragalactic, which can be attributed to eROSITA's significantly higher sensitivity. 
In addition, the eRO-ExTra catalog was thoroughly cleaned from AGN contamination, which lets us identify more unique extragalactic transients. 

Another long-term (3.5 years) extragalactic variability study was performed by \cite{2014ApJ...781..105L} with {\it XMM-Newton}. This study focused on the AGN population in the COSMOS field and, therefore, cannot be directly compared with the eRO-ExTra catalog. It showed that the majority of AGN are X-ray-variable, which is in line with the large fraction of AGN (70\,\%) that was removed from the eRASS1-eRASS2 variability sample for the selection of the eRO-ExTra catalog.

A first variability study of the X-ray sky with eROSITA was performed within the eROSITA final Equatorial-Depth Survey (eFEDS) field in November 2019 \citep{boller_efeds}. The majority of 65 sources (82\,\%) found in this study were of Galactic origin, while only 18\,\% were extragalactic sources. The extragalactic sample included 7 previously known AGN and 5 galaxies detected in X-rays with eROSITA for the first time. For the whole eROSITA\_DE sky, the analysis of the short-term variability was performed in \citet{boller2024erosita} and identified 1307 out of 128,000 variable sources in eRASS1. The majority of sources are located in the galactic plane or are consistent with stellar objects, including flare stars, young stars and active binaries, and only about 10\,\% of variable sources are extragalactic. The timescales probed by the eFEDS survey and eRASS1 variability studies are significantly shorter than the ones presented in the eRO-ExTra catalog. The eFEDS observations were performed over the course of a few days and were thus primarily sensitive for identifying short-term variability, such as stellar flares. The eRASS1 variability study was performed analysing the 4\,hr cadence light curves. eRO-ExTra, in contrast, probes the long-term variability over a 1--2 year period.

Fig.~\ref{fig:flux_distr} presents the flux distribution of the eRO-ExTra sample compared with those of the X-ray variability studies discussed above. Using WebPIMMS\footnote{https://heasarc.gsfc.nasa.gov/cgi-bin/Tools/w3pimms/w3pimms.pl}, the fluxes from \citet{Fuhrmeister} (0.1--2.4\,keV) and \citet{2014ApJ...781..105L} (0.5--2\,keV) were converted into the eROSITA energy band (0.2--2.3\,keV) assuming the same spectral model as described in Sect.~\ref{sec:varcat}. Since the eRO-ExTra catalog probes the extragalactic population, we selected 403 sources from the \citet{Fuhrmeister} catalog not associated with Galactic objects, according to the provided classification. Due to multiple selection steps (Sect.~\ref{sec:selection_all}), eRO-ExTra provides a much cleaner selection of extragalactic transients, compared to \citet{Fuhrmeister} as can be seen by the excess of high flux sources, which are likely associated with a remaining population of Galactic sources. Additionally, thanks to eROSITA’s sensitivity, the eRO-ExTra sample extends to fainter objects. To compare with the extragalactic population of the eRASS1 variability catalog \citep{boller2024erosita}, we selected only sources within the LS10 footprint in their sample and further applied a cut based on parallax, proper motion, optical and infrared photometry, and X-ray information ($\mathrm{C\_gal\_ex>0}$, see  Salvato et al. b in prep. for more details). This results in 123 objects. Only six of those are also included in the eRO-ExTra sample, emphasizing the very different and complementary selection criteria.
The {\it XMM}-COSMOS sample probes significantly lower fluxes than eRO-ExTra as a result of the deeper (pencil-beam) exposure compared to the all-sky survey of eROSITA. Overall, eRO-ExTra fills the flux gap of X-ray-detected transients and variables lying between the {\it XMM}-COSMOS and {\it ROSAT} studies.

\begin{figure}
  \includegraphics[width=\linewidth]{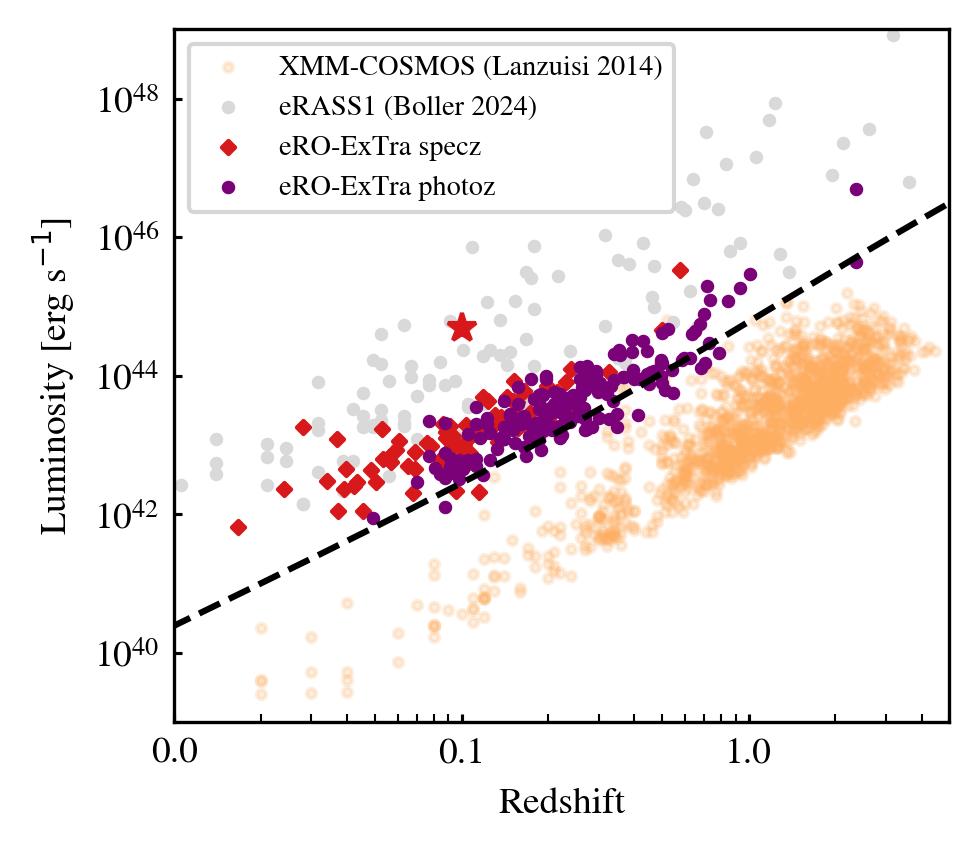}
  \caption{ Luminosity 0.2--2.3\,keV vs. redshift for all eRO-ExTra sources with reliable spectroscopic (red) or photometric (violet) redshifts. The black dashed line shows the limit where a source would satisfy the eRO-ExTra selection criteria (detection likelihood, amplitude and significance cuts) at $\mathrm{b_{ecl} = 30^{\circ}}$. A star highlights a known luminous TDE candidate eRASSt J234403.1-352640. The eRASS1 (gray, \citealt{boller2024erosita}) and {\it XMM-COSMOS} (orange, \citealt{2014ApJ...781..105L}) variability samples are shown for comparison.}
  \label{fig:L_distr}
\end{figure}

Fig.~\ref{fig:L_distr} shows the luminosity vs. spectroscopic or photometric redshift of the sources in the eRO-ExTra catalog compared to the {\it XMM}-COSMOS and {\it ROSAT} variability samples. For eRO-ExTra we used a subsample of 261 sources with reliable counterparts and redshifts. The luminosity range probed by eRO-ExTra covers $10^{42}$ to $10^{45}$ erg/s, compatible with the typical luminosities of TDEs \citep{Gezari_2021} but also low and moderate- luminosity AGN. A particularly bright TDE in the sample (eRASSt J234403.1-352640, \citealt{homan2344, Goodwin_2023}) is highlighted with a star. The {\it XMM}-COSMOS sample covers a comparable luminosity range as eRO-ExTra, but at higher redshift. In contrast, the extragalactic eRASS1 variability sample is on average more luminous. This can be attributed to their higher flux detection limit necessary to probe shorter timescales within one eroDay (4\,hr). In addition, the eRASS1 sample includes known AGN and quasars at higher luminosities.

Using a subsample of sources with credible optical counterparts and reliable redshifts, we estimated the integrated volumetric rate of eRO-ExTra transients and variables equal to $\mathrm{1.8^{+0.5}_{-0.4}\times10^{-7} Mpc^{-3}year^{-1}}$. The rate corresponds to the number of eRO-ExTra sources found according to the presented selection criteria. A direct comparison of this rate with previously published rates of specific transient populations (e.g., TDEs) cannot be done due to the heterogeneous nature of the eRO-ExTra sample. The occurrence rates of canonical TDEs from eRO-ExTra will be presented in Grotova et al. (in prep.). The computation details of the eRO-ExTra X-ray luminosity function and the rate are detailed in Appendix~\ref{sec:xlf}.

\section{Summary}
\label{sec:summary}
eROSITA let us systematically explore the variability of the X-ray sky over 2.5 years and discover numerous  transients. In this paper, we presented the catalog of 304 extragalactic transients and variables (eRO-ExTra) in the western Galactic hemisphere and contained the LS10 footprint without clear infrared AGN signatures (W1-W2<0.8mag$_{\rm vega}$) or known pre-eROSITA AGN classification. The catalog includes sources with a variability significance and a fractional amplitude larger than 4 in the first two eROSITA all-sky surveys (Nov 2019--Nov 2020). In addition, X-ray sources are identified with optical LS10/NWAY counterparts, with 93\,\% having a reliable counterpart. The catalog includes eROSITA X-ray positions, discovery dates, fluxes or upper limits for eRASS1-4(5), as well as counterpart optical LS10 coordinates and photometry.

We presented the analysis of the X-ray properties of the sample, which is useful for further classification of the sources. Firstly, we reported on the archival X-ray data from {\it Swift}, {\it XMM-Newton} and {\it ROSAT} for the sample to explore the long-term variability of the sources. More than 95\,\% of the transients have no archival detection and were discovered in X-rays with eROSITA for the first time. The archival X-ray detections and upper limits are provided in the catalog. Secondly, we analyzed the long-term X-ray eROSITA light curves and provided a classifaction in decline, flare, brightening and other. Finally, for the light curve peak in eRASS1 or eRASS2, we presented the results of the X-ray spectral modeling with a power law model. The catalog includes the best-fit values of $\Gamma$.  Furthermore, the catalog includes archival and follow-up spectroscopic or photometric redshifts, providing a redshift completeness of $>$80\,\% for the sample.

A selection of databases, including SIMBAD, TNS, RACS and VLASS, was used to classify already known sources. There are 31 sources showing archival radio, which can be either a sign of AGN, star formation or a newly launched jet. Several sources are already published transients, namely, the TDE candidates AT2019avd, eRASSt J074426.3+2916066 and eRASSt J234403.1-352640.

The catalog provides the first glimpse into the eROSITA X-ray transient sky and lets us systematically study the variability of various astronomical phenomena on months/year timescales. Looking forward, the eRO-ExTra catalog should be used to disentangle the nature of still unclassified sources via in-depth analysis and multiwavelength follow-up and to perform population studies of X-ray transients. It can furthermore guide the transient selection and identification approach to new and future X-ray surveys, such as Einstein Probe \citep{Yuan_2022}.

\section{Data availability}
The eRO-ExTra catalog is only available in electronic form at the CDS via anonymous ftp to \url{cdsarc.u-strasbg.fr} (\url{130.79.128.5}) or via \url{https://cdsarc.cds.unistra.fr/viz-bin/cat/J/A+A/693/A62}. The reduced optical spectra listed in Appendix~\ref{optical_fu} can be downloaded here: \url{https://github.com/grotova/eRO-ExTra}.

\begin{acknowledgements}
      This work is based on data from eROSITA, the soft X-ray instrument aboard SRG, a joint Russian-German science mission supported by the Russian Space Agency (Roskosmos), in the interests of the Russian Academy of Sciences represented by its Space Research Institute (IKI), and the Deutsches Zentrum f\"ur Luft- und Raumfahrt (DLR). The SRG spacecraft was built by Lavochkin Association (NPOL) and its subcontractors and is operated by NPOL with support from the Max Planck Institute for Extraterrestrial Physics (MPE). The development and construction of the eROSITA X-ray instrument was led by MPE, with contributions from the Dr. Karl Remeis Observatory Bamberg  ECAP (FAU Erlangen-Nuernberg), the University of Hamburg Observatory, the Leibniz Institute for Astrophysics Potsdam (AIP), and the Institute for Astronomy and Astrophysics of the University of T\"ubingen, with the support of DLR and the Max Planck Society. The Argelander Institute for Astronomy of the University of Bonn and the Ludwig Maximilians Universit\"at Munich also participated in the science preparation for eROSITA. The eROSITA data shown here were processed using the eSASS/NRTA software system developed by the German eROSITA consortium.
      A. Malyali acknowledges support by DLR under the grant 50 QR 2110 (XMM\_NuTra). M. Krumpe acknowledges support by DLR under grant 50 OR 2307. D. Homan acknowledges support by DLR under the grant 50 OR 2003. D. Tub\'in-Arenas acknowledges support by DLR under grant 50 OR 2203. The LCO observations have been made possible by the support of the Deutsche Forschungsgemeinschaft (DFG, German Research Foundation) under Germany’s Excellence Strategy-EXC-2094-390783311. This work was supported by the Australian government through the Australian Research Council’s Discovery Projects funding scheme (DP200102471). This paper includes data gathered with the 6.5 m Magellan Telescopes located at Las Campanas Observatory, Chile. A part of this work is based on observations made with the Southern African Large Telescope (SALT), with the Large Science Programmes on transients 2018-2-LSP-001 \& 2021-2-LSP-001 (PI: DAHB). Polish participation in SALT is funded by grant No. MEiN 2021/WK/01.This research has made use of the SIMBAD database,
operated at CDS, Strasbourg, France. This work has made use of data from the European Space Agency (ESA) mission
{\it Gaia} (\url{https://www.cosmos.esa.int/gaia}), processed by the {\it Gaia}
Data Processing and Analysis Consortium (DPAC,
\url{https://www.cosmos.esa.int/web/gaia/dpac/consortium}). Funding for the DPAC
has been provided by national institutions, in particular the institutions
participating in the {\it Gaia} Multilateral Agreement.
\end{acknowledgements}

\bibliographystyle{aa}
\bibliography{eroextra.bib}

\begin{thebibliography}{94}
\expandafter\ifx\csname natexlab\endcsname\relax\def\natexlab#1{#1}\fi

\bibitem[{{Arcodia} {et~al.}(2024){Arcodia}, {Liu}, {Merloni}, {Malyali},
  {Rau}, {Chakraborty}, {Goodwin}, {Buckley}, {Brink}, {Gromadzki},
  {Arzoumanian}, {Buchner}, {Kara}, {Nandra}, {Ponti}, {Salvato}, {Anderson},
  {Baldini}, {Grotova}, {Krumpe}, {Maitra}, {Miller-Jones}, \&
  {Ramos-Ceja}}]{2024A&A...684A..64A}
{Arcodia}, R., {Liu}, Z., {Merloni}, A., {et~al.} 2024, \aap, 684, A64

\bibitem[{Arcodia {et~al.}(2021)Arcodia, Merloni, Nandra, Buchner, Salvato,
  Pasham, Remillard, Comparat, Lamer, Ponti, Malyali, Wolf, Arzoumanian,
  Bogensberger, Buckley, Gendreau, Gromadzki, Kara, Krumpe, Markwardt,
  Ramos-Ceja, Rau, Schramm, \& Schwope}]{Arcodia_2021}
Arcodia, R., Merloni, A., Nandra, K., {et~al.} 2021, Nat, 592, 704

\bibitem[{{Arnaud}(1996)}]{1996ASPC..101...17A}
{Arnaud}, K.~A. 1996, in ASP Conf. Ser., Vol. 101, Astronomical Data Analysis
  Software and Systems V, ed. G.~H. {Jacoby} \& J.~{Barnes}, 17

\bibitem[{{Ballet} {et~al.}(2023){Ballet}, {Bruel}, {Burnett}, {Lott}, \& {The
  Fermi-LAT collaboration}}]{ballet2024}
{Ballet}, J., {Bruel}, P., {Burnett}, T.~H., {Lott}, B., \& {The Fermi-LAT
  collaboration}. 2023, arXiv e-prints, arXiv:2307.12546

\bibitem[{{Bogensberger} {et~al.}(2024){Bogensberger}, {Nandra}, \&
  {Buchner}}]{2024arXiv240117278B}
{Bogensberger}, D., {Nandra}, K., \& {Buchner}, J. 2024, \aap, 687, A21

\bibitem[{{Boller} {et~al.}(2024){Boller}, {Freyberg}, {Buchner}, {Haberl},
  {Maitra}, {Schwope}, {Robrade}, {Rau}, {Grotova}, {Waddell}, {Ni}, {Salvato},
  {Krumpe}, {Georgakakis}, {Merloni}, \& {Nandra}}]{boller2024erosita}
{Boller}, T., {Freyberg}, M., {Buchner}, J., {et~al.} 2024, arXiv e-prints,
  arXiv:2401.17280

\bibitem[{{Boller} {et~al.}(2022){Boller}, {Schmitt, J. H. M. M.}, {Buchner,
  J.}, {Freyberg, M.}, {Georgakakis, A.}, {Liu, T.}, {Robrade, J.}, {Merloni,
  A.}, {Nandra, K.}, {Malyali, A.}, {Krumpe, M.}, {Salvato, M.}, \& {Dwelly,
  T.}}]{boller_efeds}
{Boller}, T., {Schmitt, J. H. M. M.}, {Buchner, J.}, {et~al.} 2022, A\&A, 661,
  A8

\bibitem[{Brown {et~al.}(2013)Brown, Baliber, Bianco, Bowman, Burleson, Conway,
  Crellin, Depagne, Vera, Dilday, Dragomir, Dubberley, Eastman, Elphick,
  Falarski, Foale, Ford, Fulton, Garza, \& Willis}]{FLOYDS}
Brown, T., Baliber, N., Bianco, F., {et~al.} 2013, PASP, 125

\bibitem[{{Brunner} {et~al.}(2022){Brunner}, {Liu}, {Lamer}, {Georgakakis},
  {Merloni}, {Brusa}, {Bulbul}, {Dennerl}, {Friedrich}, {Liu}, {Maitra},
  {Nandra}, {Ramos-Ceja}, {Sanders}, {Stewart}, {Boller}, {Buchner}, {Clerc},
  {Comparat}, {Dwelly}, {Eckert}, {Finoguenov}, {Freyberg}, {Ghirardini},
  {Gueguen}, {Haberl}, {Kreykenbohm}, {Krumpe}, {Osterhage}, {Pacaud},
  {Predehl}, {Reiprich}, {Robrade}, {Salvato}, {Santangelo}, {Schrabback},
  {Schwope}, \& {Wilms}}]{brunner22}
{Brunner}, H., {Liu}, T., {Lamer}, G., {et~al.} 2022, \aap, 661, A1

\bibitem[{{Buchner}(2021)}]{buchner2021ultranest}
{Buchner}, J. 2021, J. Open Source Softw., 6, 3001

\bibitem[{{Buchner} {et~al.}(2014){Buchner}, {Georgakakis}, {Nandra}, {Hsu},
  {Rangel}, {Brightman}, {Merloni}, {Salvato}, {Donley}, \&
  {Kocevski}}]{2014A&A...564A.125B}
{Buchner}, J., {Georgakakis}, A., {Nandra}, K., {et~al.} 2014, \aap, 564, A125

\bibitem[{{Buckley} {et~al.}(2006){Buckley}, {Swart}, \&
  {Meiring}}]{2006SPIE.6267E..0ZB}
{Buckley}, D. A.~H., {Swart}, G.~P., \& {Meiring}, J.~G. 2006, in Society of
  Photo-Optical Instrumentation Engineers (SPIE) Conference Series, Vol. 6267,
  Ground-based and Airborne Telescopes, ed. L.~M. {Stepp}, 62670Z

\bibitem[{{Burgh} {et~al.}(2003){Burgh}, {Nordsieck}, {Kobulnicky}, {Williams},
  {O'Donoghue}, {Smith}, \& {Percival}}]{2003SPIE.4841.1463B}
{Burgh}, E.~B., {Nordsieck}, K.~H., {Kobulnicky}, H.~A., {et~al.} 2003, in
  Society of Photo-Optical Instrumentation Engineers (SPIE) Conference Series,
  Vol. 4841, Instrument Design and Performance for Optical/Infrared
  Ground-based Telescopes, ed. M.~{Iye} \& A.~F.~M. {Moorwood}, 1463--1471

\bibitem[{{Burrows} {et~al.}(2005){Burrows}, {Hill}, {Nousek}, {Kennea},
  {Wells}, {Osborne}, {Abbey}, {Beardmore}, {Mukerjee}, {Short}, {Chincarini},
  {Campana}, {Citterio}, {Moretti}, {Pagani}, {Tagliaferri}, {Giommi},
  {Capalbi}, {Tamburelli}, {Angelini}, {Cusumano}, {Br{\"a}uninger}, {Burkert},
  \& {Hartner}}]{swiftxrt}
{Burrows}, D.~N., {Hill}, J.~E., {Nousek}, J.~A., {et~al.} 2005, \ssr, 120, 165

\bibitem[{{Burrows} {et~al.}(2011){Burrows}, {Kennea}, {Ghisellini}, {Mangano},
  {Zhang}, {Page}, {Eracleous}, {Romano}, {Sakamoto}, {Falcone}, {Osborne},
  {Campana}, {Beardmore}, {Breeveld}, {Chester}, {Corbet}, {Covino},
  {Cummings}, {D'Avanzo}, {D'Elia}, {Esposito}, {Evans}, {Fugazza}, {Gelbord},
  {Hiroi}, {Holland}, {Huang}, {Im}, {Israel}, {Jeon}, {Jeon}, {Jun}, {Kawai},
  {Kim}, {Krimm}, {Marshall}, {P. M{\'e}sz{\'a}ros}, {Negoro}, {Omodei},
  {Park}, {Perkins}, {Sugizaki}, {Sung}, {Tagliaferri}, {Troja}, {Ueda},
  {Urata}, {Usui}, {Antonelli}, {Barthelmy}, {Cusumano}, {Giommi}, {Melandri},
  {Perri}, {Racusin}, {Sbarufatti}, {Siegel}, \&
  {Gehrels}}]{2011Natur.476..421B}
{Burrows}, D.~N., {Kennea}, J.~A., {Ghisellini}, G., {et~al.} 2011, Nat, 476,
  421

\bibitem[{{Buzzoni} {et~al.}(1984){Buzzoni}, {Delabre}, {Dekker}, {Dodorico},
  {Enard}, {Focardi}, {Gustafsson}, {Nees}, {Paureau}, \&
  {Reiss}}]{1984Msngr..38....9B}
{Buzzoni}, B., {Delabre}, B., {Dekker}, H., {et~al.} 1984, The Messenger, 38, 9

\bibitem[{{Cash}(1976)}]{1976cash}
{Cash}, W. 1976, \aap, 52, 307

\bibitem[{Cenko {et~al.}(2012)Cenko, Krimm, Horesh, Rau, Frail, Kennea, Levan,
  Holland, Butler, Quimby, Bloom, Filippenko, Gal-Yam, Greiner, Kulkarni, Ofek,
  Olivares~E., Schady, Silverman, Tanvir, \& Xu}]{Bradley_Cenko_2012}
Cenko, S., Krimm, H.~A., Horesh, A., {et~al.} 2012, ApJ, 753, 77

\bibitem[{{Chakraborty} {et~al.}(2021){Chakraborty}, {Kara}, {Masterson},
  {Giustini}, {Miniutti}, \& {Saxton}}]{2021Chakraborty}
{Chakraborty}, J., {Kara}, E., {Masterson}, M., {et~al.} 2021, \apjl, 921, L40

\bibitem[{{Chang} {et~al.}(2019){Chang}, {Arsioli}, {Giommi}, {Padovani}, \&
  {Brandt}}]{2019A&A...632A..77C}
{Chang}, Y.~L., {Arsioli}, B., {Giommi}, P., {Padovani}, P., \& {Brandt}, C.~H.
  2019, \aap, 632, A77

\bibitem[{{Childress} {et~al.}(2014){Childress}, {Vogt}, {Nielsen}, \&
  {Sharp}}]{Childress2014}
{Childress}, M.~J., {Vogt}, F. P.~A., {Nielsen}, J., \& {Sharp}, R.~G. 2014,
  \apss, 349, 617

\bibitem[{{Crawford} {et~al.}(2010){Crawford}, {Still}, {Schellart}, {Balona},
  {Buckley}, {Dugmore}, {Gulbis}, {Kniazev}, {Kotze}, {Loaring}, {Nordsieck},
  {Pickering}, {Potter}, {Romero Colmenero}, {Vaisanen}, {Williams}, \&
  {Zietsman}}]{2010SPIE.7737E..25C}
{Crawford}, S.~M., {Still}, M., {Schellart}, P., {et~al.} 2010, in Society of
  Photo-Optical Instrumentation Engineers (SPIE) Conference Series, Vol. 7737,
  Observatory Operations: Strategies, Processes, and Systems III, ed. D.~R.
  {Silva}, A.~B. {Peck}, \& B.~T. {Soifer}, 773725

\bibitem[{{D'Abrusco} {et~al.}(2019){D'Abrusco}, {{\'A}lvarez Crespo},
  {Massaro}, {Campana}, {Chavushyan}, {Landoni}, {La Franca}, {Masetti},
  {Milisavljevic}, {Paggi}, {Ricci}, \& {Smith}}]{2019ApJS..242....4D}
{D'Abrusco}, R., {{\'A}lvarez Crespo}, N., {Massaro}, F., {et~al.} 2019, \apjs,
  242, 4

\bibitem[{{Dey} {et~al.}(2019){Dey}, {Schlegel}, {Lang}, {Blum}, {Burleigh},
  {Fan}, {Findlay}, {Finkbeiner}, {Herrera}, {Juneau}, {Landriau}, {Levi},
  {McGreer}, {Meisner}, {Myers}, {Moustakas}, {Nugent}, {Patej}, {Schlafly},
  {Walker}, {Valdes}, {Weaver}, {Y{\`e}che}, {Zou}, {Zhou}, {Abareshi},
  {Abbott}, {Abolfathi}, {Aguilera}, {Alam}, {Allen}, {Alvarez}, {Annis},
  {Ansarinejad}, {Aubert}, {Beechert}, {Bell}, {BenZvi}, {Beutler}, {Bielby},
  {Bolton}, {Brice{\~n}o}, {Buckley-Geer}, {Butler}, {Calamida}, {Carlberg},
  {Carter}, {Casas}, {Castander}, {Choi}, {Comparat}, {Cukanovaite}, {Delubac},
  {DeVries}, {Dey}, {Dhungana}, {Dickinson}, {Ding}, {Donaldson}, {Duan},
  {Duckworth}, {Eftekharzadeh}, {Eisenstein}, {Etourneau}, {Fagrelius},
  {Farihi}, {Fitzpatrick}, {Font-Ribera}, {Fulmer}, {G{\"a}nsicke},
  {Gaztanaga}, {George}, {Gerdes}, {Gontcho}, {Gorgoni}, {Green}, {Guy},
  {Harmer}, {Hernandez}, {Honscheid}, {Huang}, {James}, {Jannuzi}, {Jiang},
  {Joyce}, {Karcher}, {Karkar}, {Kehoe}, {Kneib}, {Kueter-Young}, {Lan},
  {Lauer}, {Le Guillou}, {Le Van Suu}, {Lee}, {Lesser}, {Perreault Levasseur},
  {Li}, {Mann}, {Marshall}, {Mart{\'\i}nez-V{\'a}zquez}, {Martini}, {du Mas des
  Bourboux}, {McManus}, {Meier}, {M{\'e}nard}, {Metcalfe},
  {Mu{\~n}oz-Guti{\'e}rrez}, {Najita}, {Napier}, {Narayan}, {Newman}, {Nie},
  {Nord}, {Norman}, {Olsen}, {Paat}, {Palanque-Delabrouille}, {Peng},
  {Poppett}, {Poremba}, {Prakash}, {Rabinowitz}, {Raichoor}, {Rezaie},
  {Robertson}, {Roe}, {Ross}, {Ross}, {Rudnick}, {Safonova}, {Saha},
  {S{\'a}nchez}, {Savary}, {Schweiker}, {Scott}, {Seo}, {Shan}, {Silva},
  {Slepian}, {Soto}, {Sprayberry}, {Staten}, {Stillman}, {Stupak}, {Summers},
  {Sien Tie}, {Tirado}, {Vargas-Maga{\~n}a}, {Vivas}, {Wechsler}, {Williams},
  {Yang}, {Yang}, {Yapici}, {Zaritsky}, {Zenteno}, {Zhang}, {Zhang}, {Zhou}, \&
  {Zhou}}]{2019AJ....157..168D}
{Dey}, A., {Schlegel}, D.~J., {Lang}, D., {et~al.} 2019, \aj, 157, 168

\bibitem[{{Di Matteo}(1998)}]{1998MNRAS.299L..15D}
{Di Matteo}, T. 1998, \mnras, 299, L15

\bibitem[{{Dopita} {et~al.}(2010){Dopita}, {Rhee}, {Farage}, {McGregor},
  {Bloxham}, {Green}, {Roberts}, {Neilson}, {Wilson}, {Young}, {Firth},
  {Busarello}, \& {Merluzzi}}]{2010Ap&SS.327..245D}
{Dopita}, M., {Rhee}, J., {Farage}, C., {et~al.} 2010, \apss, 327, 245

\bibitem[{Dressler {et~al.}(2011)Dressler, Bigelow, Hare, Sutin, Thompson,
  Burley, Epps, Oemler, Bagish, Birk, Clardy, Gunnels, Kelson, Shectman, \&
  Osip}]{Dressler_2011}
Dressler, A., Bigelow, B., Hare, T., {et~al.} 2011, PASP, 123, 288

\bibitem[{{Eckart} {et~al.}(1986){Eckart}, {Witzel}, {Biermann}, {Johnston},
  {Simon}, {Schalinski}, \& {Kuhr}}]{1986A&A...168...17E}
{Eckart}, A., {Witzel}, A., {Biermann}, P., {et~al.} 1986, \aap, 168, 17

\bibitem[{Evans {et~al.}(2023)Evans, Nixon, Campana, Charalampopoulos, Perley,
  Breeveld, Page, Oates, Eyles-Ferris, Malesani, Izzo, Goad, O’Brien,
  Osborne, \& Sbarufatti}]{Evans_2023}
Evans, P.~A., Nixon, C.~J., Campana, S., {et~al.} 2023, Nat. Astron., 7,
  1368–1375

\bibitem[{{Flesch}(2023)}]{flesch2023million}
{Flesch}, E.~W. 2023, OJAp, 6, 49

\bibitem[{{Freudling} {et~al.}(2013){Freudling}, {Romaniello}, {Bramich},
  {Ballester}, {Forchi}, {Garc{\'\i}a-Dabl{\'o}}, {Moehler}, \&
  {Neeser}}]{2013A&A...559A..96F}
{Freudling}, W., {Romaniello}, M., {Bramich}, D.~M., {et~al.} 2013, \aap, 559,
  A96

\bibitem[{Fuhrmeister \& Schmitt(2003)}]{Fuhrmeister}
Fuhrmeister, B. \& Schmitt, J. H. M.~M. 2003, A\&A, 403, 247

\bibitem[{{Gaia Collaboration} {et~al.}(2016){Gaia Collaboration}, {Prusti,
  T.}, {de Bruijne, J. H. J.}, {Brown, A. G. A.}, {Vallenari, A.}, {Babusiaux,
  C.}, {Bailer-Jones, C. A. L.}, {Bastian, U.}, {Biermann, M.}, {Evans, D. W.},
  {Eyer, L.}, {Jansen, F.}, {Jordi, C.}, {Klioner, S. A.}, {Lammers, U.},
  {Lindegren, L.}, {Luri, X.}, {Mignard, F.}, {Milligan, D. J.}, {Panem, C.},
  {Poinsignon, V.}, {Pourbaix, D.}, {Randich, S.}, {Sarri, G.}, {Sartoretti,
  P.}, {Siddiqui, H. I.}, {Soubiran, C.}, {Valette, V.}, {van Leeuwen, F.},
  {Walton, N. A.}, {Aerts, C.}, {Arenou, F.}, {Cropper, M.}, {Drimmel, R.},
  {H\o{}g, E.}, {Katz, D.}, {Lattanzi, M. G.}, {O\'{}Mullane, W.}, {Grebel, E.
  K.}, {Holland, A. D.}, {Huc, C.}, {Passot, X.}, {Bramante, L.}, {Cacciari,
  C.}, {Casta\~neda, J.}, {Chaoul, L.}, {Cheek, N.}, {De Angeli, F.},
  {Fabricius, C.}, {Guerra, R.}, {Hern\'andez, J.}, {Jean-Antoine-Piccolo, A.},
  {Masana, E.}, {Messineo, R.}, {Mowlavi, N.}, {Nienartowicz, K.},
  {Ord\'o\~nez-Blanco, D.}, {Panuzzo, P.}, {Portell, J.}, {Richards, P. J.},
  {Riello, M.}, {Seabroke, G. M.}, {Tanga, P.}, {Th\'evenin, F.}, {Torra, J.},
  {Els, S. G.}, {Gracia-Abril, G.}, {Comoretto, G.}, {Garcia-Reinaldos, M.},
  {Lock, T.}, {Mercier, E.}, {Altmann, M.}, {Andrae, R.}, {Astraatmadja, T.
  L.}, {Bellas-Velidis, I.}, {Benson, K.}, {Berthier, J.}, {Blomme, R.},
  {Busso, G.}, {Carry, B.}, {Cellino, A.}, {Clementini, G.}, {Cowell, S.},
  {Creevey, O.}, {Cuypers, J.}, {Davidson, M.}, {De Ridder, J.}, {de Torres,
  A.}, {Delchambre, L.}, {Dell\'{}Oro, A.}, {Ducourant, C.}, {Fr\'emat, Y.},
  {Garc\'{\i}a-Torres, M.}, {Gosset, E.}, {Halbwachs, J.-L.}, {Hambly, N. C.},
  {Harrison, D. L.}, {Hauser, M.}, {Hestroffer, D.}, {Hodgkin, S. T.}, {Huckle,
  H. E.}, {Hutton, A.}, {Jasniewicz, G.}, {Jordan, S.}, {Kontizas, M.}, {Korn,
  A. J.}, {Lanzafame, A. C.}, {Manteiga, M.}, {Moitinho, A.}, {Muinonen, K.},
  {Osinde, J.}, {Pancino, E.}, {Pauwels, T.}, {Petit, J.-M.}, {Recio-Blanco,
  A.}, {Robin, A. C.}, {Sarro, L. M.}, {Siopis, C.}, {Smith, M.}, {Smith, K.
  W.}, {Sozzetti, A.}, {Thuillot, W.}, {van Reeven, W.}, {Viala, Y.}, {Abbas,
  U.}, {Abreu Aramburu, A.}, {Accart, S.}, {Aguado, J. J.}, {Allan, P. M.},
  {Allasia, W.}, {Altavilla, G.}, {\'Alvarez, M. A.}, {Alves, J.}, {Anderson,
  R. I.}, {Andrei, A. H.}, {Anglada Varela, E.}, {Antiche, E.}, {Antoja, T.},
  {Ant\'on, S.}, {Arcay, B.}, {Atzei, A.}, {Ayache, L.}, {Bach, N.}, {Baker, S.
  G.}, {Balaguer-N\'u\~nez, L.}, {Barache, C.}, {Barata, C.}, {Barbier, A.},
  {Barblan, F.}, {Baroni, M.}, {Barrado y Navascu\'es, D.}, {Barros, M.},
  {Barstow, M. A.}, {Becciani, U.}, {Bellazzini, M.}, {Bellei, G.}, {Bello
  Garc\'{\i}a, A.}, {Belokurov, V.}, {Bendjoya, P.}, {Berihuete, A.}, {Bianchi,
  L.}, {Bienaym\'e, O.}, {Billebaud, F.}, {Blagorodnova, N.}, {Blanco-Cuaresma,
  S.}, {Boch, T.}, {Bombrun, A.}, {Borrachero, R.}, {Bouquillon, S.}, {Bourda,
  G.}, {Bouy, H.}, {Bragaglia, A.}, {Breddels, M. A.}, {Brouillet, N.},
  {Br\"usemeister, T.}, {Bucciarelli, B.}, {Budnik, F.}, {Burgess, P.},
  {Burgon, R.}, {Burlacu, A.}, {Busonero, D.}, {Buzzi, R.}, {Caffau, E.},
  {Cambras, J.}, {Campbell, H.}, {Cancelliere, R.}, {Cantat-Gaudin, T.},
  {Carlucci, T.}, {Carrasco, J. M.}, {Castellani, M.}, {Charlot, P.}, {Charnas,
  J.}, {Charvet, P.}, {Chassat, F.}, {Chiavassa, A.}, {Clotet, M.}, {Cocozza,
  G.}, {Collins, R. S.}, {Collins, P.}, {Costigan, G.}, {Crifo, F.}, {Cross, N.
  J. G.}, {Crosta, M.}, {Crowley, C.}, {Dafonte, C.}, {Damerdji, Y.},
  {Dapergolas, A.}, {David, P.}, {David, M.}, {De Cat, P.}, {de Felice, F.},
  {de Laverny, P.}, {De Luise, F.}, {De March, R.}, {de Martino, D.}, {de
  Souza, R.}, {Debosscher, J.}, {del Pozo, E.}, {Delbo, M.}, {Delgado, A.},
  {Delgado, H. E.}, {di Marco, F.}, {Di Matteo, P.}, {Diakite, S.}, {Distefano,
  E.}, {Dolding, C.}, {Dos Anjos, S.}, {Drazinos, P.}, {Dur\'an, J.}, {Dzigan,
  Y.}, {Ecale, E.}, {Edvardsson, B.}, {Enke, H.}, {Erdmann, M.}, {Escolar, D.},
  {Espina, M.}, {Evans, N. W.}, {Eynard Bontemps, G.}, {Fabre, C.}, {Fabrizio,
  M.}, {Faigler, S.}, {Falc\~ao, A. J.}, {Farr\`as Casas, M.}, {Faye, F.},
  {Federici, L.}, {Fedorets, G.}, {Fern\'andez-Hern\'andez, J.}, {Fernique,
  P.}, {Fienga, A.}, {Figueras, F.}, {Filippi, F.}, {Findeisen, K.}, {Fonti,
  A.}, {Fouesneau, M.}, {Fraile, E.}, {Fraser, M.}, {Fuchs, J.}, {Furnell, R.},
  {Gai, M.}, {Galleti, S.}, {Galluccio, L.}, {Garabato, D.},
  {Garc\'{\i}a-Sedano, F.}, {Gar\'e, P.}, {Garofalo, A.}, {Garralda, N.},
  {Gavras, P.}, {Gerssen, J.}, {Geyer, R.}, {Gilmore, G.}, {Girona, S.},
  {Giuffrida, G.}, {Gomes, M.}, {Gonz\'alez-Marcos, A.}, {Gonz\'alez-N\'u\~nez,
  J.}, {Gonz\'alez-Vidal, J. J.}, {Granvik, M.}, {Guerrier, A.}, {Guillout,
  P.}, {Guiraud, J.}, {G\'urpide, A.}, {Guti\'errez-S\'anchez, R.}, {Guy, L.
  P.}, {Haigron, R.}, {Hatzidimitriou, D.}, {Haywood, M.}, {Heiter, U.},
  {Helmi, A.}, {Hobbs, D.}, {Hofmann, W.}, {Holl, B.}, {Holland, G.}, {Hunt, J.
  A. S.}, {Hypki, A.}, {Icardi, V.}, {Irwin, M.}, {Jevardat de Fombelle, G.},
  {Jofr\'e, P.}, {Jonker, P. G.}, {Jorissen, A.}, {Julbe, F.}, {Karampelas,
  A.}, {Kochoska, A.}, {Kohley, R.}, {Kolenberg, K.}, {Kontizas, E.}, {Koposov,
  S. E.}, {Kordopatis, G.}, {Koubsky, P.}, {Kowalczyk, A.}, {Krone-Martins,
  A.}, {Kudryashova, M.}, {Kull, I.}, {Bachchan, R. K.}, {Lacoste-Seris, F.},
  {Lanza, A. F.}, {Lavigne, J.-B.}, {Le Poncin-Lafitte, C.}, {Lebreton, Y.},
  {Lebzelter, T.}, {Leccia, S.}, {Leclerc, N.}, {Lecoeur-Taibi, I.}, {Lemaitre,
  V.}, {Lenhardt, H.}, {Leroux, F.}, {Liao, S.}, {Licata, E.}, {Lindstr\o{}m,
  H. E. P.}, {Lister, T. A.}, {Livanou, E.}, {Lobel, A.}, {L\"offler, W.},
  {L\'opez, M.}, {Lopez-Lozano, A.}, {Lorenz, D.}, {Loureiro, T.}, {MacDonald,
  I.}, {Magalh\~aes Fernandes, T.}, {Managau, S.}, {Mann, R. G.}, {Mantelet,
  G.}, {Marchal, O.}, {Marchant, J. M.}, {Marconi, M.}, {Marie, J.}, {Marinoni,
  S.}, {Marrese, P. M.}, {Marschalk\'o, G.}, {Marshall, D. J.},
  {Mart\'{\i}n-Fleitas, J. M.}, {Martino, M.}, {Mary, N.}, {Matijevic, G.},
  {Mazeh, T.}, {McMillan, P. J.}, {Messina, S.}, {Mestre, A.}, {Michalik, D.},
  {Millar, N. R.}, {Miranda, B. M. H.}, {Molina, D.}, {Molinaro, R.},
  {Molinaro, M.}, {Moln\'ar, L.}, {Moniez, M.}, {Montegriffo, P.}, {Monteiro,
  D.}, {Mor, R.}, {Mora, A.}, {Morbidelli, R.}, {Morel, T.}, {Morgenthaler,
  S.}, {Morley, T.}, {Morris, D.}, {Mulone, A. F.}, {Muraveva, T.}, {Musella,
  I.}, {Narbonne, J.}, {Nelemans, G.}, {Nicastro, L.}, {Noval, L.},
  {Ord\'enovic, C.}, {Ordieres-Mer\'e, J.}, {Osborne, P.}, {Pagani, C.},
  {Pagano, I.}, {Pailler, F.}, {Palacin, H.}, {Palaversa, L.}, {Parsons, P.},
  {Paulsen, T.}, {Pecoraro, M.}, {Pedrosa, R.}, {Pentik\"ainen, H.}, {Pereira,
  J.}, {Pichon, B.}, {Piersimoni, A. M.}, {Pineau, F.-X.}, {Plachy, E.}, {Plum,
  G.}, {Poujoulet, E.}, {Prsa, A.}, {Pulone, L.}, {Ragaini, S.}, {Rago, S.},
  {Rambaux, N.}, {Ramos-Lerate, M.}, {Ranalli, P.}, {Rauw, G.}, {Read, A.},
  {Regibo, S.}, {Renk, F.}, {Reyl\'e, C.}, {Ribeiro, R. A.}, {Rimoldini, L.},
  {Ripepi, V.}, {Riva, A.}, {Rixon, G.}, {Roelens, M.}, {Romero-G\'omez, M.},
  {Rowell, N.}, {Royer, F.}, {Rudolph, A.}, {Ruiz-Dern, L.}, {Sadowski, G.},
  {Sagrist\`a Sell\'es, T.}, {Sahlmann, J.}, {Salgado, J.}, {Salguero, E.},
  {Sarasso, M.}, {Savietto, H.}, {Schnorhk, A.}, {Schultheis, M.}, {Sciacca,
  E.}, {Segol, M.}, {Segovia, J. C.}, {Segransan, D.}, {Serpell, E.}, {Shih,
  I-C.}, {Smareglia, R.}, {Smart, R. L.}, {Smith, C.}, {Solano, E.}, {Solitro,
  F.}, {Sordo, R.}, {Soria Nieto, S.}, {Souchay, J.}, {Spagna, A.}, {Spoto,
  F.}, {Stampa, U.}, {Steele, I. A.}, {Steidelm\"uller, H.}, {Stephenson, C.
  A.}, {Stoev, H.}, {Suess, F. F.}, {S\"uveges, M.}, {Surdej, J.}, {Szabados,
  L.}, {Szegedi-Elek, E.}, {Tapiador, D.}, {Taris, F.}, {Tauran, G.}, {Taylor,
  M. B.}, {Teixeira, R.}, {Terrett, D.}, {Tingley, B.}, {Trager, S. C.},
  {Turon, C.}, {Ulla, A.}, {Utrilla, E.}, {Valentini, G.}, {van Elteren, A.},
  {Van Hemelryck, E.}, {van Leeuwen, M.}, {Varadi, M.}, {Vecchiato, A.},
  {Veljanoski, J.}, {Via, T.}, {Vicente, D.}, {Vogt, S.}, {Voss, H.}, {Votruba,
  V.}, {Voutsinas, S.}, {Walmsley, G.}, {Weiler, M.}, {Weingrill, K.}, {Werner,
  D.}, {Wevers, T.}, {Whitehead, G.}, {Wyrzykowski, L.}, {Yoldas, A.}, {Zerjal,
  M.}, {Zucker, S.}, {Zurbach, C.}, {Zwitter, T.}, {Alecu, A.}, {Allen, M.},
  {Allende Prieto, C.}, {Amorim, A.}, {Anglada-Escud\'e, G.}, {Arsenijevic,
  V.}, {Azaz, S.}, {Balm, P.}, {Beck, M.}, {Bernstein, H.-H.}, {Bigot, L.},
  {Bijaoui, A.}, {Blasco, C.}, {Bonfigli, M.}, {Bono, G.}, {Boudreault, S.},
  {Bressan, A.}, {Brown, S.}, {Brunet, P.-M.}, {Bunclark, P.}, {Buonanno, R.},
  {Butkevich, A. G.}, {Carret, C.}, {Carrion, C.}, {Chemin, L.}, {Ch\'ereau,
  F.}, {Corcione, L.}, {Darmigny, E.}, {de Boer, K. S.}, {de Teodoro, P.}, {de
  Zeeuw, P. T.}, {Delle Luche, C.}, {Domingues, C. D.}, {Dubath, P.}, {Fodor,
  F.}, {Fr\'ezouls, B.}, {Fries, A.}, {Fustes, D.}, {Fyfe, D.}, {Gallardo, E.},
  {Gallegos, J.}, {Gardiol, D.}, {Gebran, M.}, {Gomboc, A.}, {G\'omez, A.},
  {Grux, E.}, {Gueguen, A.}, {Heyrovsky, A.}, {Hoar, J.}, {Iannicola, G.},
  {Isasi Parache, Y.}, {Janotto, A.-M.}, {Joliet, E.}, {Jonckheere, A.}, {Keil,
  R.}, {Kim, D.-W.}, {Klagyivik, P.}, {Klar, J.}, {Knude, J.}, {Kochukhov, O.},
  {Kolka, I.}, {Kos, J.}, {Kutka, A.}, {Lainey, V.}, {LeBouquin, D.}, {Liu,
  C.}, {Loreggia, D.}, {Makarov, V. V.}, {Marseille, M. G.}, {Martayan, C.},
  {Martinez-Rubi, O.}, {Massart, B.}, {Meynadier, F.}, {Mignot, S.}, {Munari,
  U.}, {Nguyen, A.-T.}, {Nordlander, T.}, {Ocvirk, P.}, {O\'{}Flaherty, K. S.},
  {Olias Sanz, A.}, {Ortiz, P.}, {Osorio, J.}, {Oszkiewicz, D.}, {Ouzounis,
  A.}, {Palmer, M.}, {Park, P.}, {Pasquato, E.}, {Peltzer, C.}, {Peralta, J.},
  {P\'eturaud, F.}, {Pieniluoma, T.}, {Pigozzi, E.}, {Poels, J.}, {Prat, G.},
  {Prod\'{}homme, T.}, {Raison, F.}, {Rebordao, J. M.}, {Risquez, D.},
  {Rocca-Volmerange, B.}, {Rosen, S.}, {Ruiz-Fuertes, M. I.}, {Russo, F.},
  {Sembay, S.}, {Serraller Vizcaino, I.}, {Short, A.}, {Siebert, A.}, {Silva,
  H.}, {Sinachopoulos, D.}, {Slezak, E.}, {Soffel, M.}, {Sosnowska, D.},
  {Straizys, V.}, {ter Linden, M.}, {Terrell, D.}, {Theil, S.}, {Tiede, C.},
  {Troisi, L.}, {Tsalmantza, P.}, {Tur, D.}, {Vaccari, M.}, {Vachier, F.},
  {Valles, P.}, {Van Hamme, W.}, {Veltz, L.}, {Virtanen, J.}, {Wallut, J.-M.},
  {Wichmann, R.}, {Wilkinson, M. I.}, {Ziaeepour, H.}, \& {Zschocke,
  S.}}]{gaia2016}
{Gaia Collaboration}, {Prusti, T.}, {de Bruijne, J. H. J.}, {et~al.} 2016,
  A\&A, 595, A1

\bibitem[{{Gaia Collaboration} {et~al.}(2023){Gaia Collaboration}, {Vallenari},
  {Brown}, {Prusti}, {de Bruijne}, {Arenou}, {Babusiaux}, {Biermann},
  {Creevey}, {Ducourant}, {Evans}, {Eyer}, {Guerra}, {Hutton}, {Jordi},
  {Klioner}, {Lammers}, {Lindegren}, {Luri}, {Mignard}, {Panem}, {Pourbaix},
  {Randich}, {Sartoretti}, {Soubiran}, {Tanga}, {Walton}, {Bailer-Jones},
  {Bastian}, {Drimmel}, {Jansen}, {Katz}, {Lattanzi}, {van Leeuwen}, {Bakker},
  {Cacciari}, {Casta{\~n}eda}, {De Angeli}, {Fabricius}, {Fouesneau},
  {Fr{\'e}mat}, {Galluccio}, {Guerrier}, {Heiter}, {Masana}, {Messineo},
  {Mowlavi}, {Nicolas}, {Nienartowicz}, {Pailler}, {Panuzzo}, {Riclet}, {Roux},
  {Seabroke}, {Sordo}, {Th{\'e}venin}, {Gracia-Abril}, {Portell}, {Teyssier},
  {Altmann}, {Andrae}, {Audard}, {Bellas-Velidis}, {Benson}, {Berthier},
  {Blomme}, {Burgess}, {Busonero}, {Busso}, {C{\'a}novas}, {Carry}, {Cellino},
  {Cheek}, {Clementini}, {Damerdji}, {Davidson}, {de Teodoro}, {Nu{\~n}ez
  Campos}, {Delchambre}, {Dell'Oro}, {Esquej}, {Fern{\'a}ndez-Hern{\'a}ndez},
  {Fraile}, {Garabato}, {Garc{\'\i}a-Lario}, {Gosset}, {Haigron}, {Halbwachs},
  {Hambly}, {Harrison}, {Hern{\'a}ndez}, {Hestroffer}, {Hodgkin}, {Holl},
  {Jan{\ss}en}, {Jevardat de Fombelle}, {Jordan}, {Krone-Martins}, {Lanzafame},
  {L{\"o}ffler}, {Marchal}, {Marrese}, {Moitinho}, {Muinonen}, {Osborne},
  {Pancino}, {Pauwels}, {Recio-Blanco}, {Reyl{\'e}}, {Riello}, {Rimoldini},
  {Roegiers}, {Rybizki}, {Sarro}, {Siopis}, {Smith}, {Sozzetti}, {Utrilla},
  {van Leeuwen}, {Abbas}, {{\'A}brah{\'a}m}, {Abreu Aramburu}, {Aerts},
  {Aguado}, {Ajaj}, {Aldea-Montero}, {Altavilla}, {{\'A}lvarez}, {Alves},
  {Anders}, {Anderson}, {Anglada Varela}, {Antoja}, {Baines}, {Baker},
  {Balaguer-N{\'u}{\~n}ez}, {Balbinot}, {Balog}, {Barache}, {Barbato},
  {Barros}, {Barstow}, {Bartolom{\'e}}, {Bassilana}, {Bauchet}, {Becciani},
  {Bellazzini}, {Berihuete}, {Bernet}, {Bertone}, {Bianchi}, {Binnenfeld},
  {Blanco-Cuaresma}, {Blazere}, {Boch}, {Bombrun}, {Bossini}, {Bouquillon},
  {Bragaglia}, {Bramante}, {Breedt}, {Bressan}, {Brouillet}, {Brugaletta},
  {Bucciarelli}, {Burlacu}, {Butkevich}, {Buzzi}, {Caffau}, {Cancelliere},
  {Cantat-Gaudin}, {Carballo}, {Carlucci}, {Carnerero}, {Carrasco},
  {Casamiquela}, {Castellani}, {Castro-Ginard}, {Chaoul}, {Charlot}, {Chemin},
  {Chiaramida}, {Chiavassa}, {Chornay}, {Comoretto}, {Contursi}, {Cooper},
  {Cornez}, {Cowell}, {Crifo}, {Cropper}, {Crosta}, {Crowley}, {Dafonte},
  {Dapergolas}, {David}, {David}, {de Laverny}, {De Luise}, {De March}, {De
  Ridder}, {de Souza}, {de Torres}, {del Peloso}, {del Pozo}, {Delbo},
  {Delgado}, {Delisle}, {Demouchy}, {Dharmawardena}, {Di Matteo}, {Diakite},
  {Diener}, {Distefano}, {Dolding}, {Edvardsson}, {Enke}, {Fabre}, {Fabrizio},
  {Faigler}, {Fedorets}, {Fernique}, {Fienga}, {Figueras}, {Fournier},
  {Fouron}, {Fragkoudi}, {Gai}, {Garcia-Gutierrez}, {Garcia-Reinaldos},
  {Garc{\'\i}a-Torres}, {Garofalo}, {Gavel}, {Gavras}, {Gerlach}, {Geyer},
  {Giacobbe}, {Gilmore}, {Girona}, {Giuffrida}, {Gomel}, {Gomez},
  {Gonz{\'a}lez-N{\'u}{\~n}ez}, {Gonz{\'a}lez-Santamar{\'\i}a},
  {Gonz{\'a}lez-Vidal}, {Granvik}, {Guillout}, {Guiraud},
  {Guti{\'e}rrez-S{\'a}nchez}, {Guy}, {Hatzidimitriou}, {Hauser}, {Haywood},
  {Helmer}, {Helmi}, {Sarmiento}, {Hidalgo}, {Hilger}, {H{\l}adczuk}, {Hobbs},
  {Holland}, {Huckle}, {Jardine}, {Jasniewicz}, {Jean-Antoine Piccolo},
  {Jim{\'e}nez-Arranz}, {Jorissen}, {Juaristi Campillo}, {Julbe}, {Karbevska},
  {Kervella}, {Khanna}, {Kontizas}, {Kordopatis}, {Korn}, {K{\'o}sp{\'a}l},
  {Kostrzewa-Rutkowska}, {Kruszy{\'n}ska}, {Kun}, {Laizeau}, {Lambert},
  {Lanza}, {Lasne}, {Le Campion}, {Lebreton}, {Lebzelter}, {Leccia}, {Leclerc},
  {Lecoeur-Taibi}, {Liao}, {Licata}, {Lindstr{\o}m}, {Lister}, {Livanou},
  {Lobel}, {Lorca}, {Loup}, {Madrero Pardo}, {Magdaleno Romeo}, {Managau},
  {Mann}, {Manteiga}, {Marchant}, {Marconi}, {Marcos}, {Marcos Santos},
  {Mar{\'\i}n Pina}, {Marinoni}, {Marocco}, {Marshall}, {Martin Polo},
  {Mart{\'\i}n-Fleitas}, {Marton}, {Mary}, {Masip}, {Massari},
  {Mastrobuono-Battisti}, {Mazeh}, {McMillan}, {Messina}, {Michalik}, {Millar},
  {Mints}, {Molina}, {Molinaro}, {Moln{\'a}r}, {Monari}, {Mongui{\'o}},
  {Montegriffo}, {Montero}, {Mor}, {Mora}, {Morbidelli}, {Morel}, {Morris},
  {Muraveva}, {Murphy}, {Musella}, {Nagy}, {Noval}, {Oca{\~n}a}, {Ogden},
  {Ordenovic}, {Osinde}, {Pagani}, {Pagano}, {Palaversa}, {Palicio},
  {Pallas-Quintela}, {Panahi}, {Payne-Wardenaar}, {Pe{\~n}alosa Esteller},
  {Penttil{\"a}}, {Pichon}, {Piersimoni}, {Pineau}, {Plachy}, {Plum}, {Poggio},
  {Pr{\v{s}}a}, {Pulone}, {Racero}, {Ragaini}, {Rainer}, {Raiteri}, {Rambaux},
  {Ramos}, {Ramos-Lerate}, {Re Fiorentin}, {Regibo}, {Richards}, {Rios Diaz},
  {Ripepi}, {Riva}, {Rix}, {Rixon}, {Robichon}, {Robin}, {Robin}, {Roelens},
  {Rogues}, {Rohrbasser}, {Romero-G{\'o}mez}, {Rowell}, {Royer}, {Ruz Mieres},
  {Rybicki}, {Sadowski}, {S{\'a}ez N{\'u}{\~n}ez}, {Sagrist{\`a} Sell{\'e}s},
  {Sahlmann}, {Salguero}, {Samaras}, {Sanchez Gimenez}, {Sanna},
  {Santove{\~n}a}, {Sarasso}, {Schultheis}, {Sciacca}, {Segol}, {Segovia},
  {S{\'e}gransan}, {Semeux}, {Shahaf}, {Siddiqui}, {Siebert}, {Siltala},
  {Silvelo}, {Slezak}, {Slezak}, {Smart}, {Snaith}, {Solano}, {Solitro},
  {Souami}, {Souchay}, {Spagna}, {Spina}, {Spoto}, {Steele},
  {Steidelm{\"u}ller}, {Stephenson}, {S{\"u}veges}, {Surdej}, {Szabados},
  {Szegedi-Elek}, {Taris}, {Taylor}, {Teixeira}, {Tolomei}, {Tonello}, {Torra},
  {Torra}, {Torralba Elipe}, {Trabucchi}, {Tsounis}, {Turon}, {Ulla}, {Unger},
  {Vaillant}, {van Dillen}, {van Reeven}, {Vanel}, {Vecchiato}, {Viala},
  {Vicente}, {Voutsinas}, {Weiler}, {Wevers}, {Wyrzykowski}, {Yoldas}, {Yvard},
  {Zhao}, {Zorec}, {Zucker}, \& {Zwitter}}]{gaia2023}
{Gaia Collaboration}, {Vallenari}, A., {Brown}, A.~G.~A., {et~al.} 2023, \aap,
  674, A1

\bibitem[{Gezari(2021)}]{Gezari_2021}
Gezari, S. 2021, ARA\&A, 59, 21

\bibitem[{{Giustini} {et~al.}(2020){Giustini}, {Miniutti}, \&
  {Saxton}}]{2020A&A...636L...2G}
{Giustini}, M., {Miniutti}, G., \& {Saxton}, R.~D. 2020, \aap, 636, L2

\bibitem[{Goodwin {et~al.}(2023)Goodwin, Alexander, Miller-Jones, Bietenholz,
  van Velzen, Anderson, Berger, Cendes, Chornock, Coppejans, Eftekhari,
  Gezari, Laskar, Ramirez-Ruiz, \& Saxton}]{Goodwin_2023}
Goodwin, A.~J., Alexander, K.~D., Miller-Jones, J. C.~A., {et~al.} 2023, MNRAS,
  522, 5084–5097

\bibitem[{{Guolo} {et~al.}(2024){Guolo}, {Gezari}, {Yao}, {van Velzen},
  {Hammerstein}, {Cenko}, \& {Tokayer}}]{guolo2024systematic}
{Guolo}, M., {Gezari}, S., {Yao}, Y., {et~al.} 2024, \apj, 966, 160

\bibitem[{Hawkins(2002)}]{hawkins2002}
Hawkins, M. 2002, MNRAS, 329, 76

\bibitem[{{HI4PI Collaboration:} {et~al.}(2016){HI4PI Collaboration:}, {Ben
  Bekhti, N.}, {Fl\"oer, L.}, {Keller, R.}, {Kerp, J.}, {Lenz, D.}, {Winkel,
  B.}, {Bailin, J.}, {Calabretta, M. R.}, {Dedes, L.}, {Ford, H. A.}, {Gibson,
  B. K.}, {Haud, U.}, {Janowiecki, S.}, {Kalberla, P. M. W.}, {Lockman, F. J.},
  {McClure-Griffiths, N. M.}, {Murphy, T.}, {Nakanishi, H.}, {Pisano, D. J.},
  \& {Staveley-Smith, L.}}]{HI4PICollaboration}
{HI4PI Collaboration:}, {Ben Bekhti, N.}, {Fl\"oer, L.}, {et~al.} 2016, \aap,
  594, A116

\bibitem[{{Hills}(1975)}]{1975Natur.254..295H}
{Hills}, J.~G. 1975, Nat, 254, 295

\bibitem[{{Homan} {et~al.}(2023){Homan}, {Krumpe, M.}, {Markowitz, A.}, {Saha,
  T.}, {Gokus, A.}, {Partington, E.}, {Lamer, G.}, {Malyali, A.}, {Liu, Z.},
  {Rau, A.}, {Grotova, I.}, {Cackett, E. M.}, {Buckley, D. A. H.}, {Ciroi, S.},
  {Di Mille, F.}, {Gendreau, K.}, {Gromadzki, M.}, {Krishnan, S.}, {Schramm,
  M.}, \& {Steiner, J. F.}}]{homan2344}
{Homan}, D., {Krumpe, M.}, {Markowitz, A.}, {et~al.} 2023, A\&A, 672, A167

\bibitem[{Hotan {et~al.}(2021)Hotan, Bunton, Chippendale, Whiting, Tuthill,
  Moss, McConnell, Amy, Huynh, Allison, Anderson, Bannister, Bastholm,
  Beresford, Bock, Bolton, Chapman, Chow, Collier, Cooray, Cornwell, Diamond,
  Edwards, Feain, Franzen, George, Gupta, Hampson, Harvey-Smith, Hayman,
  Heywood, Jacka, Jackson, Jackson, Jeganathan, Johnston, Kesteven, Kleiner,
  Koribalski, Lee-Waddell, Lenc, Lensson, Mackay, Mahony, McClure-Griffiths,
  McConigley, Mirtschin, Ng, Norris, Pearce, Phillips, Pilawa, Raja, Reynolds,
  Roberts, Roxby, Sadler, Shields, Schinckel, Serra, Shaw, Sweetnam, Troup,
  Tzioumis, Voronkov, \& Westmeier}]{Hotan_2021}
Hotan, A.~W., Bunton, J.~D., Chippendale, A.~P., {et~al.} 2021, PASA, 38

\bibitem[{{Itoh} {et~al.}(2020){Itoh}, {Utsumi}, {Inoue}, {Ohta}, {Doi},
  {Morokuma}, {Kawabata}, \& {Tanaka}}]{2021yCat..19010003I}
{Itoh}, R., {Utsumi}, Y., {Inoue}, Y., {et~al.} 2020, \apj, 901, 3

\bibitem[{{Kennicutt}(1983)}]{1983ApJ...272...54K}
{Kennicutt}, R.~C., J. 1983, \apj, 272, 54

\bibitem[{{Komossa} \& {Bade}(1999)}]{1999A&A...343..775K}
{Komossa}, S. \& {Bade}, N. 1999, \aap, 343, 775

\bibitem[{{K{\"o}nig} {et~al.}(2022{\natexlab{a}}){K{\"o}nig}, {Saxton},
  {Kretschmar}, {Angelini}, {Belanger}, {Evans}, {Freyberg}, {Savchenko},
  {Traulsen}, \& {Wilms}}]{2022A&C....3800529K}
{K{\"o}nig}, O., {Saxton}, R.~D., {Kretschmar}, P., {et~al.}
  2022{\natexlab{a}}, Astron. Comput., 38, 100529

\bibitem[{{K{\"o}nig} {et~al.}(2022{\natexlab{b}}){K{\"o}nig}, {Wilms},
  {Arcodia}, {Dauser}, {Dennerl}, {Doroshenko}, {Haberl}, {H{\"a}mmerich},
  {Kirsch}, {Kreykenbohm}, {Lorenz}, {Malyali}, {Merloni}, {Rau}, {Rauch},
  {Sala}, {Schwope}, {Suleimanov}, {Weber}, \& {Werner}}]{2022Natur.605..248K}
{K{\"o}nig}, O., {Wilms}, J., {Arcodia}, R., {et~al.} 2022{\natexlab{b}}, Nat,
  605, 248

\bibitem[{{Kraft} {et~al.}(1991){Kraft}, {Burrows}, \&
  {Nousek}}]{1991ApJ...374..344}
{Kraft}, R.~P., {Burrows}, D.~N., \& {Nousek}, J.~A. 1991, \apj, 374, 344

\bibitem[{{Lacy} {et~al.}(2020){Lacy}, {Baum}, {Chandler}, {Chatterjee},
  {Clarke}, {Deustua}, {English}, {Farnes}, {Gaensler}, {Gugliucci},
  {Hallinan}, {Kent}, {Kimball}, {Law}, {Lazio}, {Marvil}, {Mao}, {Medlin},
  {Mooley}, {Murphy}, {Myers}, {Osten}, {Richards}, {Rosolowsky}, {Rudnick},
  {Schinzel}, {Sivakoff}, {Sjouwerman}, {Taylor}, {White}, {Wrobel},
  {Andernach}, {Beasley}, {Berger}, {Bhatnager}, {Birkinshaw}, {Bower},
  {Brandt}, {Brown}, {Burke-Spolaor}, {Butler}, {Comerford}, {Demorest}, {Fu},
  {Giacintucci}, {Golap}, {G{\"u}th}, {Hales}, {Hiriart}, {Hodge}, {Horesh},
  {Ivezi{\'c}}, {Jarvis}, {Kamble}, {Kassim}, {Liu}, {Loinard}, {Lyons},
  {Masters}, {Mezcua}, {Moellenbrock}, {Mroczkowski}, {Nyland}, {O'Dea},
  {O'Sullivan}, {Peters}, {Radford}, {Rao}, {Robnett}, {Salcido}, {Shen},
  {Sobotka}, {Witz}, {Vaccari}, {van Weeren}, {Vargas}, {Williams}, \&
  {Yoon}}]{Lacy2020}
{Lacy}, M., {Baum}, S.~A., {Chandler}, C.~J., {et~al.} 2020, \pasp, 132, 035001

\bibitem[{{Lanzuisi} {et~al.}(2014){Lanzuisi}, {Ponti}, {Salvato}, {Hasinger},
  {Cappelluti}, {Bongiorno}, {Brusa}, {Lusso}, {Nandra}, {Merloni},
  {Silverman}, {Trump}, {Vignali}, {Comastri}, {Gilli}, {Schramm},
  {Steinhardt}, {Sanders}, {Kartaltepe}, {Rosario}, \&
  {Trakhtenbrot}}]{2014ApJ...781..105L}
{Lanzuisi}, G., {Ponti}, G., {Salvato}, M., {et~al.} 2014, \apj, 781, 105

\bibitem[{{Liu} {et~al.}(2022){Liu}, {Buchner}, {Nandra}, {Merloni}, {Dwelly},
  {Sanders}, {Salvato}, {Arcodia}, {Brusa}, {Wolf}, {Georgakakis}, {Boller},
  {Krumpe}, {Lamer}, {Waddell}, {Urrutia}, {Schwope}, {Robrade}, {Wilms},
  {Dauser}, {Comparat}, {Toba}, {Ichikawa}, {Iwasawa}, {Shen}, \&
  {Medel}}]{2022A&A...661A...5L}
{Liu}, T., {Buchner}, J., {Nandra}, K., {et~al.} 2022, \aap, 661, A5

\bibitem[{Liu {et~al.}(2023)Liu, Malyali, Krumpe, Homan, Goodwin, Grotova,
  Kawka, Rau, Merloni, Anderson, Miller-Jones, Markowitz, Ciroi, Mille,
  Schramm, Tang, Buckley, Gromadzki, Jin, \& Buchner}]{Liu_2023}
Liu, Z., Malyali, A., Krumpe, M., {et~al.} 2023, A\&A, 669, A75

\bibitem[{{Malyali} {et~al.}(2023){Malyali}, {Liu}, {Merloni}, {Rau},
  {Buchner}, {Ciroi}, {Di Mille}, {Grotova}, {Dwelly}, {Nandra}, {Salvato},
  {Homan}, \& {Krumpe}}]{Malyali_2023}
{Malyali}, A., {Liu}, Z., {Merloni}, A., {et~al.} 2023, \mnras, 520, 4209

\bibitem[{Malyali {et~al.}(2023)Malyali, Liu, Rau, Grotova, Merloni, Goodwin,
  Anderson, Miller-Jones, Kawka, Arcodia, Buchner, Nandra, Homan, \&
  Krumpe}]{malyali_rep}
Malyali, A., Liu, Z., Rau, A., {et~al.} 2023, MNRAS, 520, 3549

\bibitem[{{Malyali} {et~al.}(2021){Malyali}, {Rau}, {Merloni}, {Nandra},
  {Buchner}, {Liu}, {Gezari}, {Sollerman}, {Shappee}, {Trakhtenbrot}, {Arcavi},
  {Ricci}, {van Velzen}, {Goobar}, {Frederick}, {Kawka}, {Tartaglia}, {Burke},
  {Hiramatsu}, {Schramm}, {van der Boom}, {Anderson}, {Miller-Jones}, {Bellm},
  {Drake}, {Duev}, {Fremling}, {Graham}, {Masci}, {Rusholme}, {Soumagnac}, \&
  {Walters}}]{Malyali_AT2019avd}
{Malyali}, A., {Rau}, A., {Merloni}, A., {et~al.} 2021, \aap, 647, A9

\bibitem[{{Markowitz} {et~al.}(2024){Markowitz}, {Krumpe}, {Homan},
  {Gromadzki}, {Schramm}, {Boller}, {Krishnan}, {Saha}, {Wilms}, {Gokus},
  {Haemmerich}, {Winkler}, {Buchner}, {Buckley}, {Brogan}, \&
  {Reichart}}]{2024A&A...684A.101M}
{Markowitz}, A., {Krumpe}, M., {Homan}, D., {et~al.} 2024, \aap, 684, A101

\bibitem[{Marocco {et~al.}(2021)Marocco, Eisenhardt, Fowler, Kirkpatrick,
  Meisner, Schlafly, Stanford, Garcia, Caselden, Cushing, Cutri, Faherty,
  Gelino, Gonzalez, Jarrett, Koontz, Mainzer, Marchese, Mobasher, Schlegel,
  Stern, Teplitz, \& Wright}]{catwise2020}
Marocco, F., Eisenhardt, P. R.~M., Fowler, J.~W., {et~al.} 2021, ApJS, 253, 8

\bibitem[{{Massaro} {et~al.}(2015){Massaro}, {Maselli}, {Leto}, {Marchegiani},
  {Perri}, {Giommi}, \& {Piranomonte}}]{2016yCat.7274....0M}
{Massaro}, E., {Maselli}, A., {Leto}, C., {et~al.} 2015, \apss, 357, 75

\bibitem[{{Masterson} {et~al.}(2024){Masterson}, {De}, {Panagiotou}, {Kara},
  {Arcavi}, {Eilers}, {Frostig}, {Gezari}, {Grotova}, {Liu}, {Malyali},
  {Meisner}, {Merloni}, {Newsome}, {Rau}, {Simcoe}, \& {van
  Velzen}}]{masterson2024new}
{Masterson}, M., {De}, K., {Panagiotou}, C., {et~al.} 2024, \apj, 961, 211

\bibitem[{Matt {et~al.}(2003)Matt, Guainazzi, \& Maiolino}]{matt_2003}
Matt, G., Guainazzi, M., \& Maiolino, R. 2003, MNRAS, 342, 422

\bibitem[{McConnell {et~al.}(2020)McConnell, Hale, Lenc, Banfield, Heald,
  Hotan, Leung, Moss, Murphy, O’Brien, \& et~al.}]{RACS}
McConnell, D., Hale, C.~L., Lenc, E., {et~al.} 2020, PASA, 37, e048

\bibitem[{Merloni {et~al.}(2024)Merloni, Lamer, Liu, Ramos-Ceja, Brunner,
  Bulbul, Dennerl, Doroshenko, Freyberg, Friedrich, Gatuzz, Georgakakis,
  Haberl, Igo, Kreykenbohm, Liu, Maitra, Malyali, Mayer, Nandra, Predehl,
  Robrade, Salvato, Sanders, Stewart, Tubín-Arenas, Weber, Wilms, Arcodia,
  Artis, Aschersleben, Avakyan, Aydar, Bahar, Balzer, Becker, Berger, Boller,
  Bornemann, Brüggen, Brusa, Buchner, Burwitz, Camilloni, Clerc, Comparat,
  Coutinho, Czesla, Dannhauer, Dauner, Dauser, Dietl, Dolag, Dwelly, Egg, Ehl,
  Freund, Friedrich, Gaida, Garrel, Ghirardini, Gokus, Grünwald, Grandis,
  Grotova, Gruen, Gueguen, Hämmerich, Hamaus, Hasinger, Haubner, Homan,
  Ider~Chitham, Joseph, Joyce, König, Kaltenbrunner, Khokhriakova, Kink,
  Kirsch, Kluge, Knies, Krippendorf, Krumpe, Kurpas, Li, Liu, Locatelli,
  Lorenz, Müller, Magaudda, Mannes, McCall, Meidinger, Michailidis, Migkas,
  Muñoz-Giraldo, Musiimenta, Nguyen-Dang, Ni, Olechowska, Ota, Pacaud, Pasini,
  Perinati, Pires, Pommranz, Ponti, Poppenhaeger, Pühlhofer, Rau, Reh,
  Reiprich, Roster, Saeedi, Santangelo, Sasaki, Schmitt, Schneider, Schrabback,
  Schuster, Schwope, Seppi, Serim, Shreeram, Sokolova-Lapa, Starck, Stelzer,
  Stierhof, Suleimanov, Tenzer, Traulsen, Trümper, Tsuge, Urrutia, Veronica,
  Waddell, Willer, Wolf, Yeung, Zainab, Zangrandi, Zhang, Zhang, \&
  Zheng}]{Merloni_2024}
Merloni, A., Lamer, G., Liu, T., {et~al.} 2024, A\&A, 682, A34

\bibitem[{{Miniutti} {et~al.}(2023){Miniutti}, {Giustini}, {Arcodia}, {Saxton},
  {Read}, {Bianchi}, \& {Alexander}}]{2023miniuttu}
{Miniutti}, G., {Giustini}, M., {Arcodia}, R., {et~al.} 2023, \aap, 670, A93

\bibitem[{{Miniutti} {et~al.}(2019){Miniutti}, {Saxton}, {Giustini},
  {Alexander}, {Fender}, {Heywood}, {Monageng}, {Coriat}, {Tzioumis}, {Read},
  {Knigge}, {Gandhi}, {Pretorius}, \&
  {Ag{\'\i}s-Gonz{\'a}lez}}]{2019Natur.573..381M}
{Miniutti}, G., {Saxton}, R.~D., {Giustini}, M., {et~al.} 2019, Nat, 573, 381

\bibitem[{{Murphy} {et~al.}(2011){Murphy}, {Condon}, {Schinnerer}, {Kennicutt},
  {Calzetti}, {Armus}, {Helou}, {Turner}, {Aniano}, {Beir{\~a}o}, {Bolatto},
  {Brandl}, {Croxall}, {Dale}, {Donovan Meyer}, {Draine}, {Engelbracht},
  {Hunt}, {Hao}, {Koda}, {Roussel}, {Skibba}, \& {Smith}}]{2011ApJ...737...67M}
{Murphy}, E.~J., {Condon}, J.~J., {Schinnerer}, E., {et~al.} 2011, \apj, 737,
  67

\bibitem[{Noda \& Done(2018)}]{noda2018}
Noda, H. \& Done, C. 2018, MNRAS, 480, 3898

\bibitem[{{Paggi} {et~al.}(2020){Paggi}, {Bonato}, {Raiteri}, {Villata}, {De
  Zotti}, \& {Carnerero}}]{2020yCat..36410062P}
{Paggi}, A., {Bonato}, M., {Raiteri}, C.~M., {et~al.} 2020, \aap, 641, A62

\bibitem[{Paolillo {et~al.}(2017)Paolillo, Papadakis, Brandt, Luo, Xue, Tozzi,
  Shemmer, Allevato, Bauer, Comastri, Gilli, Koekemoer, Liu, Vignali, Vito,
  Yang, Wang, \& Zheng}]{10.1093/mnras/stx1761}
Paolillo, M., Papadakis, I., Brandt, W.~N., {et~al.} 2017, MNRAS, 471, 4398

\bibitem[{{Planck Collaboration} {et~al.}(2016){Planck Collaboration}, {Ade},
  {Aghanim}, {Arnaud}, {Ashdown}, {Aumont}, {Baccigalupi}, {Banday},
  {Barreiro}, {Bartlett}, {Bartolo}, {Battaner}, {Battye}, {Benabed},
  {Beno{\^\i}t}, {Benoit-L{\'e}vy}, {Bernard}, {Bersanelli}, {Bielewicz},
  {Bock}, {Bonaldi}, {Bonavera}, {Bond}, {Borrill}, {Bouchet}, {Boulanger},
  {Bucher}, {Burigana}, {Butler}, {Calabrese}, {Cardoso}, {Catalano},
  {Challinor}, {Chamballu}, {Chary}, {Chiang}, {Chluba}, {Christensen},
  {Church}, {Clements}, {Colombi}, {Colombo}, {Combet}, {Coulais}, {Crill},
  {Curto}, {Cuttaia}, {Danese}, {Davies}, {Davis}, {de Bernardis}, {de Rosa},
  {de Zotti}, {Delabrouille}, {D{\'e}sert}, {Di Valentino}, {Dickinson},
  {Diego}, {Dolag}, {Dole}, {Donzelli}, {Dor{\'e}}, {Douspis}, {Ducout},
  {Dunkley}, {Dupac}, {Efstathiou}, {Elsner}, {En{\ss}lin}, {Eriksen},
  {Farhang}, {Fergusson}, {Finelli}, {Forni}, {Frailis}, {Fraisse},
  {Franceschi}, {Frejsel}, {Galeotta}, {Galli}, {Ganga}, {Gauthier}, {Gerbino},
  {Ghosh}, {Giard}, {Giraud-H{\'e}raud}, {Giusarma}, {Gjerl{\o}w},
  {Gonz{\'a}lez-Nuevo}, {G{\'o}rski}, {Gratton}, {Gregorio}, {Gruppuso},
  {Gudmundsson}, {Hamann}, {Hansen}, {Hanson}, {Harrison}, {Helou},
  {Henrot-Versill{\'e}}, {Hern{\'a}ndez-Monteagudo}, {Herranz}, {Hildebrandt},
  {Hivon}, {Hobson}, {Holmes}, {Hornstrup}, {Hovest}, {Huang}, {Huffenberger},
  {Hurier}, {Jaffe}, {Jaffe}, {Jones}, {Juvela}, {Keih{\"a}nen}, {Keskitalo},
  {Kisner}, {Kneissl}, {Knoche}, {Knox}, {Kunz}, {Kurki-Suonio}, {Lagache},
  {L{\"a}hteenm{\"a}ki}, {Lamarre}, {Lasenby}, {Lattanzi}, {Lawrence}, {Leahy},
  {Leonardi}, {Lesgourgues}, {Levrier}, {Lewis}, {Liguori}, {Lilje},
  {Linden-V{\o}rnle}, {L{\'o}pez-Caniego}, {Lubin}, {Mac{\'\i}as-P{\'e}rez},
  {Maggio}, {Maino}, {Mandolesi}, {Mangilli}, {Marchini}, {Maris}, {Martin},
  {Martinelli}, {Mart{\'\i}nez-Gonz{\'a}lez}, {Masi}, {Matarrese}, {McGehee},
  {Meinhold}, {Melchiorri}, {Melin}, {Mendes}, {Mennella}, {Migliaccio},
  {Millea}, {Mitra}, {Miville-Desch{\^e}nes}, {Moneti}, {Montier}, {Morgante},
  {Mortlock}, {Moss}, {Munshi}, {Murphy}, {Naselsky}, {Nati}, {Natoli},
  {Netterfield}, {N{\o}rgaard-Nielsen}, {Noviello}, {Novikov}, {Novikov},
  {Oxborrow}, {Paci}, {Pagano}, {Pajot}, {Paladini}, {Paoletti}, {Partridge},
  {Pasian}, {Patanchon}, {Pearson}, {Perdereau}, {Perotto}, {Perrotta},
  {Pettorino}, {Piacentini}, {Piat}, {Pierpaoli}, {Pietrobon}, {Plaszczynski},
  {Pointecouteau}, {Polenta}, {Popa}, {Pratt}, {Pr{\'e}zeau}, {Prunet},
  {Puget}, {Rachen}, {Reach}, {Rebolo}, {Reinecke}, {Remazeilles}, {Renault},
  {Renzi}, {Ristorcelli}, {Rocha}, {Rosset}, {Rossetti}, {Roudier},
  {Rouill{\'e} d'Orfeuil}, {Rowan-Robinson}, {Rubi{\~n}o-Mart{\'\i}n},
  {Rusholme}, {Said}, {Salvatelli}, {Salvati}, {Sandri}, {Santos},
  {Savelainen}, {Savini}, {Scott}, {Seiffert}, {Serra}, {Shellard}, {Spencer},
  {Spinelli}, {Stolyarov}, {Stompor}, {Sudiwala}, {Sunyaev}, {Sutton},
  {Suur-Uski}, {Sygnet}, {Tauber}, {Terenzi}, {Toffolatti}, {Tomasi},
  {Tristram}, {Trombetti}, {Tucci}, {Tuovinen}, {T{\"u}rler}, {Umana},
  {Valenziano}, {Valiviita}, {Van Tent}, {Vielva}, {Villa}, {Wade}, {Wandelt},
  {Wehus}, {White}, {White}, {Wilkinson}, {Yvon}, {Zacchei}, \&
  {Zonca}}]{2016A&A...594A..13P}
{Planck Collaboration}, {Ade}, P.~A.~R., {Aghanim}, N., {et~al.} 2016, \aap,
  594, A13

\bibitem[{{Predehl} {et~al.}(2021){Predehl}, {Andritschke, R.}, {Arefiev, V.},
  {Babyshkin, V.}, {Batanov, O.}, {Becker, W.}, {B\"ohringer, H.}, {Bogomolov,
  A.}, {Boller, T.}, {Borm, K.}, {Bornemann, W.}, {Br\"auninger, H.},
  {Br\"uggen, M.}, {Brunner, H.}, {Brusa, M.}, {Bulbul, E.}, {Buntov, M.},
  {Burwitz, V.}, {Burkert, W.}, {Clerc, N.}, {Churazov, E.}, {Coutinho, D.},
  {Dauser, T.}, {Dennerl, K.}, {Doroshenko, V.}, {Eder, J.}, {Emberger, V.},
  {Eraerds, T.}, {Finoguenov, A.}, {Freyberg, M.}, {Friedrich, P.}, {Friedrich,
  S.}, {F\"urmetz, M.}, {Georgakakis, A.}, {Gilfanov, M.}, {Granato, S.},
  {Grossberger, C.}, {Gueguen, A.}, {Gureev, P.}, {Haberl, F.}, {H\"alker, O.},
  {Hartner, G.}, {Hasinger, G.}, {Huber, H.}, {Ji, L.}, {Kienlin, A. v.},
  {Kink, W.}, {Korotkov, F.}, {Kreykenbohm, I.}, {Lamer, G.}, {Lomakin, I.},
  {Lapshov, I.}, {Liu, T.}, {Maitra, C.}, {Meidinger, N.}, {Menz, B.},
  {Merloni, A.}, {Mernik, T.}, {Mican, B.}, {Mohr, J.}, {M\"uller, S.},
  {Nandra, K.}, {Nazarov, V.}, {Pacaud, F.}, {Pavlinsky, M.}, {Perinati, E.},
  {Pfeffermann, E.}, {Pietschner, D.}, {Ramos-Ceja, M. E.}, {Rau, A.},
  {Reiffers, J.}, {Reiprich, T. H.}, {Robrade, J.}, {Salvato, M.}, {Sanders,
  J.}, {Santangelo, A.}, {Sasaki, M.}, {Scheuerle, H.}, {Schmid, C.}, {Schmitt,
  J.}, {Schwope, A.}, {Shirshakov, A.}, {Steinmetz, M.}, {Stewart, I.},
  {Str\"uder, L.}, {Sunyaev, R.}, {Tenzer, C.}, {Tiedemann, L.}, {Tr\"umper,
  J.}, {Voron, V.}, {Weber, P.}, {Wilms, J.}, \& {Yaroshenko,
  V.}}]{predehl_2021}
{Predehl}, P., {Andritschke, R.}, {Arefiev, V.}, {et~al.} 2021, A\&A, 647, A1

\bibitem[{{Quintin} {et~al.}(2023){Quintin}, {Webb}, {Guillot}, {Miniutti},
  {Kammoun}, {Giustini}, {Arcodia}, {Soucail}, {Clerc}, {Amato}, \&
  {Markwardt}}]{2023Quintin}
{Quintin}, E., {Webb}, N.~A., {Guillot}, S., {et~al.} 2023, \aap, 675, A152

\bibitem[{{Rees}(1988)}]{1988Natur.333..523R}
{Rees}, M.~J. 1988, Nat, 333, 523

\bibitem[{{Ricci} {et~al.}(2020){Ricci}, {Kara}, {Loewenstein}, {Trakhtenbrot},
  {Arcavi}, {Remillard}, {Fabian}, {Gendreau}, {Arzoumanian}, {Li}, {Ho},
  {MacLeod}, {Cackett}, {Altamirano}, {Gandhi}, {Kosec}, {Pasham}, {Steiner},
  \& {Chan}}]{2020ApJ...898L...1R}
{Ricci}, C., {Kara}, E., {Loewenstein}, M., {et~al.} 2020, \apjl, 898, L1

\bibitem[{{Risaliti} {et~al.}(2002){Risaliti}, {Elvis}, \&
  {Nicastro}}]{2002ApJ...571..234R}
{Risaliti}, G., {Elvis}, M., \& {Nicastro}, F. 2002, \apj, 571, 234

\bibitem[{Ruiz {et~al.}(2022)Ruiz, Georgakakis, Gerakakis, Saxton, Kretschmar,
  Akylas, \& Georgantopoulos}]{ruiz2022}
Ruiz, A., Georgakakis, A., Gerakakis, S., {et~al.} 2022, MNRAS, 511, 4265

\bibitem[{{Salvato} {et~al.}(2018){Salvato}, {Buchner}, {Budav{\'a}ri},
  {Dwelly}, {Merloni}, {Brusa}, {Rau}, {Fotopoulou}, \& {Nandra}}]{Salvato2018}
{Salvato}, M., {Buchner}, J., {Budav{\'a}ri}, T., {et~al.} 2018, \mnras, 473,
  4937

\bibitem[{Salvato {et~al.}(2022)Salvato, Wolf, Dwelly, Georgakakis, Brusa,
  Merloni, Liu, Toba, Nandra, Lamer, Buchner, Schneider, Freund, Rau, Schwope,
  Nishizawa, Klein, Arcodia, Comparat, Musiimenta, Nagao, Brunner, Malyali,
  Finoguenov, Anderson, Shen, Ibarra-Medel, Trump, Brandt, Urry, Rivera,
  Krumpe, Urrutia, Miyaji, Ichikawa, Schneider, Fresco, Boller, Haase,
  Brownstein, Lane, Bizyaev, \& Nitschelm}]{Salvato_2022}
Salvato, M., Wolf, J., Dwelly, T., {et~al.} 2022, A\&A, 661, A3

\bibitem[{Saxton {et~al.}(2012)Saxton, Soria, Wu, \& Kuin}]{Saxton_2012}
Saxton, C.~J., Soria, R., Wu, K., \& Kuin, N. P.~M. 2012, MNRAS, 422,
  1625–1639

\bibitem[{{Saxton} {et~al.}(2020){Saxton}, {Komossa}, {Auchettl}, \&
  {Jonker}}]{2020SSRv..216...85S}
{Saxton}, R., {Komossa}, S., {Auchettl}, K., \& {Jonker}, P.~G. 2020, \ssr,
  216, 85

\bibitem[{{Saxton} {et~al.}(2022){Saxton}, {K{\"o}nig}, {Descalzo}, {Belanger},
  {Kretschmar}, {Gabriel}, {Evans}, {Ibarra}, {Colomo}, {Sarmiento}, {Salgado},
  {Agrafojo}, \& {Kuulkers}}]{2022A&C....3800531S}
{Saxton}, R.~D., {K{\"o}nig}, O., {Descalzo}, M., {et~al.} 2022, Astron.
  Comput., 38, 100531

\bibitem[{Saxton {et~al.}(2008)Saxton, Read, Esquej, Freyberg, Altieri, \&
  Bermejo}]{Saxton_2008_xmmslew}
Saxton, R.~D., Read, A.~M., Esquej, P., {et~al.} 2008, A\&A, 480, 611

\bibitem[{{Schmidt}(1968)}]{1968ApJ...151..393S}
{Schmidt}, M. 1968, \apj, 151, 393

\bibitem[{{Seppi} {et~al.}(2022){Seppi}, {Comparat}, {Bulbul}, {Nandra},
  {Merloni}, {Clerc}, {Liu}, {Ghirardini}, {Liu}, {Salvato}, {Sanders},
  {Wilms}, {Dwelly}, {Dauser}, {K{\"o}nig}, {Ramos-Ceja}, {Garrel}, \&
  {Reiprich}}]{seppi2022}
{Seppi}, R., {Comparat}, J., {Bulbul}, E., {et~al.} 2022, \aap, 665, A78

\bibitem[{{Simmonds} {et~al.}(2018){Simmonds}, {Buchner}, {Salvato}, {Hsu}, \&
  {Bauer}}]{2018A&A...618A..66S}
{Simmonds}, C., {Buchner}, J., {Salvato}, M., {Hsu}, L.~T., \& {Bauer}, F.~E.
  2018, \aap, 618, A66

\bibitem[{{Stern} {et~al.}(2012){Stern}, {Assef}, {Benford}, {Blain}, {Cutri},
  {Dey}, {Eisenhardt}, {Griffith}, {Jarrett}, {Lake}, {Masci}, {Petty},
  {Stanford}, {Tsai}, {Wright}, {Yan}, {Harrison}, \& {Madsen}}]{stern2012}
{Stern}, D., {Assef}, R.~J., {Benford}, D.~J., {et~al.} 2012, \apj, 753, 30

\bibitem[{{Sunyaev} {et~al.}(2021){Sunyaev}, {Arefiev}, {Babyshkin},
  {Bogomolov}, {Borisov}, {Buntov}, {Brunner}, {Burenin}, {Churazov},
  {Coutinho}, {Eder}, {Eismont}, {Freyberg}, {Gilfanov}, {Gureyev}, {Hasinger},
  {Khabibullin}, {Kolmykov}, {Komovkin}, {Krivonos}, {Lapshov}, {Levin},
  {Lomakin}, {Lutovinov}, {Medvedev}, {Merloni}, {Mernik}, {Mikhailov},
  {Molodtsov}, {Mzhelsky}, {M{\"u}ller}, {Nandra}, {Nazarov}, {Pavlinsky},
  {Poghodin}, {Predehl}, {Robrade}, {Sazonov}, {Scheuerle}, {Shirshakov},
  {Tkachenko}, \& {Voron}}]{2021A&A...656A.132S}
{Sunyaev}, R., {Arefiev}, V., {Babyshkin}, V., {et~al.} 2021, \aap, 656, A132

\bibitem[{{Tub{\'\i}n-Arenas} {et~al.}(2024){Tub{\'\i}n-Arenas}, {Krumpe},
  {Lamer}, {Haase}, {Sanders}, {Brunner}, {Homan}, {Schwope}, {Georgakakis},
  {Poppenhaeger}, {Traulsen}, {K{\"o}nig}, {Merloni}, {Gueguen}, {Strong}, \&
  {Liu}}]{tubinarenas2024erosita}
{Tub{\'\i}n-Arenas}, D., {Krumpe}, M., {Lamer}, G., {et~al.} 2024, \aap, 682,
  A35

\bibitem[{{Valenti} {et~al.}(2013){Valenti}, {Graham}, {Howell}, {Sand},
  {Parrent}, {Hadjiyska}, {Walker}, {Rabinowitz}, {Baltay}, {Ellman},
  {McKinnon}, {Feindt}, \& {Nugent}}]{2013ATel.4958....1V}
{Valenti}, S., {Graham}, M.~L., {Howell}, D.~A., {et~al.} 2013, The
  Astronomer's Telegram, 4958, 1

\bibitem[{{Voges} {et~al.}(1999){Voges}, {Aschenbach}, {Boller},
  {Br{\"a}uninger}, {Briel}, {Burkert}, {Dennerl}, {Englhauser}, {Gruber},
  {Haberl}, {Hartner}, {Hasinger}, {K{\"u}rster}, {Pfeffermann}, {Pietsch},
  {Predehl}, {Rosso}, {Schmitt}, {Tr{\"u}mper}, \&
  {Zimmermann}}]{voges1999rosat}
{Voges}, W., {Aschenbach}, B., {Boller}, T., {et~al.} 1999, \aap, 349, 389

\bibitem[{{Wenger} {et~al.}(2000){Wenger}, {Ochsenbein}, {Egret}, {Dubois},
  {Bonnarel}, {Borde}, {Genova}, {Jasniewicz}, {Lalo{\"e}}, {Lesteven}, \&
  {Monier}}]{wenger_simbad}
{Wenger}, M., {Ochsenbein}, F., {Egret}, D., {et~al.} 2000, \aaps, 143, 9

\bibitem[{{Wilkins} \& {Gallo}(2015)}]{2015MNRAS.449..129W}
{Wilkins}, D.~R. \& {Gallo}, L.~C. 2015, \mnras, 449, 129

\bibitem[{{Wilms} {et~al.}(2020){Wilms}, {Kreykenbohm}, {Weber}, {Falkner},
  {Dauser}, {Knies}, {Koenig}, {Malyali}, {Rau}, {Merloni}, {Bogensberger},
  {Brunner}, {Buchner}, {Carpano}, {Freyberg}, {Haberl}, {Maitra}, {Salvato},
  {Doroshenko}, {Ducci}, {Ji}, {Schmitt}, \& {Schwope}}]{2020ATel13416....1W}
{Wilms}, J., {Kreykenbohm}, I., {Weber}, P., {et~al.} 2020, The Astronomer's
  Telegram, 13416, 1

\bibitem[{{Yuan} {et~al.}(2022){Yuan}, {Zhang}, {Chen}, \& {Ling}}]{Yuan_2022}
{Yuan}, W., {Zhang}, C., {Chen}, Y., \& {Ling}, Z. 2022, in Handbook of X-ray
  and Gamma-ray Astrophysics, 86

\end{thebibliography}

\begin{appendix}
    
\section{Upper limit calculation}
\label{appendix_ul}

The eROSITA data was reduced using the eROSITA Science Analysis Software (\texttt{eSASS}) version 020 \citep{brunner22}. We used Eventfiles merged for all seven telescopes in the full band "0" (0.2--10.0 keV) and other necessary files (images, background images, exposure maps, and detection masks) in the energy band "4" (0.2--2.3 keV) also merged for all seven telescopes. The upper limit was then computed at the position of the brightest detection in another eRASS. 

First, we used the task \texttt{apetool} to generate PSF maps for the energy band "4", which were used to calculate aperture photometry as well as sensitivity maps and obtain total photon counts in the source region and background counts. The extraction apertures are defined in units of Encircled Energy Fraction (EEF), namely, a radius that includes a certain fraction of the total source photons. We set the source region radius to $\mathrm{re = 0.8}$. The estimation of the expected background level within the extraction region was done using SourceMap products generated by \texttt{ERMLDET}, which are the images of the model background plus a model of the detected sources. In addition, in cases where the local background is contaminated, \texttt{apetool} removes model sources from the SourceMap if an input position is closer than  radius $\mathrm{rr = 0.7}$.

Second, to calculate the upper limits, we used the Bayesian approach by \citet{1991ApJ...374..344} to estimate the photon count upper limit at a given confidence level. We chose a single-side confidence level of 0.9987 \citep{ruiz2022}, corresponding to a double-side confidence level of 0.997, to estimate the 3$\sigma$ upper limit. The measurement was considered an upper limit if the Poisson false probability that measured gross counts in the source region are due to the background noise being more than $\mathrm{4\times10^{-6}}$ (a typical value for eROSITA). Otherwise, the measurement was considered a detection, and we used 1$\mathrm{\sigma}$ lower and upper limit to estimate the detection counts with an error. We extracted the vignetting corrected exposure time at the X-ray position from the exposure map in the energy band 0.2--2.3 keV and calculated the count rate and subsequently correct for the EEF.  

Finally, the source photon count rate was converted into flux. We used the standard energy conversion factor ($\mathrm{ECF = 1.074\times 10^{12}}$) from 1B eROSITA source catalog for 0.2--2.3 keV for the straightforward comparison with the source catalog fluxes. The spectral model assumed for ECF is an absorbed power law model with fixed $\mathrm{\Gamma = 2}$, $\mathrm{N_{\rm H} = 3\times 10^{20}cm^{-2}}$. A similar approach of the eROSITA upper limit computation is described in \citet{tubinarenas2024erosita}, which published upper limits for the public DR1 data release.

\section{Optical follow-up of X-ray transients}
\label{optical_fu}

The optical follow-up of eRO-ExTra sources was performed with NTT/EFOSC2, ANU/WiFeS, LCO/FLOYDS, Las Campanas/IMACS, and SALT. The summary of the observations and measured redshifts are presented in Table~\ref{tab:optical_f_up}, where (RA, DEC) are coordinates of observed optical counterparts. 

NTT: The sources were observed with the ESO Faint Object
Spectrograph and Camera v.2 (EFOSC2; \citealt{1984Msngr..38....9B}) mounted on the ESO New Technology Telescope (NTT) in La Silla, Chile (proposal IDs: 109.23JL.001 and 108.225J.001, PI Grotova; 106.21RU.001, 110.2465.001/002 PI Malyali). We used grism 13 and the 1.2\arcsec slit, providing a wavelength range of $\mathrm{3685-9315\AA}$ with a dispersion of $\mathrm{2.77\AA/pixel}$. The data were reduced and calibrated using the esoreflex pipeline \citep[][v2.11.5]{2013A&A...559A..96F}. The He+Ar arcs were used to obtain the wavelength calibration, and the standard stars were used for flux calibration, which were observed with the same grism and the same slit oriented along the parallactic angle at corresponding observation nights.

SALT: The spectra were obtained using the RSS instrument \citep{2003SPIE.4841.1463B} on the Southern African Large Telescope (SALT; \citealt{2006SPIE.6267E..0ZB}). The PyRAF-based PySALT package\footnote{\url{https://astronomers.salt.ac.za/software/pysalt-documentation}} \citep{2010SPIE.7737E..25C} was used for spectral reduction, which includes corrections for gain and cross talk and performs bias subtraction. The spectrum was extracted using standard IRAF\footnote{\url{https://iraf.net}} tasks, including wavelength calibration (neon and argon calibration lamp exposures were taken, one immediately before and one immediately after the science spectra, respectively), background subtraction, and one-dimensional spectrum extraction. The pupil (i.e., the view of the mirror from the tracker) moves during all SALT observations, causing the effective area of the telescope to change during exposures. Therefore, no absolute flux calibration is possible. However, by observing spectro-photometric standards during twilight, we were able to obtain relative flux calibration, allowing the recovery of
the correct spectral shape and relative line strengths.

WiFeS: We obtained spectra with the Wide
field Spectrograph (WiFeS; \citealt{2010Ap&SS.327..245D}) mounted on the ANU 2.3m telescope at Siding Spring Observatory on (proposal ID 1210124, 2210131 PI: Miller-Jones). The data were reduced using the PyWiFeS reduction pipeline \citep{Childress2014}. The pipeline produces three-dimensional sets consisting of spatially resolved, bias subtracted, flat-fielded, wavelength- and flux-calibrated spectra for each slitlet. We then extracted background-subtracted spectra from the slitlets that provided the most significant flux using the task apall in IRAF.

FLOYDS: We obtained spectra with FLOYDS spectrographs \citep{FLOYDS} mounted on the Las Cumbres Observatory 2m telescopes at Haleakala, Hawaii, and Siding Spring, Australia. The spectra were reduced using PyRAF\footnote{\url{https://lco.global/documentation/data/FLOYDS-pipeline}} tasks as described in \citep{2013ATel.4958....1V}.

Magellan: We obtained spectra with the IMACS Short Camera \citep{Dressler_2011} mounted on the 6.5m Baade Magellan Telescope located at Las Campanas Observatory, Chile. The spectra were reduced with IRAF following the usual procedure of overscan subtraction, flat-field correction, and wavelength calibration by means of a He-Ne-Ar lamp.

\begin{table*}
\small
\let\center\empty
\let\endcenter\relax
\centering
\caption{Optical spectroscopic follow-up summary of eRO-ExTra catalog.}
\label{tab:optical_f_up}

\begin{tabular}{lrrllll}
\hline
\hline
  \multicolumn{1}{c}{ERO\_NAME} &
  \multicolumn{1}{c}{RA} &
  \multicolumn{1}{c}{DEC} &
  \multicolumn{1}{c}{Redshift} &
  \multicolumn{1}{c}{Date} &
  \multicolumn{1}{c}{Telescope} &
  \multicolumn{1}{c}{Instrument} \\
\hline
eRASSU J001308.6-462524 & 3.2844 & -46.4220 & 0.198 & 2022-07-03 & La Silla/NTT & EFOSC2\\
1eRASS J004058.3-683816 & 10.2434 & -68.6388 & 0.151 & 2023-12-25 & La Silla/NTT & EFOSC2\\
eRASSU J011430.8-593654 & 18.6292 & -59.6156 & 0.156 & 2020-12-12 & Las Campanas/Baade & IMACS\\
 1eRASS J014133.4-443413 & 25.3891 & -44.5706 & 0.091 & 2021-12-06 & La Silla/NTT & EFOSC2\\
1eRASS J015752.7-520716 & 29.4716 & -52.1208 & 0.499 & 2023-12-23 & La Silla/NTT & EFOSC2\\
1eRASS J022756.0-840730 & 36.9827 & -84.1246 & 0.102 & 2022-12-31 & Siding Spring/LCO & FLOYDS\\
1eRASS J024140.3-422435 & 40.4171 & -42.4102 & 0.214 & 2023-02-03 &La Silla/NTT & EFOSC2\\
1eRASS J024930.1-274958 & 42.3771 & -27.8327 &0.089  & 2023-12-23  & La Silla/NTT & EFOSC2\\
eRASSU J030334.1-544438 & 45.8917 & -54.7442 & 0.091 & 2020-10-14 & Sutherland/SALT & RSS\\
eRASSU J034425.8-332719 & 56.1084 & -33.4553 & 0.092 & 2020-08-22 & Siding Spring/ANU 2.3m & WiFeS\\
eRASSU J040131.8-512541 & 60.3834 & -51.4278 & 0.088 & 2020-09-07 & Sutherland/SALT & RSS\\
1eRASS J042703.1-261156 & 66.7628 & -26.1968 & 0.173 & 2023-12-25 & La Silla/NTT & EFOSC2\\
1eRASS J043541.5-702302 & 68.9224 & -70.3840 & 0.067 & 2023-12-23 & La Silla/NTT & EFOSC2\\
1eRASS J043959.6-651403 & 69.9985 & -65.2342 & 0.152 & 2023-12-24 & La Silla/NTT & EFOSC2\\
1eRASS J045816.7-425956 & 74.5698 & -42.9991 & 0.146 & 2023-12-23  & La Silla/NTT & EFOSC2\\
1eRASS J051041.0-384512 & 77.6701 & -38.7533 & 0.088 & 2022-12-10 & Siding Spring/LCO & FLOYDS\\
1eRASS J051902.9-512630 & 79.7624 & -51.4424 & 0.106 & 2023-02-02 & La Silla/NTT & EFOSC2\\
1eRASS J052001.1-561254 & 80.0066 & -56.2151 & 0.095 & 2023-02-02 & La Silla/NTT & EFOSC2\\
1eRASS J053952.9-272523 & 84.9711 & -27.4215 & 0.231 & 2023-02-03 & La Silla/NTT & EFOSC2\\
1eRASS J054236.2-591856 & 85.6504 & -59.3154 & 0.1 & 2023-12-25 & La Silla/NTT & EFOSC2\\
1eRASS J055648.8-211921 & 89.2050 & -21.3207 & 0.050 & 2023-02-04 & La Silla/NTT & EFOSC2\\
1eRASS J055701.1-451409 & 89.2556 & -45.2360 & 0.248 & 2023-12-24 & La Silla/NTT & EFOSC2\\
eRASSU J060829.4-435320 & 92.1226 & -43.8892 & 0.076 & 2022-11-28 & Siding Spring/LCO & FLOYDS\\
eRASSU J063413.3-713908 & 98.5558 & -71.6518 & 0.134 & 2021-03-28 & La Silla/NTT & EFOSC2\\
eRASSU J064402.2-574247 & 101.0098 & -57.7123 & 0.207 & 2023-02-03 & La Silla/NTT & EFOSC2\\
1eRASS J064449.4-603704 & 101.2052 & -60.6178 & 0.115 & 2023-04-03 & La Silla/NTT & EFOSC2\\
1eRASS J065708.6-682354 & 104.2875 & -68.3975 & 0.097 & 2023-12-24 & La Silla/NTT & EFOSC2\\
1eRASS J071217.2-703056 & 108.0719 & -70.5160 & 0.037 & 2023-12-25 & La Silla/NTT & EFOSC2\\
eRASSt J074426.3+291606 & 116.1089 & 29.2688 & 0.039 & 2021-03-31 & La Silla/NTT & EFOSC2\\
1eRASS J075031.8+093330 & 117.6331 & 9.5588 & 0.060 & 2023-12-24 & La Silla/NTT & EFOSC2\\
1eRASS J080622.8+070247 & 121.5952 & 7.0468 & 0.053 & 2023-02-02 & La Silla/NTT & EFOSC2\\
eRASSU J082055.8+192538 & 125.2324 & 19.4270 & 0.124 & 2023-04-02 & La Silla/NTT & EFOSC2\\
1eRASS J082336.8+042303 & 125.9042 & 4.3842 & 0.028 & 2021-02-18 & Siding Spring/ANU 2.3m & WiFeS\\
1eRASS J083748.4-062335 & 129.4516 & -6.3937 & 0.094 & 2023-12-23 & La Silla/NTT & EFOSC2\\
1eRASS J091657.8+060955 & 139.2407 & 6.1659 & 0.091 & 2021-12-10 & La Silla/NTT & EFOSC2\\
1eRASS J093546.2-835823 & 143.9404 & -83.9732 & 0.100 & 2023-02-04 & La Silla/NTT & EFOSC2\\
1eRASS J103656.4-144922 & 159.2351 & -14.8228 & 0.076 & 2023-04-04 & La Silla/NTT & EFOSC2\\
1eRASS J110936.6-083725 & 167.4021 &-8.6241  & 0.24 & 2023-04-02  & La Silla/NTT & EFOSC2\\
1eRASS J121115.1-223501 & 182.8138 & -22.5835 & 0.084 & 2023-02-03 & La Silla/NTT & EFOSC2\\
1eRASS J122529.2-215250 & 186.3716 & -21.8800 & 0.034 & 2023-02-02 & La Silla/NTT & EFOSC2\\
1eRASS J130804.7+040128 & 197.0196 & 4.0246 & 0.164 & 2023-02-03 & La Silla/NTT & EFOSC2\\  
1eRASS J142140.3-295325 & 215.4210 & -29.8880  & 0.058 & 2021-03-29 & La Silla/NTT & EFOSC2\\
1eRASS J143308.1-772938 & 218.2832 & -77.4949 & 0.102 & 2023-02-04 & La Silla/NTT & EFOSC2\\
1eRASS J143915.1-270227 & 219.8133 & -27.0413 & 0.065 & 2023-02-04 & La Silla/NTT & EFOSC2\\
eRASSU J145622.8-283853 & 224.0958 & -28.6487 & 0.093 & 2021-02-18 & Siding Spring/ANU 2.3m & WiFeS\\
eRASSU J145954.5-260822 & 224.9780 & -26.1409 & 0.090 & 2023-04-02 & La Silla/NTT & EFOSC2\\  
eRASSU J164649.4-692539 & 251.7043 & -69.4272 & 0.016 & 2021-06-05 & Siding Spring/ANU 2.3m & WiFeS\\
1eRASS J190146.6-552200 & 285.4461 & -55.3661 & 0.059 & 2021-03-27 & La Silla/NTT & EFOSC2\\
1eRASS J192128.0-502746 & 290.3667 & -50.4651 & 0.056 & 2021-03-31 & La Silla/NTT & EFOSC2\\
1eRASS J201206.8-442838 & 303.0293 & -44.4783 & 0.119 & 2021-06-05 & Siding Spring/ANU 2.3m & WiFeS\\
1eRASS J231004.4-453925 & 347.5210 & -45.6575 & 0.056 & 2020-08-25 & Siding Spring/ANU 2.3m & WiFeS\\
eRASSt J234403.1-352640 & 356.0126 & -35.4450 & 0.1 & 2021-05-05 & Siding Spring/ANU 2.3m & WiFeS\\ 
\hline\end{tabular}
\tablefoot{RA and DEC are coordinates of the optical counterpart [deg].}
\end{table*}

\FloatBarrier

\section{Description of columns of the eRO-ExTra catalog}
\label{column_descr}

The full description of columns of the eRO-ExTra catalog is provided in Table~\ref{tab:columns} as well as in the available README file (in addition, includes column numbers and notes about several sources). Example columns for the first five sources in the eRO-ExTra catalog are shown in Table~\ref{tab:cat_example}.

\begin{table*}[ht]
    \centering
    \caption{eRO-ExTra catalog column description}
    \label{tab:columns}
    \begin{tabular}{lll}
    \hline
    \hline
         \text{Column} &  \text{Unit}&  \text{Description}\\
        \hline
      \text{ERO\_NAME} &  & \text{source name from DR1 or in the format eRASSU Jhhmmss.s-ddmmss}\\ 
      \text{ERO\_RA} &  \text{deg} & \text{eROSITA Right Ascension (ICRS) from eRASS:5}\\ 
      \text{ERO\_DEC} &  \text{deg} & \text{eROSITA Declination (ICRS) from eRASS:5}\\
      \text{ERO\_FLUX\_ERASS<N>} &  \text{erg/cm$^{2}$/s} & \text{source flux in 0.2--2.3 keV in eRASSN, N =1,2,3,4,(5, if available)}\\
      \text{ERO\_FLUX\_ERR\_ERASS<N>} &  \text{erg/cm$^{2}$/s} & \text{error flux (68 \,\%) in 0.2--2.3 keV in eRASSN, N =1,2,3,4,(5, if available);}\\
      \text{} &  \text{} & \text{$-1 =$ flux upper limit (99.7\,\%)}\\
      \text{Amplitude} &  & \text{amplitude of variability between eRASS1 and eRASS2}\\
      \text{Significance} &  & \text{significance of variability between eRASS1 and eRASS2}\\
       \text{ERO\_LCCLASS} & \text{} & \text{eROSITA light curve class: $1=$decline, $2=$flare, $3=$brightening, $4=$other} \\
      \text{ERO\_DATE} & \text{mjd} & \text{dates of eROSITA observations: (MJD\_1,..,MJD\_<N>),}\\
     \text{} &  \text{} & \text{N =1,2,3,4,(5, if available)}\\

       \text{NH} &  \text{$\mathrm{cm^{-2}}$} & \text{Galactic equivalent neutral hydrogen column density in the line of sight}\\
       \text{ERO\_GAMMA} &  \text{} & \text{photon index for the peak eRASS}\\ 
      \text{ERO\_GAMMA\_ERR} &  \text{} & \text{error photon index (68 \,\%) for the peak eRASS}\\ 
      \text{ERO\_CSTAT\_DOF} &  \text{} & \text{C-statistic per degree of freedom, used to assess the goodness of fit}\\ 
      \text{ERO\_FLUX\_MOD} & \text{erg/cm$^{2}$/s} & \text{source flux in energy band 0.2--2.3 keV for the peak eRASS} \\
      \text{ERO\_FLUX\_MOD\_ERR} & \text{erg/cm$^{2}$/s} & \text{error flux (68\,\%) in energy band 0.2--2.3 keV for the peak eRASS} \\
\text{ARCH\_FLAG} & \text{} & \text{archival X-ray constraints: } \\
      \text{} & \text{} & \text{$0=$not constraining UL, $1=$constraining UL, $2=$previously detected} \\ 
      \text{ARCH\_DATE} & \text{mjd} & \text{dates of N archival observations: (X\_1,..,X\_<N>)}\\
      \text{ARCH\_FLUX} & \text{erg/cm$^{2}$/s} & \text{source flux of N archival observations:(F\_1,..,F\_<N>) in 0.2--2.0 keV} \\
      \text{ARCH\_FLUX\_ERR} & \text{erg/cm$^{2}$/s} & \text{error flux (68\,\%) of N archival observations:(F\_err\_1,..,F\_err\_<N>)} \\
      \text{} & \text{} & \text{in 0.2--2.0 keV; $-1$=flux upper limit (99.7\,\%)} \\
    \text{ARCH\_INSTRUMENT} & \text{} & \text{instruments of N archival observations (Instrument\_1,..,Instrument\_<N>)} \\

\text{NWAY\_bias\_LS10\_Xray\_proba} &  \text{} & \text{probability weighting intruduced by Xray\_proba prior \citep{Salvato2018} }\\
\text{NWAY\_dist\_bayesfactor} &  \text{} & \text{logarithm of the ratio of the prior and posterior from separation,}\\
\text{} &  \text{} & \text{positional error, and number density \citep{Salvato2018} }\\
\text{NWAY\_dist\_post} &  \text{} & \text{distance posterior probability comparing this association vs. no}\\
\text{} &  \text{} & \text{association \citep{Salvato2018}}\\
\text{NWAY\_p\_single} &  \text{} & \text{same as dist\_post, but weighted by the prior \citep{Salvato2018}}\\
\text{NWAY\_p\_any} &  \text{} & \text{for each entry in the X-ray catalog, the probability that there is}\\
\text{} &  \text{} & \text{a counterpart \citep{Salvato2018}}\\
\text{NWAY\_p\_i} &  \text{} & \text{relative probability of the eROSITA/LS8 match \citep{Salvato2018} }\\

\text{LS10\_RELEASE} &  \text{} & \text{integer denoting the camera and filter set used in LS10,}\\ 
\text{} &  \text{} & \text{which is unique for a given processing run of the data}\\ 
\text{LS10\_BRICKID} &  \text{} & \text{LS10 Brick ID}\\ 
\text{LS10\_OBJID} &  \text{} & \text{catalog object number within this LS10 brick}\\ 
      \text{LS10\_RA} &  \text{deg} & \text{J2000 Right Ascension of the LS10 counterpart}\\ 
      \text{LS10\_DEC} &  \text{deg} & \text{J2000 Declination of the LS10 counterpart}\\
      \text{LS10\_RA\_IVAR} &  \text{$\mathrm{1/deg^{2}}$} & \text{inverse variance of RA , excluding astrometric calibration errors}\\
       \text{LS10\_DEC\_IVAR} &  \text{$\mathrm{1/deg^{2}}$} & \text{inverse variance of DEC, excluding astrometric calibration errors}\\
       \text{LS10\_FLUX\_<X>} &  \text{nanomaggy} & \text{LS10 model flux <X>, where X= {\it g,r,i,z},W1,W2}\\
       \text{LS10\_FLUX\_IVAR\_<X>} &  \text{$\mathrm{1/nanomaggy^{2}}$} & \text{Inverse variance of flux <X>, where X= {\it g,r,i,z},W1,W2}\\
       \text{LS10\_MW\_TRANSMISSION\_<X>} &  \text{} & \text{Galactic transmission in <X> filter in linear units [0, 1],}\\
       \text{} &  \text{} & \text{where X= {\it g,r,i,z},W1,W2}\\
       \text{LS10\_TYPE} &  \text{} & \text{morphological model from LS10}\\
      \text{Z\_REDSHIFT} &  \text{} & \text{redshift}\\
      \text{Z\_TYPE} &  \text{} & \text{redshift type: photoz = photometric, specz = spectroscopic}\\
      \text{Z\_REFERENCE} &  \text{} & \text{redshift origin}\\
      \text{SIMBAD\_NAME} &  \text{} & \text{SIMBAD source ID}\\
      \text{SIMBAD\_RA} &  \text{deg} & \text{SIMBAD J2000 Right Ascension}\\
      \text{SIMBAD\_DEC} &  \text{deg} & \text{SIMBAD J2000 Declination}\\
      \text{SIMBAD\_TYPE} &  \text{} & \text{SIMBAD main type}\\
       \text{TNS\_NAME} &  \text{} & \text{Transient Name Server (TNS) source name}\\
      \text{TNS\_RA} &  \text{deg} & \text{TNS Right Ascension}\\
      \text{TNS\_DEC} &  \text{deg} & \text{TNS Declination}\\
      \text{TNS\_DATE} &  \text{mjd} & \text{TNS discovery date}\\

       \text{VLASS\_NAME} &  \text{} & \text{The Very Large Array Sky Survey (VLASS) source name}\\
      \text{VLASS\_RA} &  \text{deg} & \text{VLASS Right Ascension}\\
      \text{VLASS\_DEC} &  \text{deg} & \text{VLASS Declination}\\
      \text{VLASS\_FLUX} &  \text{mJy} & \text{VLASS total flux density at 3GHz}\\

       \text{RACS\_NAME} &  \text{} & \text{Rapid ASKAP Continuum Survey (RACS) source name}\\
      \text{RACS\_RA} &  \text{deg} & \text{RACS Right Ascension}\\
      \text{RACS\_DEC} &  \text{deg} & \text{RACS Declination}\\
      \text{RACS\_FLUX} &  \text{mJy} & \text{RACS total flux density at 0.88GHz}\\
      
      \hline
    \end{tabular}
\end{table*}

\pagestyle{empty}

\begin{landscape}

\begin{table}[]
    \centering
    \caption{Example selection of 11/79 columns for first 5/304 rows of eRO-ExTra catalog. }
\label{tab:cat_example}
\small{
\begin{tabular}{lcccccccccc}

\hline
\hline
  \multicolumn{1}{c}{ ERO\_NAME} &
  \multicolumn{1}{c}{ ERO\_RA} &
  \multicolumn{1}{c}{ ERO\_DEC} &
  \multicolumn{1}{c}{ ERO\_FLUX\_ERASS1} &
  \multicolumn{1}{c}{ Amplitude} &
  \multicolumn{1}{c}{ Significance} &
  \multicolumn{1}{c}{ NWAY\_p\_any} &
  \multicolumn{1}{c}{ LS10\_RA} &
  \multicolumn{1}{c}{ LS10\_DEC} &
  \multicolumn{1}{c}{ Z\_REDSHIFT} &
  \multicolumn{1}{c}{ Z\_TYPE} \\
\hline

  1eRASS J000258.3-401950 & 0.7411 & -40.3308 & 1.70  & 4.67 & 6.47  & 0.99 & 0.7409 & -40.3301 & 0.144 & specz\\
  1eRASS J001053.3-334021 & 2.7223 & -33.6724 & 6.00  & 4.98 & 5.05  & 0.88 & 2.7227 & -33.6723 & 0.297 & photoz\\
  eRASSU J001308.6-462524 & 3.2856 & -46.4232 & 0.64 & 9.12 & 6.08  & 0.33 & 3.2844 & -46.4220 & 0.198 & specz\\
  1eRASS J003902.6-611323 & 9.7615 & -61.2232 &  2.41 & 6.20 & 4.21 & 0.99 & 9.7618 & -61.2234 & 0.301 & photoz\\
  1eRASS J004058.3-683816 & 10.2425 & -68.63816 & 6.46 & 11.06 & 8.71 &  0.99 & 10.2434 & -68.6388 & 0.151 & specz\\
\hline\end{tabular}}

\tablefoot{The table includes eROSITA names (ERO\_NAME) with the X-ray detection coordinates (ERO\_RA, ERO\_DEC), eRASS1 flux ERO\_FLUX\_ERASS1 (in the units of $10^{-13}$ erg $\mathrm{cm^{-2}s^{-1}}$) and Significance and Amplitude of variability between eRASS1 and eRASS2. The table also shows several columns describing NWAY/LS10 optical counterparts: NWAY/LS10 coordinates (LS10\_RA, LS10\_DEC) with their probabilities (NWAY\_p\_any), redshifts (Z\_REDSHIFT) and redshift types (Z\_TYPE).  The full list of columns and their description can be found in Table~\ref{tab:columns}.}
\end{table}

\end{landscape}

\pagestyle{plain}

\section{Radio luminosity calculation}
\label{appendix_sfr}
To estimate radio luminosities at 1.4GHz, firstly, we calculated K-corrected radio luminosity at a frequency available in used radio surveys ( $\mathrm{\nu_{RACS}=0.88}$GHz, $\mathrm{\nu_{VLASS}=3.0}$GHz):
\begin{equation}
    \mathrm{L_{\nu} = 4\pi D_L^{2}S_{\nu}(1+z)^{-\alpha -1}},
\end{equation}
where radio spectral index $\alpha = -0.7$ for galaxies \citep{1986A&A...168...17E}, $\mathrm{D_L}$ is luminosity distance, z is redshift and $\mathrm{S_{\nu}}$ is flux density. Then, we converted $\mathrm{L_{\nu}}$ to $\mathrm{L_{1.4GHz}}$:
\begin{equation}
    \mathrm{L_{1.4GHz} = L_{\nu} (1.4GHz/\nu)^{\alpha}}.
\end{equation}
Finally, the threshold for a radio luminosity corresponding to the expectation from star formation (<20 M\textsubscript{\(\odot\)}$\mathrm{yr^{-1}}$; \citealt{1983ApJ...272...54K}) was estimated using the $\mathrm{SFR-L_{1.4GHz}}$ relation from \citet{2011ApJ...737...67M}:
\begin{equation}
    \mathrm{\frac{L_{1.4GHz}}{ergs^{-1}Hz^{-1}}=1.5\times10^{28}\frac{SFR_{1.4GHz}}{M_{\sun}yr^{-1}}}.
\end{equation}

\section{X-ray luminosity function and rates}
\label{sec:xlf}

The selected sample can be used to statistically describe the population of non-AGN transients and variables. We used the classical maximum observable volume ($\mathrm{1/V_{max}}$) approach \citep{1968ApJ...151..393S} to estimate the X-ray luminosity function (XLF). 

In order to estimate the comoving volume $\mathrm{V_{max}}$, first, we estimated the maximum observable luminosity distance $\mathrm{D_{max}}$ for each source, corresponding to the maximum distance at which a given transient can be detected. Several factors affect the detectability of sources in the eROSITA all-sky survey, such as the position in the sky, spectral shape, redshift and intrinsic luminosity. 

eROSITA has a nonhomogeneous sensitivity, which reaches the maximum at the ecliptic poles, as shown in Fig.~\ref{fig:hemispheres1}. Therefore, the detectability depends on the ecliptic latitude ($\mathrm{b_{ecl}}$). To account for this, for each candidate, we computed the minimum flux $\mathrm{F_{min}(b_{ecl}})$  required for a detection at a certain latitude according to the applied selection criteria, namely, amplitude cut ($\mathrm{A > 4}$) and detection likelihood cut (det\_like>15). For this, we created sensitivity maps using eSASS command \texttt{ersensmap} in the  energy band 0.2--2.3 keV for det\_like = 15 and also det\_like = 5, which is the minimum detection likelihood of sources included in eROSITA source catalogs. The former map was directly used to extract the minimum flux for a source to be detected with det\_like = 15. The minimum flux needed to satisfy the amplitude cut was estimated as $\mathrm{A \times F_{min}(b_{ecl})_{det\_like=5}}$ for sources with detections in both eRASS1 and eRASS2, and $\mathrm{A \times F(b_{ecl})_{UL}}$, where $\mathrm{F(b_{ecl})_{UL}}$ is the upper limit of eRASS1 or eRASS2. The minimum flux required for the amplitude cut is more constraining than that for the detection likelihood cur for all sources, and was, therefore, used for further steps.

Next, we estimated the maximum redshift $\mathrm{z_{max}}$ and the resulting $\mathrm{D_{max}}$, at which a TDE with a given intrinsic L (rest-flame 0.2--6.0 keV) in an X-ray spectral shape could be detected according to the selection criteria. With pyxspec, we faked a spectrum with the best-fit parameters of the \texttt{tbabs*zpow} model of the peak eROSITA epoch with the corresponding redshift, ARF and RMF files. We fixed the intrinsic luminosity using the normalization of the model to find the largest $\mathrm{z_{max}}$ for a detection above $\mathrm{F_{min}}$. For a fraction of sources, only photometric redshifts are available (see Sect.~\ref{sec:redshifts}). Therefore, it is necessary to account for their errors and photo-z probability distribution function (PDZ) shapes, as they may significantly affect the values of the intrinsic luminosity used in the computation. For this, for each source we drew a sample of 100 redshift values from the corresponding PDZ and performed the $\mathrm{D_{max,z}}$ estimation for each value. The final $\mathrm{D_{max}}$ was calculated as a mean of $\mathrm{D_{max,z}}$.

Finally, to calculate the XLF, we summed the derived $\mathrm{1/V_{max}}$ values of individual sources into luminosity bins. We chose 10 equal bins between 41.5 and 46.5. The uncertainty for each bin is estimated as $\mathrm{\sqrt{\Sigma(1/V_{max})^2_i}}$ with the summation over the objects in a particular bin. Finally, the function should have been corrected by a factor of 2.6, since the sample was selected from the LS10 covered area (76\%, see Fig.~\ref{fig:hemispheres}) in the eROSITA\_DE hemisphere. The search identified sources peaking in eRASS1 or eRASS2, which accounts for a year of observations. Therefore, no correction was needed to evaluate the volume per year.

\begin{figure}
\centering
    \includegraphics[width=\linewidth]{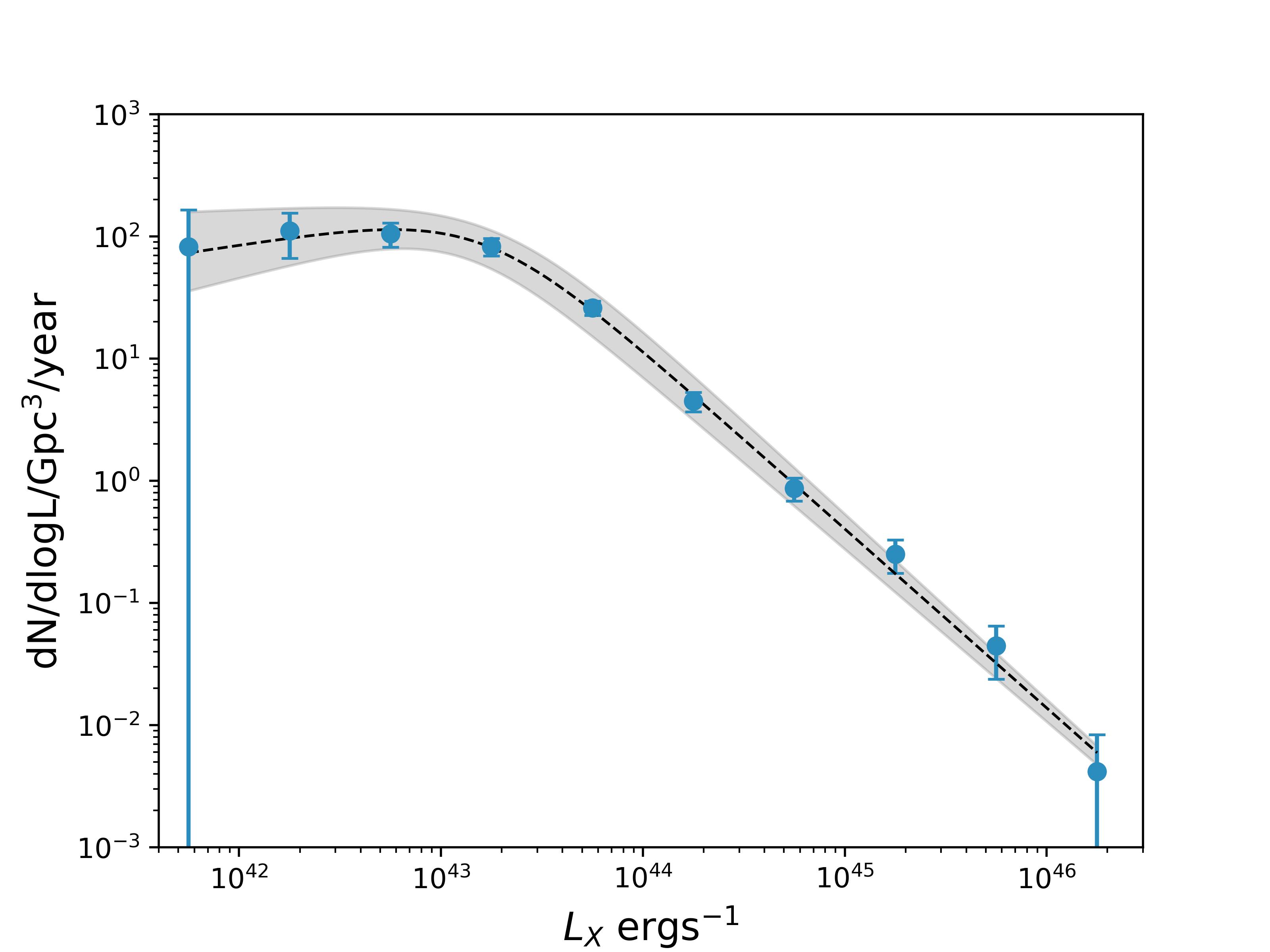}  
\caption{XLF function for eRO-ExTra catalog in terms of rest frame 0.2-6.0 keV luminosity. The dashed black line shows the double power law fit (Eq.~\ref{eq:xlf}) with the 1-$\sigma$ uncertainty shown in gray. }
\label{fig:xlf}
\end{figure}

The final XLF was computed for the eRO-ExTra subsample of 281 sources, which have credible counterparts ($\mathrm{p\_any>0.17}$) and photo-z measurements with at least 2 LS10 parameters $\mathrm{ANYMASK\_X=0}$, where X = {\it g,r,i,z}. The XLF is presented in Fig.~\ref{fig:xlf}. Fitting the XLF with a double power law model, we obtained:
\begin{equation}
\label{eq:xlf}
    f(L) = (1.7^{\pm 0.4})10^{-7} Mpc^{-3}year^{-1} \times \left( \left( \frac{L}{L_{\text{br}}} \right)^{-0.25^{\pm0.18}} + \left( \frac{L}{L_{\text{br}}} \right)^{1.47^{\pm0.05}} \right)^{-1},
\end{equation}
where $\mathrm{L_{br} = (1.6 \pm 0.4)\times 10^{43}ergs^{-1}}$.
Using the equation above, the integrated volumetric rate of eRO-ExTra transients and variables is $\mathrm{1.8^{+0.5}_{-0.4}\times10^{-7} Mpc^{-3}year^{-1}}$.

\end{appendix}
\end{document}